\documentclass[12pt]{article}
\usepackage[dvips]{graphicx}
\usepackage{amssymb,amsmath}
\usepackage[english]{babel}
\usepackage{bm}
\usepackage{epsfig}
\usepackage{graphicx,colordvi}

\newcommand{\der}[2]{{\frac{d #1}{d #2}}}

\def\be{\begin{equation}}
\def\ee{\end{equation}}
\def\bea{\begin{eqnarray}}
\def\eea{\end{eqnarray}}
\def\l{\label}
\def\bfr{{\bf r}}
\def\p{{\bf p}}

\def\om{\omega}

\def\siml{\;\hbox{\kern.1em \lower.7ex \hbox{$\sim$} \kern-1.12em
 \raise.5ex \hbox{$<$} \kern.1em}}
\def\simg{\;\hbox{\kern.1em \lower.7ex \hbox{$\sim$} \kern-1.12em
 \raise.5ex \hbox{$>$} \kern.1em}}

\def\d{\hbox{d}}
\def\vareps{E}

\def\l{\label}
\def\bfr{{\bf r}}
\def\p{{\bf p}}

\def\om{\omega}

\def\Re{\mathop{\rm Re}}

\def\erf{\mathop{\rm erf}\nolimits}

\def\eq#1{(\ref{#1})}

\newcommand{\cA}{\mathcal{A}}
\newcommand{\cK}{\mathcal{K}}

\newcommand{\cZ}{\mathcal{Z}}

\begin{document}

\centerline{\Large \bf SHELL STRUCTURE AND ORBIT BIFURCATIONS}
\centerline{\Large \bf IN FINITE FERMION SYSTEMS}

\bigskip

\centerline{\bf {A. G. Magner$^{1}$, I. S. Yatsyshyn$^{1}$,
 K. Arita$^{2}$, M. Brack$^{3}$}}

\bigskip

{\it $^{1}$Institute for Nuclear Research, 03680, Kyiv, Ukraine}

{\it $^{2}$Physic Department, Nagoya Institute
of Technology, 466-8555, Nagoya, Japan}

{\it $^{3}$Institute for Theoretical Physics,
 University of Regensburg, D-93040, Regensburg, Germany}

\begin{abstract}

We first give an overview of the shell-correction method 
which was developed by V. M. Strutinsky as a practicable and efficient 
approximation to the general selfconsistent theory of finite fermion systems 
suggested by A. B. Migdal and collaborators. Then we present in more 
detail a semiclassical theory of shell effects, 
also developed by Strutinsky following original ideas of M. Gutzwiller. We emphasize,
in particular, the influence of orbit bifurcations on shell structure. We first give 
a short overview of semiclassical trace formulae, which connect the shell oscillations 
of a quantum system with a sum over periodic orbits of the corresponding classical 
system, in what is usually called the ``periodic orbit theory''. We then present
a case study in which the gross features of a typical double-humped nuclear fission 
barrier, including the effects of mass asymmetry, can be obtained in terms of the 
shortest periodic orbits of a cavity model with realistic deformations relevant for 
nuclear fission. Next we investigate shell structures in a spheroidal cavity 
model which is integrable and allows for far-going analytical computation. We show, 
in particular, how period-doubling bifurcations are closely connected to the existence 
of the so-called ``superdeformed'' energy minimum which corresponds to the fission isomer
of actinide nuclei. Finally, we present a general class of radial power-law potentials 
which approximate well the shape of a Woods-Saxon potential in the bound region, give 
analytical trace formulae for it and discuss various limits (including the harmonic 
oscillator and the spherical box potentials).
  
\end{abstract}



\bigskip

\centerline{\today}

\newpage

\section{INTRODUCTION}

This paper is devoted to the memory of A. B. Migdal. Our first aim here 
is to review the shell-correction method (SCM) which was first introduced 
by Strutinsky on a phenomenological basis \cite{str2} and then
microscopically founded \cite{buna} on Migdal's theory for strongly 
interacting finite fermion systems \cite{migdal}.  Our second aim is the 
discussion of a semiclassical theory of shell effects, using the so-called 
periodic orbit theory (POT) (see \cite{book} for an introductory text book). 
It provides us with a nice tool for answering, sometimes even analytically, 
some fundamental questions asked \cite{strd,strphyschem_iaea_ws,strpramana} 
by Strutinsky: Why are nuclei deformed? 
What are the physical origins of the double-humped fission barrier and, 
in particular, of the existence of the isomer minimum? His idea was to use 
the POT for a deeper understanding, based on classical pictures, of the 
origin of nuclear shell structure and its relation to a possible chaotic 
nature of the nucleons' dynamics. We shall present some applications of the 
POT to nuclear deformation energies and discuss in more detail the relation 
of bifurcations of periodic orbits with pronounced shell effects.

\begin{figure}
\begin{center}
\epsfig{
figure = 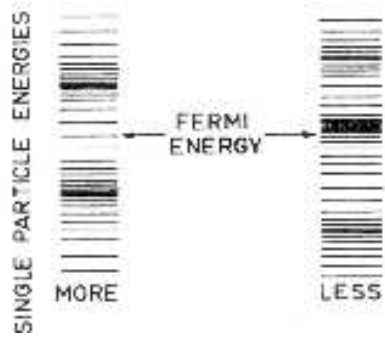,width = 0.3\textwidth,clip}
\end{center}
\caption{
Schematic picture showing the relation of local level density
at the Fermi energy to the binding of the system.
Left side: Low level density -- more bound; right side: high level
density -- less bound.
}
\label{figspec}
\end{figure}

\begin{figure}
\begin{center}
\epsfig{
figure = 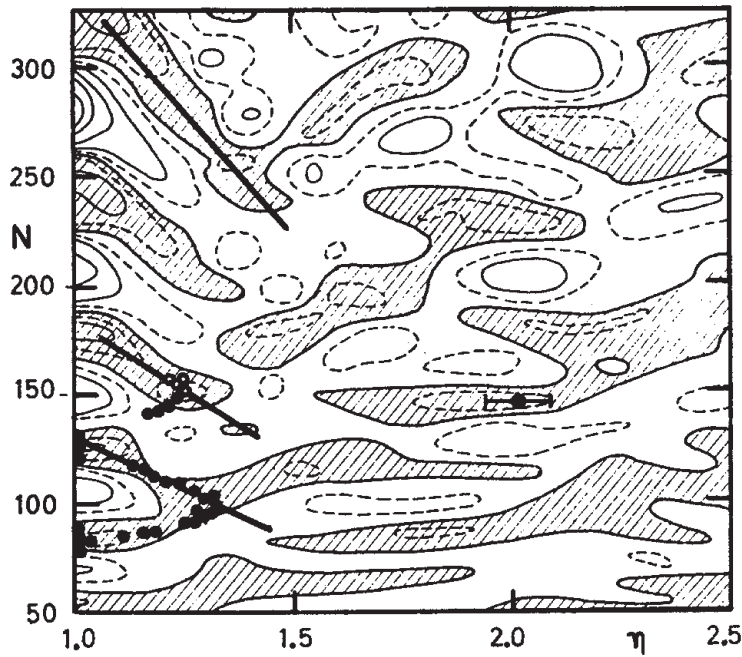,width = 0.55\textwidth,clip}
\end{center}
\caption{
Neutron energy shell-correction $\delta E_n(N,\eta)$ for a realistic nuclear
Woods-Saxon potential (with spin-orbit term), versus neutron
number $N$ and axis ratio $\eta$ of the spheroidally deformed potential
(equidistance: 2.5 MeV, areas with negative values are shaded). Dots indicate
experimental deformations. The heavy bars are the semiclassical predictions for
the loci of the ground-state minima using the leading classical periodic orbit
families in a spheroidal cavity.
}
\label{figshells}
\end{figure}

According to the SCM, the oscillating part of the total energy of a finite 
fermion system, the so-called shell-correction energy $\delta E$, is associated 
with an inhomogeneity of the single-particle (s.p.) energy levels near the Fermi 
surface. Its existence in dense fermion systems is a basic point of Landau's 
quasi-particle theory of infinite Fermi liquids \cite{landau,abrkhal}, as 
extended to self-consistent finite fermion systems by Migdal and  collaborators 
\cite{migdal,khodsaper}. This is schematically illustrated in Fig.\ \ref{figspec}, 
where the s.p.\ level spectrum of a bound nucleus is shown in two extremal 
situations. Depending on the level density at the Fermi energy -- and with 
it the shell-correction energy $\delta E$ -- being a maximum or a minimum, 
the nucleus is particularly unstable or stable, respectively. This
situation varies with particle numbers and deformations of the 
nucleus. In consequence, the shapes of stable nuclei depend strongly 
on particle numbers and deformations. This is illustrated in Fig.\
\ref{figshells}. Here the shell correction $\delta E$ of the neutrons is 
shown as function of the neutron number $N$ and the deformation parameter 
$\eta$ of a Woods-Saxon potential \cite{dws} with spheroidal shape,
$\eta$ being the ratio of the semi-axes. If we fix the neutron number 
$N$, e.g.\ $N=150$, and increase the deformation $\eta$, we meet the 
first minimum (ground state) at about $\eta \sim 1.25$ and the next one 
(isomeric state) at much larger deformations $\eta \sim 1.9-2.1$. 
The experimental data corresponding to these deformations are shown in 
Fig.\ \ref{figshells} by the heavy dots.   

The SCM was successfully used to describe nuclear masses and deformation
energies and, in particular, fission barriers of heavy nuclei. For an 
early review by Strutinsky's group, in which also the miscroscopic 
foundations of the SCM are discussed, see Ref.\ \cite{fuhi}. 
(We refer to Sect.\ 3.2 for a further discussion of fission barriers.)

In Sect.\ 2, we will give a short review of the SCM and its foundation on
the basis of a selfconsistent theory of finite interacting fermion systems.

Sect.\ 3 is devoted to the semiclassical theory of shell effects.
The POT is based on Gutzwiller's semiclassical trace formula for the level 
density for a Hamiltonian system with isolated orbits \cite{gutz} and its 
extensions to systems with continuous symmetries 
\cite{book,strd,babl,strm,bt,crlit}. It allows one to relate both 
the oscillating part of the level density and the shell-correction 
energy of a quantum system to the shortest periodic orbits (POs) 
of the corresponding classical Hamiltonian system. Thus, one can often
explain pronounced shell effects by the role of particular short POs.
As an early example, taken from Ref.\ \cite{strd}, the heavy bars in Fig.\ 
\ref{figshells} are the predictions of the POT for the loci of the 
ground-state minima, using the shortest POs in a spheroidal cavity.
Bifurcations of POs under the variation of a deformation parameter or the
(Fermi) energy can have noticeable effects for the shell structure 
\cite{strd,strphyschem_iaea_ws,strpramana,ellipse,spheroid_pr,spheroid,maf}.

In Sect.\ 3.1, we will present the structure of semiclassical trace formula 
and discuss a general method of treating bifurcations in the POT, using the 
catastrophe theory of Fedoryuk and Maslov for caustic and turning-point 
problems \cite{fedoryuk_pr,maslov,fedoryuk_book1,fedoryuk_book2}.

In Sect.\ 3.2, we review the semiclassical description \cite{brreisie} of a 
typical nuclear fission barrier in terms of the shortest periodic orbits, 
employing a cavity model with the realistic shape parameterization developed 
in \cite{fuhi}. In particular, the effect of left-right asymmetric deformations 
on the height of the outer fission barrier will be discussed. Isochronous 
bifurcations of the shortest orbits are treated here in a uniform approximation 
employing a suitable normal form for the action function.

In Sect.\ 3.3, we use the spheroidal cavity \cite{strd,spheroid_pr,spheroid,spheroidgrsh}
as a simple integrable model that allows to study semiclassically 
the shell structure related to the 'super-deformed' energy minimum 
which in realistic actinide nuclei corresponds to the fission isomers. 
Strutinsky's prediction concerning the importance of the enhancement of the 
shell structure owing to period-doubling bifurcations of three-dimensional 
POs from simple equatorial (EQ) orbits in the spheroidal cavity model 
\cite{strd} will be discussed. 

In Sect.\ 3.4, we shall study a radial power-law potential \cite{aritapap}, 
which is a good approximation to the familiar Woods-Saxon potential for nuclei 
in the spatial domain where the particles are bound. We shall establish 
generalised trace formulae for this potential and discuss various limits to 
other known potentials. 

The paper is summarized in Sect.\ 4, where we also present some conclusions and
plans for future research. Some technical details of our POT calculations are 
given in the Appendix.

\newpage

\section{THE SHELL-CORRECTION METHOD}

In 1966, Strutinsky achieved a far-reaching break-through \cite{str1}, 
following basically Migdal's theory of finite fermion systems\cite{buna,migdal}. 
Until then, many attempts had been made to incorporate quantum shell effects 
in the calculation of nuclear deformation energies. But they all
failed in reproducing the fission barriers of actinide nuclei and details such 
as, e.g., the left-right asymmetry of the nascent fission fragments 
\cite{joha}. Summing the s.p.\ energies of a deformed shell model (like the 
Nilsson model \cite{nils}) up to the Fermi energy failed at larger deformations. 
There was a need to renormalize the wrong average part of the s.p.\ energy 
sum. Knowing that the smooth part of the nuclear binding energy could be well 
described by the phenomenological LDM \cite{ldm,cosw,wilf} (or droplet 
model \cite{dm}), Strutinsky wrote the total nuclear energy as \cite{str2,fuhi}
\begin{equation}
E_{tot} = E_{LDM} + \delta E\,,
\label{etot}
\end{equation}
where $E_{LDM}$ is the LDM energy and $\delta E$ the so-called
{\it ``shell-correction energy''} which contains the fluctuating part 
of the s.p.\ energy sums for the neutrons:
\begin{equation}
\delta E = \sum_{n=1}^{N/2} E_n - \langle \sum_{n=1}^{N/2} E_n \rangle\,,
\label{dele}
\end{equation}
and similarly for the protons. 
Both parts of the total energy (\ref{etot}) depend on the neutron and proton
numbers $N$ and $Z$, as well as on the nuclear deformation which has to be 
suitably parameterized both in the LDM and the shell model. While it had
been a wide-spread belief that the shell-correction $\delta E$ was important 
only for spherical nuclei and would vanish at larger deformations, Strutinsky
was convinced that shell effects play an important role also at larger
deformations and lead, in fact, to new magic numbers corresponding
to deformed systems with increased local stability \cite{str2}. 

For the determination of the smooth parts $\langle \sum_{n=1}^{N/2} E_n 
\rangle$ in (\ref{dele}), Strutinsky designed a very ingenious averaging method 
\cite{str2,str1} which has been termed the {\it ``Strutinsky averaging} 
(or {\it smoothing}) {\it method''}. It consists of a Gaussian convolution of 
the s.p.\ energy spectrum, modified by the so-called {\it ``curvature corrections''} 
in such a way that the result does not depend on the energy averaging width 
$\gamma$ (at least within a finite interval of $\gamma$ of the order of the 
main shell spacing) and at the same time reproduces the true average 
level density which is found, e.g., from the extended Thomas-Fermi (ETF) model, 
or from the Weyl expansion in the case of billiard or cavity models (see, e.g., 
\cite{book}, Chap.\ 4, for details).

Strutinsky applied the SCM to the calculation of fission barriers\cite{str2,str1}, 
employing 
the Nilsson model. For a typical actinide nucleus he obtained a second minimum 
in the deformation energy, lying above the ground state by about 3 MeV, at a 
deformation much larger than that of the ground state. This was, in fact, the 
physical explanation of the {\it fission isomer} which had been known 
experimentally \cite{poli} since 1962, but not understood theoretically. 
Strutinsky presented this result at the Symposium {\it ``Nuclides 
far off the stability line''} in Lysekil (Sweden) \cite{lyse} in 1966 and, 
being a member of Migdal's theory group, he immediately became famous. He was 
then invited to the NBI in Copenhagen, in order to extend his calculations of 
fission barriers on a larger scale, employing more realistic shell-model 
potentials, which led to the team work published in Ref.\ \cite{fuhi}.

The SCM was soon taken up by many groups of the international scientific community 
and often also called the {\it ``microscopic-macroscopic''} method \cite{dm,swia}. 
Various combinations of nuclear shell models and liquid drop(let) models were
used. Still today, the shell-correction method is being used world wide
for calculations of nuclear masses and deformation energies, and it persists
to yield the most accurate nuclear mass tables, ground-state and isomeric 
deformations and, in particular, fission barriers.

The Strutinsky averaging used for the second term in (\ref{dele}) is done in 
energy space and leads, within a sufficient numerical accuracy, to the same 
results as the ETF model (\cite{sob}, see also \cite{book}, Ch.\ 4.7).
However, it is formally not completely consistent with the particle-number
averaging that is implicitly done in the standard least-square fits to the
LDM which defines the smooth part of the total nuclear energy (\ref{etot}).
Already early, alternative particle-number averagings were investigated 
in \cite{ivan}. Using semiclassical POT arguments, the difference between 
energy and particle-number averaging was understood \cite{pomo,schuck} as a 
symmetry correction which becomes especially significant for spherical nuclear 
shapes (particularly in the harmonic oscillator model). Smaller discrepancies 
for the deformed Fermi systems in the resulting shell-correction energies persist, 
however, as discussed in \cite{pomo,schuck,lero}. This point is thus still an 
object of current debate.

The decomposition of the total nuclear energy into a smooth and an oscillating
part in (\ref{etot}) may at first glance look like a rather phenomenological 
ansatz. In particular, since the oscillating part is taken from the sum of occupied 
s.p.\ energies of the nucleons, one may argue that it cannot be correct, 
since this sum is well known not to represent the total energy in a self-consistent 
microscopic theory (where it double-counts the potential energy if the interaction 
does not depend on the density). However, it was soon realized that (\ref{etot}) is 
nevertheless correct even within a self-consistent microscopic theory. This was 
pointed out, amongst others, by Bethe \cite{beth} who therefore termed (\ref{etot}) 
the {\it ``Strutinsky energy theorem''}. It can be rigorously proved that the 
oscillating part of the correct total energy is, indeed, contained in the 
s.p.\ sums; the proof is simply based on the variational principle that 
governs the self-consistent mean-field theories (see, e.g., \cite{book}, 
App.\ A.3), and it applies also to density-dependent nuclear interactions \cite{beth}. 
This was soon demonstrated by the Strutinsky group \cite{buna} to hold also 
within the Migdal theory \cite{migdal}. At this point it may be worth mentioning
that the Migdal theory has been used explicitly to calculate energy
shell corrections for doubly magic nuclei, investigating the so-called 
``lead anomaly'' (i.e., in particular for nuclei around $^{208}$Pb) \cite{werner}.

The Strutinsky energy theorem (\ref{etot}) was later tested numerically using the 
Hartree-Fock (HF) approach with effective Skyrme interactions. By extracting a 
self-consistently Strutinsky-averaged part of the total HF energy (which is 
ideally represented by the LDM energy $E_{LDM}$), the first-order oscillating 
term was, indeed, found to be correctly represented by the shell-correction energy 
(\ref{dele}) evaluated in terms of the energy spectrum of the {\it averaged} 
self-consistent HF potentials (which are ideally represented by the shell-model 
potentials). The remaining higher-order terms were found to be relatively small, 
less than $\sim$ 1 MeV in all (not too small) nuclei \cite{bq}. 

The shell-correction method is thus a well-established practical approximation 
to a self-consistent microscopic theory and applicable to any bound many-fermion 
system. It has, e.g., been applied to metal clusters by many groups (see 
\cite{yala,reif,frpa,pame} for a few representative references). For semiconductor 
quantum dots, the Strutinsky energy theorem and shell-corrections to the Coulomb 
interaction energy were discussed in \cite{qdot}.

\section{SEMICLASSICAL THEORY OF SHELL STRUCTURE}

\subsection{SEMICLASSICAL TRACE FORMULAE}

In the semiclassical trace formula of Gutzwiller \cite{gutz} that
connects the quantum-mechanical 
density of states with a sum over POs of the classical system, divergences 
arise at critical deformations where bifurcations of POs occur or 
where symmetry breaking (or restoring) transitions take place. At these points
the standard stationary-phase method (standard SPM or SSPM), 
used in the semiclassical 
(asymptotical) evaluation of the trace integrals, breaks down. Various ways of
avoiding these divergences have been suggested using uniform 
approximations (see \cite{book}, Sec.\ 6). Presently, we shall discuss
the evaluation of trace integrals in the phase-space representation, 
following mainly the presentations in \cite{ellipse,spheroid_pr,spheroid,maf}.

The essence of a semiclassical trace formula is to relate the oscillating
part of the level density of a quantum Hamiltonian system to the periodic 
orbits of the corresponding classical system. The level density 
$g(\vareps)$ can be obtained from the semiclassical Green's function by 
taking the imaginary part of its trace in phase space variables, 
see \cite{maf}, also the references therein,
\bea\l{pstrace}
g(\vareps)&=& \sum_n\delta\left(\vareps-
\vareps^{}_n\right)\simeq
\Re\sum^{}_{CT} \int \frac{\d\bfr'\d\p''}{(2\pi\hbar)^{3}}\,
 \delta\left[\vareps-H(\bfr'',\p'')\right]
\nonumber\\
&\times& \left|{\cal J}^{}_{CT}\right|^{1/2}\,
 \exp\left[\frac{i}{\hbar}\,\Phi^{}_{CT}
-i\frac{\pi}{2} \mu^{}_{CT}  \right],
\eea
where the delta function imposes the energy conservation of the particle 
motion along the classical trajectory (CT) in the potential well of the 
Hamiltonian $H(\bfr,\p)$. $\Phi^{}_{CT}$ is the action phase,
\be\l{actionphase}
\Phi^{}_{CT} = S^{}_{CT}(\bfr',\bfr'',\vareps)
   -\p''\cdot(\bfr''-\bfr')\;;\qquad
 S_{CT}(\bfr',\bfr'',\vareps)
 = \int_{\bfr'}^{\bfr''}\d\bfr\cdot\p(\bfr)\,.
\ee
The second equation yields the action along the CT. The Maslov phase 
$\mu_{CT}$ is 
determined by the number of caustic and turning points within the catastrophe 
theory by Fedoryuk and Maslov \cite{maf,fedoryuk_pr,maslov,fedoryuk_book1}. In 
\eq{pstrace},
${\cal J}^{}_{CT}(\p_\perp',\p_\perp'')$ is the Jacobian of the transformation 
from the momentum perpendicular to the CT $\p_\perp'$ at 
the initial point $\bfr'$ to the perpendicular momentum $\p_\perp''$ at the
final point $\bfr''$ of the CT. One may divide the
level density into a smooth and an oscillating part:
\be\l{glev}
g(\vareps) = {\widetilde g}\,(\vareps) + \delta g(\vareps)\,.
\ee
The smooth part ${\widetilde g}\,(\vareps)$ is, to leading order in $1/\hbar$, 
given \cite{bermo} by the direct (zero-time or zero-length) trajectories which
yield the Thomas-Fermi (TF) results. Except for one-dimensional systems, higher 
$\hbar$ corrections contribute to it which may be calculated within the 
ETF model (or the Weyl expansion for billiards) \cite{book}, so that
we can make the identification
\be\l{denssplit}
{\widetilde g}\,(\vareps) = g^{}_{ETF}(\vareps)\,. 
\ee

For the calculation of the oscillating component $\delta g(\vareps)$,
we apply the stationary phase method to the  integration over the phase 
space variables. The stationary phase condition reads:
\be\l{spmcond}
 \left(\frac{\partial \Phi^{}_{CT}}{\partial \p''}\right)^{\ast}
 \equiv  \left(\bfr'-\bfr''\right)^\ast=0, \qquad
\left(\frac{\partial \Phi^{}_{CT}}{\partial \bfr'}\right)^{\ast}
 \equiv -\left(\p'-\p''\right)^{\ast} =0\;,
\ee
and is nothing but a condition for CT to be periodic. In the presence
of continuous symmetries, the stationary points form a family of
periodic orbits (POs) which cover a $(\cK$$+$$1)$-dimensional submanifold 
$\Upsilon_{po}$ of phase space (denoted $\Gamma_{po}$ in
Ref.\ \cite{book} and further references cited therein), whereby
$\cK$ is the degeneracy of the PO family.  
The integration over $\Upsilon_{po}$ must be performed exactly.
In integrable systems, it is advantageous to transform phase space variables 
from Cartesian to action-angle variables (see, e.g., \cite{bt,crlit}).
Then the action $\Phi$ is a function only of the action variables, 
and the integrations over the cyclic angle variables are 
exactly carried out.
Integrating over the remaining action variables using the standard SPM
(under the existence of additional symmetries like SU(3) or O(4), one of 
which can also be performed exactly), one obtains the so called Berry-Tabor 
trace formula \cite{bt}.

For solving bifurcation problems in integrable and 
non-integrable systems, more exact integrations are required.
In the SPM, after performing the exact integrations over 
$\Upsilon_{po}$, one uses an expansion of
the action phase $\Phi_{\rm CT}$ in phase space variables
$\xi=\{\bfr',\p''\}_\perp$ perpendicular to $\Upsilon_{po}$ in the
integrand of (\ref{pstrace}) over $\xi$ near the stationary point
$\xi^\ast$, 
\be\l{ispmexp}
\Phi_{\rm CT}(\xi)=\Phi^{}_{po} + \frac12\Phi_{\rm po}''(\xi^\ast) 
(\xi-\xi^{\ast})^2 + \frac16\Phi_{po}'''(\xi^\ast) (\xi-\xi^{\ast})^3+
\dots,\quad
\Phi_{po}=\Phi^\ast_{\rm CT}=\Phi_{\rm CT}(\xi^\ast)\;.
\ee
To demonstrate the key point of our derivations of the trace formula, 
we consider here only one (one-dimensional) variable, called $\xi$
again, from the  phase space integration variables in (\ref{pstrace}), 
on which we meet a bifurcation (catastrophe) point in applying the SPM.
(We shall give comments if this might lead to a misunderstanding.)
In the standard SPM, the above expansion is truncated at the 2nd order term
and the integration over the variable $\xi$ is extended to $\pm\infty$.
The integration can be performed analytically and yields a Fresnel integral,
see e.g.\ \cite{book}, Sect.\ 2.7, Eq.\ (2.169).

However, one meets singularities using the standard SPM which are related to 
zeros or infinities of $\Phi_{po}''(\xi^\ast)$ 
(or of eigenvalues of the corresponding 
matrix in the case of several integration variables $\xi$) 
while $\Phi_{po}'''(\xi^\ast)$ remains finite in the simplest case. 
These singularities occur when a PO (isolated or family) 
undergoes a bifurcation at the stationary point $\xi^\ast$ under the 
variation of some parameter (e.g., energy or deformation). The 
Fresnel integrals 
of the standard SPM sketched above will then diverge. In order to avoid such 
singularities, we observe that the bifurcation problem is similar to the 
caustic 
singularity considered by Fedoryuk within the catastrophe theory 
\cite{fedoryuk_pr,fedoryuk_book1}, adopted to its specific position at the 
edge 
of the phase-space volume accessible for the classical motion 
(see also Appendix 
A in \cite{maf}). Therefore, we employ what we call the ``improved SPM'', in 
short: ISPM \cite{ellipse,spheroid_pr,spheroid,maf}. 
Hereby the integration over $\xi$ in (\ref{pstrace}) 
is restricted 
to the {\it finite limits} defined by the classically allowed phase space 
region through the energy-conserving delta function  
in the integrand of \eq{pstrace}. The expansion 
\eq{ispmexp} of the action phases and similarly of the amplitudes in (3) 
is generally used up to the second- and zero-order 
terms, respectively, and if necessary, to higher order terms 
in $\xi -\xi^*$.

In the simplest version of ISPM, the expansion of the phase is truncated
at 2nd order, keeping the finite integration limits $\xi_{-}$ and $\xi_{+}$ 
given by the accessible region of the classical motion in (\ref{pstrace}).  
It will lead to a factor like 
\begin{gather}
e^{i\Phi_{\rm po}/\hbar}
\int_{\xi_{-}}^{\xi_{+}} \exp\left[\frac{i}{2\hbar}\Phi_{po}''\; 
(\xi-\xi^{\ast})^2\right]
\mbox{d} \xi 
\propto \frac{1}{
\sqrt{\Phi_{po}''}}\; 
e^{i\Phi_{po}/\hbar}\;
\erf\left[\cZ_{-},\cZ_{+}\right],
\label{errfuns}
\end{gather}
where $\erf(z_1,z_2)$ is the generalized error function with complex arguments
\begin{equation}
\erf(z_1,z_2)=\frac{2}{\sqrt{\pi}}\int_{z_1}^{z_2}e^{-z^2}{\mbox d}z, \quad
\mathcal{Z}_{\pm}=\left(\xi_{\pm}-\xi^*\right)
\sqrt{-\frac{i}{2\hbar}\Phi_{po}''}\;.
\label{ispmlimits}
\end{equation}
Note that the above expression (\ref{errfuns}) has no divergence at the 
bifurcation point where $\Phi_{po}''(\xi^\ast)=0$, since the error function 
(\ref{ispmlimits}) also goes to zero linearly in $\sqrt{\Phi_{po}''}$
[cf.\ the second equation in 
(\ref{ispmlimits})], which keeps the result finite. 
[For the case of several variables $\xi$ for which we find zeros or 
infinities of eigenvalues of the matrix $\Phi_{po}''(\xi^\ast)$, we 
diagonalize this matrix and reduce the Fresnel-like integrals to products of 
error functions similar to (\ref{errfuns}).]

This procedure is proved to be valid in the semiclassical limit by the
Maslov-Fedoryuk theorem \cite{fedoryuk_pr,maslov,fedoryuk_book1}.
In this way, we can derive contributions from each periodic orbit free of
divergences at any bifurcation point, and the oscillating part of the
level density can be approximated by the following {\it semiclassical 
trace formula}: 
\begin{eqnarray}
\delta g(\vareps) &\simeq& \delta g_{scl}(\vareps)
=\sum_{po} \delta g^{\rm scl}_{po}(\vareps)\,,\qquad\qquad\mbox{with}\qquad\label{dgsc}\\
 \delta g^{\rm scl}_{po}(\vareps) & = &
 \Re \left\{ {\cal A}_{po}(\vareps)
 \exp\,\left[\frac{i}{\hbar}S_{po}(\vareps)
                           -i\sigma_{po}\pi/2-i\phi_d\right]\right\}.\nonumber
\end{eqnarray}
The sum is over all periodic orbits (isolated or families)
($po$) of the classical system (besides of the POs which belong to the 
same family on which the summation was already performed). 
$S_{po}(\vareps)=\oint{\bf p}\cdot d{\bf r}$ are their action integrals, 
the amplitude $\cA_{po}(\vareps)$ (which in general is complex) is 
of the order of the phase space volume occupied by CT, and 
the factor given in Eq.\ (\ref{errfuns}) which depends on the degeneracies
and stabilities of the POs, respectively. $\sigma_{po}$ is called the Maslov 
index. $\phi_d$ is an extra phase that depends on the dimensionality of the 
system and degeneracy $\cK$ of the PO manifold. 
($\phi_d$ is zero when all orbits are isolated ($\cK$=0), 
as defined in \cite{gutz}). The sum in (\ref{dgsc}) is an 
asymptotic one, correct to leading order in $1/\hbar^{1/2}$, and in 
non-integrable systems it is hampered by convergence  problems \cite{gutz}. 
For systems in which all orbits are isolated in phase space, Gutzwiller 
\cite{gutz} expressed the amplitudes $\cA_{po}(\vareps)$ (which are real in 
this case) explicitly in terms of the periods and stability matrices of the 
$PO$s, see some examples below. His trace formula has become famous,
in particular in connection with {\it ``quantum chaos''} \cite{gutz}.
Notice that according to (\ref{errfuns}), any more exact integration
in (\ref{pstrace}) over a bifurcation/catastrophe variable $\xi$ of the 
improved SPM leads to an enhancement of the amplitude $\cA_{po}$
in the form of a maximum in the transition interval
from the bifurcation point to the region of the asymptotic 
(SSPM) behaviour of $\cA_{po}$. 
The hight of this maximum is of order $1/\hbar^{1/2}$ as compared to the 
result of the standard SPM integration 
(integrable or non-integrable system; see more 
specific examples in Sects. 3.3 and 3.4).

The trace formula (\ref{dgsc}) thus relates the quantum oscillations 
in the level 
density to quantities that are determined purely by the classical system.
Strutinsky, in his search for simple physical explanations of shell effects,
realized that this kind of approach could help to understand the shell effects
in terms of classical pictures. However, in the application to nuclear physics,
Gutzwiller's expression for the amplitudes $\cA_{po}(\vareps)$ could not 
be used,
because they diverge when the POs are not isolated in phase space. This happens
whenever a system has continuous (e.g., rotational) symmetries, and hence for 
most typical shell-model potentials (except in non-axially deformed 
situations).
Gutzwiller's theory was extended to systems with continuous symmetries in
\cite{strm} (see also Chap.\ 6 of \cite{book} for details).

Trace formulae for systems with all kinds of mixed symmetries, 
including the integrable cases, were also developed later by various 
other authors. 
The treatment of bifurcations is still an on-going subject of current research.
Uniform approximations were constructed for orbit bifurcations and 
symmetry breaking under the variation of the energy or a potential parameter; 
references to most of these developments are given in \cite{book}, Chap.\ 6.

Strutinsky has, however, not only the merit of extending Gutzwiller's 
approach to
realistic shell-model potentials, but he and his collaborators also extended 
the semiclassical approach to the description of bound many-fermion systems in 
the mean-field approach. In \cite{strm} it was shown that for such systems the 
shell-correction energy $\delta E$ (\ref{dele}) (for one kind of particles) is 
given semiclassically by a similar-looking trace formula:
\begin{equation}
\delta E \simeq \delta E_{scl}=
              \Re \sum_{po} (\hbar/t_{po})^2 {\cA}_{po}(\vareps_F)\;
              \exp\,\left[\frac{i}{\hbar}S_{po}(\vareps_F)-
              i\sigma_{po}\pi/2-i\phi_d\right].
\label{desc}
\end{equation}
The difference to the trace formula (\ref{dgsc}) for the level density is that
here the amplitudes and actions are to be evaluated at the Fermi energy 
$\vareps_F$ of the considered particles, and the appearance of an extra factor
$(\hbar/t_{po})^2$, where $t_{po}$ are the times periods of the particle 
motion 
along the PO with $M$ repetitions, $t_{po}=M T_{po}$, $T_{po}$ being 
the primitive 
($M=1$) period of PO. This extra factor brings a natural convergence 
to the sum, 
different from that in (\ref{dgsc}): orbits with longer periods contribute 
less 
to the shell-correction energy (besides of amplitude enhancement of shell 
structure discussed above and further below). To obtain the energy 
shell correction $\delta E$ (\ref{desc}) as function of the particle number
$N$, one can use the standard relation of the Fermi energy $\vareps_F$
to the given particle number $N$,
\be\l{timepo}
N=2\int_0^{\vareps^{}_{F}} \d \vareps\:g(\vareps)\,.
\ee
The level density  $g(\vareps)$ can be approximated here with 
Eqs.\ (\ref{glev}) and (\ref{denssplit}) using the trace formula Eq.\ 
(\ref{dgsc}) for its oscillating part.\footnote{For practical purposes, 
we use the coarse-grained expression defined in Eq.\ (\ref{deltadenavbil})
below with a small value of $\gamma$ to which $\vareps_F$ is quite insensitive.}

It is the beauty of this approach that gross-shell effects in $\delta E$ 
often can be explained semiclassically 
in terms of a few of the shortest POs (with smallest periods) in the 
system. The earliest application of this idea by Strutinsky's group was 
given in 
\cite{strd}, where the correct slopes of the stability valleys $\eta(N)$ 
of nuclei in a plot of $\delta E$ versus particle number $N$ and deformation 
$\eta$ was correctly reproduced using the shortest orbits in a spheroidally
deformed cavity at small $\eta$, as shown in Fig.\ \ref{figshells}. 
A further example will be given in Sect.\ 3.2. However, if one wants to 
study finer shell structures or some specific situations
at large deformations, longer orbits have to be included. Hereby, bifurcations
of POs can play a crucial role, as it will be exemplified in Sects. 3.3 and 3.4.

For billiard systems, it is advantageous to express the level density
as a function not of energy $E$ but of the wave number
$k=\sqrt{2mE/\hbar^2}$, where $m$ is the mass of the particle.
By virtue of the simple $k$ dependence of the phase function $S_{po}/\hbar$
in \eq{dgsc} for such systems through the product of wave number $k$ and orbit 
length $L_{po}$, the Fourier transform of the level density $g(k)$ 
with respect 
to $k$ yields directly its length spectrum and thus 
provides a nice tool to examine the quantum-classical correspondence.
The Fourier transform of the semiclassical level density gives, indeed,
\begin{equation}
F(L)=\int g(k)e^{ikL}dk
\simeq F_0(L)+\sum_{po}\widetilde{\cA}_{po}\,\delta(L-L_{po})
\label{Fourier_cavity}
\end{equation}
which has peaks exactly at the lengths of the periodic orbits $L=L_{po}$ 
with heights proportional to their amplitudes $\cA_{po}$.  $F_0(L)$ 
represents a Fourier
transform of the smooth TF level density and has a peak at $L=0$.
Taking the Fourier transform of the exact quantum spectrum of the
Hamiltonian thus reveals the classical periodic orbits, and the heights
of the Fourier peaks give information about their semiclassical amplitudes.

It is often instructive to study a {\it coarse-grained level density}
which is defined by convoluting the exact sum of delta functions in
\eq{pstrace} with a normalized Gaussian function of finite width $\Gamma$:
\be
g^{}_\Gamma(E) 
 = \frac{1}{\Gamma\sqrt{\pi}}\int_{-\infty}^\infty \d E' g(E')\, 
\exp[-(E-E')^2/\Gamma^2]
 = \frac{1}{\Gamma\sqrt{\pi}} \sum_n \exp[-(E-E_n)^2/\Gamma^2]\,.
\ee
Convoluting the semiclassical trace formula \eq{dgsc},
one may expand the action $S_{po}(E')$ in the phases up to linear (and the 
amplitudes ${\cal A}_{po}(E')$ up to zero) order terms in $E'-E$, as long 
as $\Gamma$ is small with respect to the distance between the gross shells 
near the Fermi surface, see \cite{strm}. In this way, one obtains the 
Gaussian-averaged trace formula
\be\l{deltadenav}
\delta g^{}_{\Gamma,{\rm scl}}(\vareps)=
\sum_{po} \delta g^{\rm scl}_{po}(\vareps)\,
\exp\left\{-\left(\frac{\Gamma t^{}_{po}}{2\hbar}\right)^{\!2}\right\}.
\ee
For billiards, a corresponding Gaussian averaging with a width $\gamma$ 
over the wave number $k$ leads (with better accuracy owing to the
linear dependence of the action on $k$) to the averaged trace formula
\be\l{deltadenavbil}
\delta g_{\gamma,{\rm scl}}(k)=\sum_{po} \delta g^{\rm scl}_{po}(k)\,
\exp\left\{-\left(\frac{\gamma L^{}_{po}}{2}\right)^2\right\},
\ee
where $L_{po}$ is the length of the PO, and the relation between the two
smoothing widths is $\Gamma\simeq 2\gamma E/k$. We see that, depending on the 
smoothing width $\Gamma$ (or $\gamma$), longer orbits are automatically 
suppressed in the above expressions and the PO sum converges -- which it 
usually does not \cite{gutz} for non-integrable systems in the limit 
$\Gamma$ (or $\gamma$) $\to 0$.
Thus, one can emphasize the {\it gross-shell structure} in the level density
using a smoothing width which is much larger than the s.p.\ level spacing but 
smaller than the main shell spacing (the distance between gross shells) near 
the Fermi surface. Alternatively, finer shell structures can
be studied using essentially smaller smoothing widths. 

We note that the POT can also be applied to open systems above the
continuum threshold, where the level density is dominated by resonances;
an example has been given in \cite{kaibrawi}. As a first step towards
dynamics, the oscillating parts of nuclear moments of inertia have been 
studied semiclassically in terms of POs in \cite{frau,sacr}.

Finally, we mention applications of the POT to finite fermion systems
from other domains of physics such as mesoscopic semiconductor structures
(e.g., quantum dots), metal clusters, and trapped fermionic atoms, 
described in \cite{book} (Chapter 8) and \cite{bridge,disk,zphD,fkms,trap}.

\subsection{SEMICLASSICAL CALCULATION OF A NUCLEAR FISSION\\ BARRIER}

\begin{figure}
\begin{center}
\includegraphics[width=0.35\columnwidth,clip=true,angle=-90]{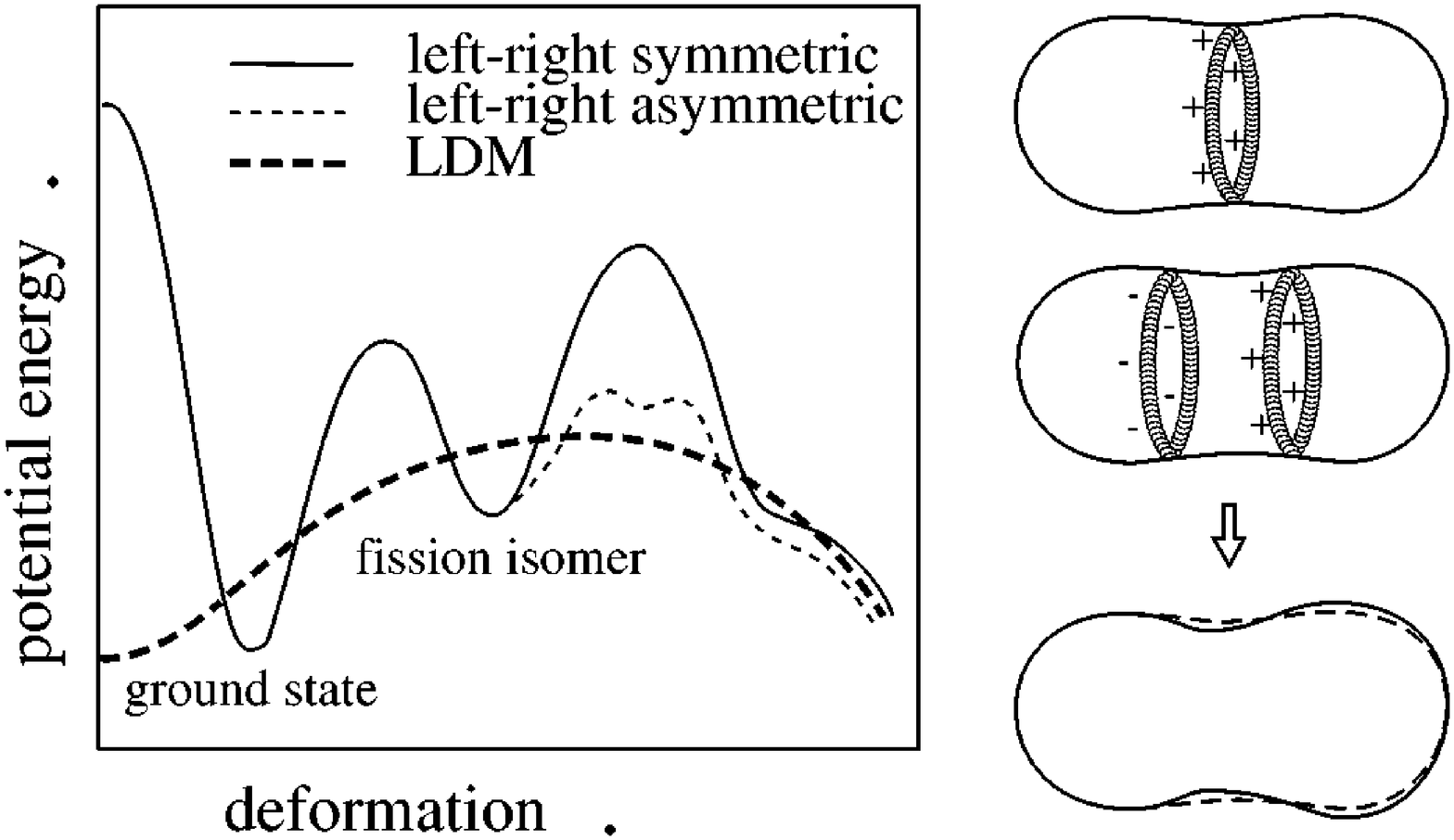}
\vspace{0.2cm}\caption{
{\it Left:} 
Schematic double-humped fission barrier of a typical actinide nucleus. 
Note the lowering of the outer barrier due to left-right asymmetric shapes. 
{\it Right:} 
Maximum probability amplitudes (schematic) of the two leading s.p.\ states 
responsible for the asymmetry effect.
}
\label{asymbar} 
\end{center}
\end{figure}

One prominent feature in the fission of actinide nuclei (isotopes of
U, Pu, etc.) is that their fragment distributions are asymmetric with a
most probable ratio of fragment masses of $\sim$ 1.3 - 1.5 (cf.\ \cite{fuhi}). 
This is an effect that cannot be described within the LDM which always 
favours the highest possible symmetries. It was one of the big successes 
of the SCM to explain the mass asymmetry of the fission fragments.
The fragment distribution is, of course, a result of nuclear dynamics.
However, already in static calculations of fission barriers, the onset
of the mass (or left-right) asymmetry at the outer fission barrier
was found in SCM calculations with realistic nuclear shell models 
\cite{asymn,asyplb}. On the l.h.s.\ of Fig.\ \ref{asymbar}, we show a 
schematic picture of the deformation energy of a typical actinide nucleus, 
plotted versus a suitably chosen deformation parameter (see below for
a specific choice of deformations). The heavy dashed line is the average 
deformation energy obtained in the LDM; the thin lines are the results 
obtained when the shell-correction energy $\delta E$ is included. They 
exhibit the characteristic deformation effects of the shell structure 
in these nuclei: a deformed ground state and the characteristic double-humped 
fission barrier, split by a second minimum corresponding to the fission
isomer. The solid line is obtained when only left-right symmetric
deformations are used; the dashed thin line is obtained when one allows
for left-right asymmetric shapes. As we see, the asymmetric shapes  
lower the outer fission barrier considerably. All shapes here are
taken to be axially symmetric.

The mass asymmetry in nuclear fission was therefore understood as a quantum 
shell effect. In a detailed microscopical study \cite{gumni} of the Lund 
group using the Nilsson model, it was shown that those single-particle states 
which are most sensitive to the left-right asymmetric deformations are 
pairs of states with opposite parity, having the nodes and extrema of 
their wave functions on parallel planes perpendicular to
the symmetry axis at and near the waist-line of the fissioning nucleus, 
as shown on the r.h.s.\ of Fig.\ \ref{asymbar}. Under the effect of the 
neck constriction one of these s.p.\ levels, which for actinides is just 
lying below the Fermi energy, is further lowered when the mass asymmetry 
is turned on. As a consequence, the asymmetry leads to a lowering of the 
total shell-correction energy and hence of the outer fission barrier,
the LDM part of the energy being much less sensitive to the mass asymmetry.
(As a historical note, we should emphasize that the essential role of 
such pairs of s.p.\ states for the mass asymmetry had already been 
recognized in 1962 by S.A.E. Johansson \cite{joha}. Unfortunately, at 
that time the SCM was not available and hence no realistic fission 
barriers could be calculated, as reported in Sect.\ 2.)

In this section, we want to show that the POT is able to reproduce this 
quantum shell effect, at least qualitatively, in s semiclassical description
using the POT. We will focus here only on the gross-shell structure, like 
that seen in the qualitative picture of the fission barrier in Fig.\ 
\ref{asymbar}. 

\begin{figure}
\hspace{0.5cm}
\includegraphics[width=0.95\columnwidth,clip=true]{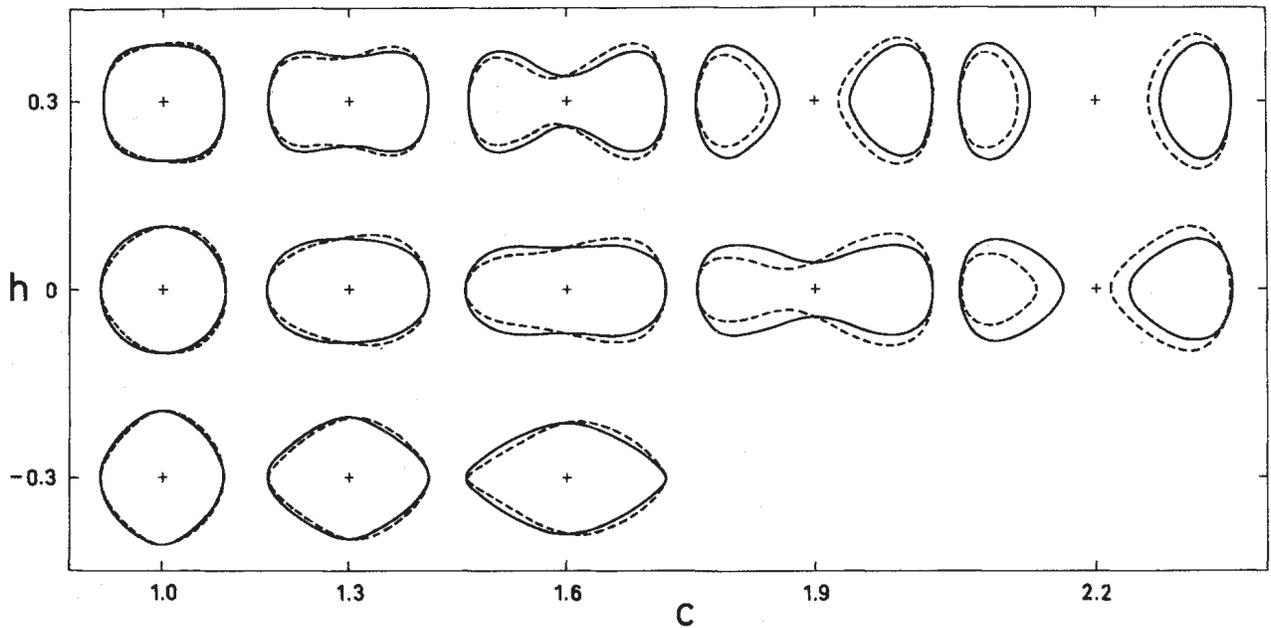}
\vspace{-0.5cm}\caption{
Axially symmetric nuclear shapes in the $(c,h,\alpha)$ parameterization of
\cite{fuhi}. Dashed lines for $\alpha\neq 0$; the sequence with $h=\alpha=0$
corresponds to the shapes obtained in the LDM \cite{cosw}.
}
\label{chalpha} 
\end{figure}

The spheroidal cavity model used in \cite{strd} and discussed in Sect.\ 3.3
does not allow one to describe nuclear fission, since an ellipsoidal 
deformation is not sufficient to yield a finite barrier towards fission. In 
\cite{brreisie,tsuk,chaofi}, a simple but more realistic ``fission cavity 
model'' was used. It consists of a cavity with the $(c,h,\alpha)$ shape 
parametrization that was used both for the LDM and for the deformed 
Woods-Saxon type shell-model potentials in the SCM calculations of 
\cite{fuhi}. These axially symmetric shapes are shown in Fig.\ \ref{chalpha}. 
The parameter $c$ describes the elongation of the nucleus (in units of the 
radius $R_0$ 
of a sphere containing $N=\rho_0 4\pi R_0^3/3$ particles, where $\rho_0$ is 
the bulk particle density), $h$ is a necking parameter, and $\alpha\neq 0$ 
describes left-right asymmetric shapes shown by the dotted lines. The 
sequence of shapes with $h=\alpha=0$ reproduces the optimized shapes  of 
the LDM \cite{cosw} (see \cite{fuhi} for details). Like in \cite{strd}, 
spin-orbit and pairing interactions were 
neglected in \cite{brreisie,tsuk,chaofi} and, for simplicitly, only one kind 
of nucleons (without Coulomb interaction) was used. The only parameter in 
the fission cavity model, the Fermi wave number $k_F=12.1/R_0$, was adjusted 
to yield the second minimum at the deformation $h=\alpha=0,\,c=1.42$ which 
is that of the fission isomer obtained in \cite{fuhi} for the nucleus 
$^{240}$Pu. This corresponds here to a particle number $N\simeq 180$, i.e., 
to $N^{1/3}\simeq 5.65$ when the spin-orbit interaction is neglected. 

This procedure is justified by the observation that, to a first 
approximation, the spin-orbit and Coulomb interactions essentially lead to 
a shift of the magic numbers, preserving the relative shell structures in 
the energy shell-correction. This shift can be simulated by a shift of the
Fermi energy as in \cite{strd}. The procedure works, however, only locally 
in a limited region of deformations and particle numbers. The results shown
in Fig.\ \ref{contours} below
suggest that it is successful in the region $1.3 \siml c \siml 1.65$; the 
ground-state deformations would, e.g., not be reproduced correctly with the 
same Fermi energy. [Note that, in principle, spin-orbit effects can be 
included in the POT \cite{lifl,boke,plet} (see also \cite{bapz} for a short 
overview). However, in non-integrable systems one is met with lots of 
bifurcations under the variation of the spin-orbit strength \cite{cham}, 
which makes the POT with spin-orbit interactions very cumbersome. Similarly,
the pairing interactions can also be included in the POT \cite{olof,braroc},
but this has not been done for nuclear deformations energies so far.]

In \cite{brreisie}, the shortest POs in the $(c,h,\alpha)$ cavity were 
found to dominate the gross-shell features of the double-humped fission 
barrier. For the deformations around the barriers ($c \simg 1.3$), the 
shortest POs are the primitive diagonal and regular polygonal orbits in 
planes perpendicular to the nuclear symmetry axis, situated at the extrema 
of the cavity shape function (see the l.h.s.\ of Fig.\ \ref{sembar} below). 
At the onset of the neck ($c=1.49$ for $h=\alpha=0$), the orbits in the 
central equatorial plane become unstable with respect to small perturbations
perpendicular to the equatorial plane and give birth to new stable orbits 
lying in planes parallel to the equatorial plane. In the restricted 
deformation space with $\alpha=0$, these bifurcations are of pitchfork
type; they are isochronous due to the reflection symmetry with respect
to the equatorial plane (cf.\ \cite{brta}, Sect.\ II.B.2, for a discussion of 
this generic bifurcation type). When the asymmetry $\alpha\neq 0$ is turned
on in the presence of a neck, the bifurcation is of a more complicated
type. These bifurcations were treated in a uniform approximation employing a 
suitable normal form for the action function (see \cite{brreisie} for details). 
(Note that with respect to small perturbations within the equatorial plane,
all these orbits are marginally stable, forming degenerate families with 
degeneracy $\cK=1$ due to the axial symmetry of the cavity.)

Before summarizing the results of \cite{brreisie}, let us study the general 
trends of the shell effects obtained in the fisson cavity model and try to 
understand them in terms of the leading POs.

\begin{figure}
\hspace{0.5cm}
\begin{center}
\includegraphics[width=0.9\columnwidth,clip=true]{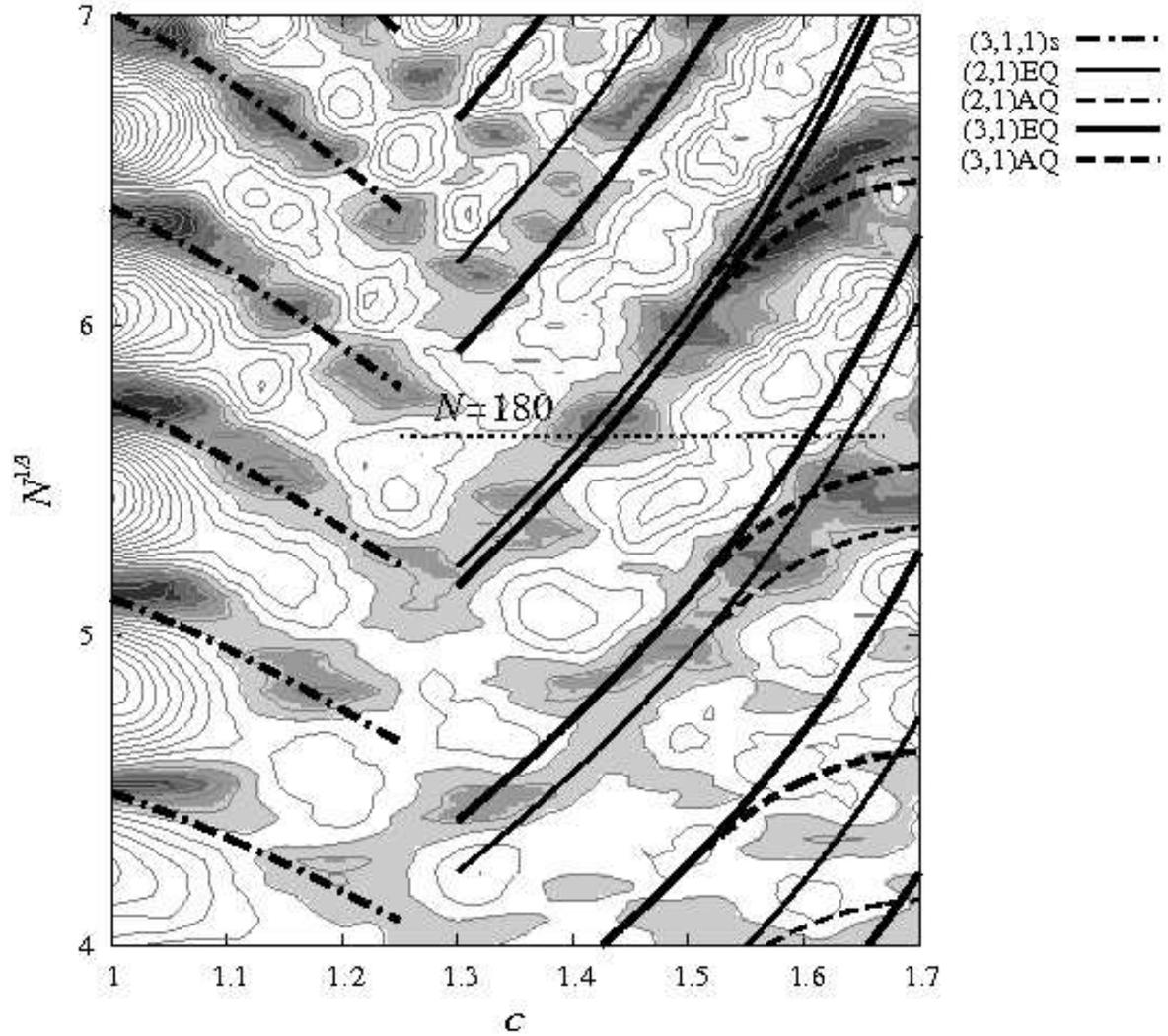}
\caption{
Plot of quantum-mechanical shell-correction energy $\delta E$
versus cube-root of particle number, $N^{1/3}$, and elongation $c$ 
(along $h=\alpha=0$) in the cavity model.
The contours lines are for constant values of $\delta E$ ({\it white:} 
positive values, {\it gray to black:} negative values.
The heavy lines indicate the loci of constant actions of the leading
POs. {\it Dashed-dotted lines:} meridional triangular orbits (3,1,1)s; 
{\it narrow lines:} diameter orbits in equatorial planes (2,1)EQ (solid) 
and in parallel perpendicular planes (2,1)AQ (dashed); 
{\it broad lines:} triangular orbits in equatorial planes (3,1)EQ (solid) 
and in parallel perpendicular planes (3,1)AQ (dashed). The 
horizontal dotted line at $N\simeq 180$ corresponds to the situation 
with the isomer minimum at the correct deformation \cite{fuhi} 
$c\simeq 1.42$ of the real nucleus $^{240}$Pu.
}
\label{dele2} 
\end{center}
\end{figure}

In Fig.\ \ref{dele2} we show a contour plot of the quantum-mechanical 
shell-correction energy $\delta E$ calculated from the s.p.\ energy
spectrum of the fission cavity model, shown versus the cube-root 
of the particle number $N^{1/3}$ and the elongation parameter $c$ along
$h=\alpha=0$ (white: positive values, gray to black: negative values). 
The horizontal dotted line for $N\simeq 180$ (i.e., $k_F=12.1/R_0$)
corresponds to the situation where the isomer minimum lies at $c\simeq 
1.4$ and the outer (symmetric) barrier is peaked around $c\simeq 1.55$, 
as seen in Figs.\ \ref{dgfis} - \ref{contours} below. The heavy lines 
give the loci of constant actions of the leading POs (3,1,1)s: meridional 
triangles; (2,1)EQ: equatorial diameters; (2,1)AQ: diameter orbits in  
planes parallel to the equator plane. We see that these lines follow the 
valleys of minimal shell-correction energy. For the valleys corresponding to 
the ground-state deformations, the situation is like in Figs.\ \ref{figshells}
(Sect.\ 1) and \ref{sce_spheroida} (Sect.\ 3.3 below) obtained for the 
spheroidal models, but here for the more realistic fission cavity model; 
in all cases the meridional orbits dominate the
ground-state valleys. The valleys corresponding to the fission isomers, 
starting around $c \sim 1.3$, are determined by the shortest POs in the 
planes perpendicular to the symmetry axis: up to $c \sim 1.5$, these are 
the equatorial orbits EQ; after their bifurcation at $c=1.49$, the 
valleys are seen to curve down towards smaller values of $N^{1/3}$, 
following the constant-action lines of the stable POs in the planes 
parallel to the equator plane (dashed lines, AQ) which have larger 
semiclassical amplitudes than the equatorial orbits (EQ) that for $c>1.49$ 
have become unstable. The fact that the quantum-mechanically obtained 
stability valleys follow the (dashed) lines AQ after their branching 
from the lines EQ is a remarkable quantum signature of the classical 
bifurcation effect. 

The most striking feature of the gross-shell structure
seen in Fig.\ \ref{dele2}, namely the opposite slopes of the ground-state 
valleys ($1 \siml c \siml 1.25$) and the isomer valleys ($c \simg 1.3$), 
are thus understood semiclassically in terms of the opposite deformation
dependence of the dominating meridional POs in the former valleys and the 
POs in planes perpendicular to the symmetry axis in the latter valleys, 
respectively.

\begin{figure}
\begin{center}
\includegraphics[width=0.75\columnwidth,clip=true]{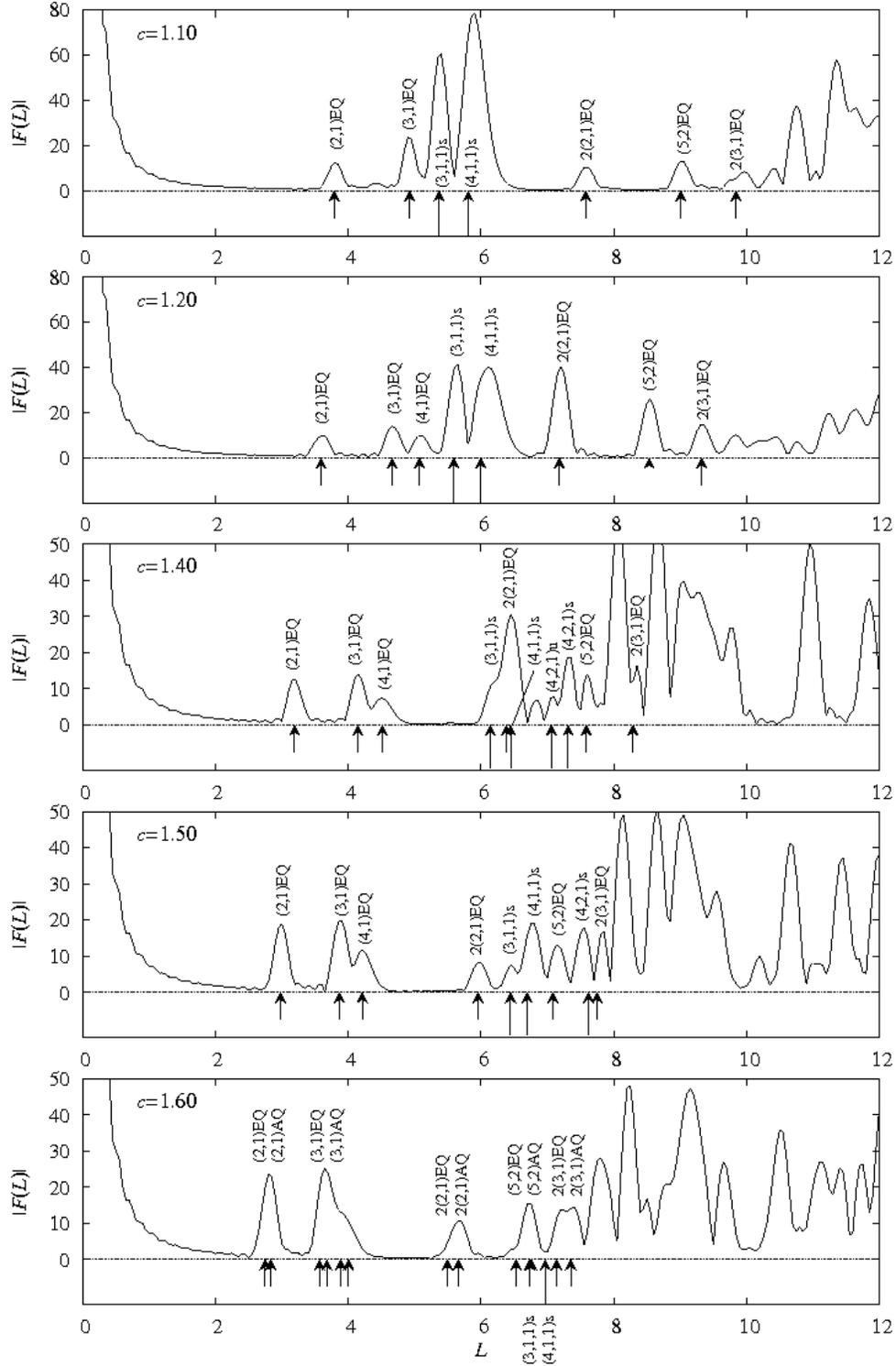}
\caption{Fourier spectra of the fission cavity model with
$h=\alpha=0$ for five values of $c$:
amplitude of Fourier transform of the quantum
spectrum versus length $L$ (in units of $R_0$) of the 
classical POs. {\it Short arrows:} POs lying in planes 
orthogonal to symmetry axis; {\it long arrows:} POs lying 
in meridional planes containing the symmetry axis (labels as in
Fig.\ \ref{dele2}; see text for more details).
}
\label{fourspec} 
\end{center}
\end{figure}

These results can be further elucidated by looking at the Fourier spectra 
in Fig.\ \ref{fourspec} for the five values (from top to bottom) $c=1.1,\,
1.2,\,1.4,\,1.5$, and 1.6 (all for symmetric shapes with $h=\alpha=0$). The short
arrows underneath the Fourier peaks indicate the lengths of the equatorial
orbits: diameter (2,1)EQ and its second repetition 2(2,1)EQ, triangle (3,1)EQ,
etc., and (for $c=1.6$) the corresponding orbits AQ in the planes parallel
to the equator plane. The long arrows correspond to the meridional
orbits: triangle (3,1,1)s and quadrangle (4,1,1)s. We see that for the
small deformations $c=1.1$ and 1.2, the meridional orbits have the 
strongest amplitudes and hence dominate the shell structure in yielding
the ground-state deformation valleys seen in Fig.\ \ref{dele2}. The 
equatorial orbits EQ and their bifurcated partners AQ have the largest 
amplitudes for $c=1.4-1.6$, which explains their dominance in yielding 
the isomer valleys.

Since the trace formula (\ref{desc}) for the shell-correction energy 
$\delta E$ contains the squares $(\hbar/t_{po})^2$ of the inverse
periods of the POs, apart from the amplitudes ${\cA}_{po}$ of the 
semiclassical level density \eq{dgsc}, there is a natural suppression 
of longer orbits contributing to $\delta E$. This ensures the convergence 
of the PO sum, particularly in non-integrable systems (like the one 
considered here) where the PO sum for the level density \eq{dgsc} usually 
does not converge \cite{gutz}. This suppression is particularly effective
amongst orbits with comparable amplitudes ${\cA}_{po}$. It explains why 
already at $c=1.4$, where the meridional orbits (3,1,1)s and (4,1,1)s still 
have similar amplitudes as the EQ orbits,\footnote{The strong Fourier peak 
near $L/R_0\sim 6.3$ for $c=1.4$ in Fig.\ \ref{fourspec} contains the 
combined amplitudes of the meridional quadrangle (4,1,1)s and the second 
repetition of the equatorial diameter orbit, 2(2,1)EQ. Although the two 
cannot be disentangled, we estimate that both these orbits have comparable
amplitudes.} the latter dominate the shell structure (by a factor 
$\sim 4$ in case of the EQ2 orbit), as suggested by Fig.\ \ref{dele2}.

In Fig.\ \ref{fourspec} we have marked some of the peaks around
$7 \siml L \siml 8.5$ for $c=1.4$ and around $6.5 \siml L \siml 8$ for
$c=1.5$ and 1.6. They correspond to orbits born from the equatorial 
orbits in period-doubling bifurcations; some of them are 3-dimensional 
orbits. Similarly, there are many other peaks at $L \simg 8$, some of 
which correspond to orbits born in higher $m$-tupling ($m\geq 3$) 
bifurcations. The contributions of all these orbits 
to the gross-shell structure is, however, practically negligible due 
to their long periods. They have therefore not been included in the 
results presented below. They might, however, become noticeable in POT 
calculations with higher resolution of the shell structure (see the
discussions in Sects.\ 3.3 and 4).

We should also recall the fact that in realistic SCM calculations, the 
pairing interactions are known to reduce the amplitude of $\delta E$ by 
up to $\sim$ 30\% (see, e.g., \cite{fuhi}). In the POT, the pairing 
effects \cite{olof,braroc} yield, indeed, an extra smoothing factor in
the semiclassical amplitudes, which further suppresses the contributions 
of longer orbits.

Fig.\ \ref{dgfis} shows a comparison of the level density in the fission 
cavity model, taken at the Fermi wave number $k_F=12.1/R_0$ and plotted 
versus $c$ along $h=\alpha=0$ in the region covering the isomer minimum
and the outer fission barrier. [As noted above, the model without
spin-orbit interaction and one fixed Fermi energy does not work at
all deformations. For this reason the deformations $c < 1.25$ are not
shown.] The solid line is the quantum-mechanical result and the dashed 
line the result of the semiclassical trace formula (with uniform
approximation \cite{brreisie} for the bifurcation at $c=1.49$) including 
all primitive POs with up to 20 reflections at the boundary in the planes 
perpendicular to the symmetry axis. Both results have been coarse-grained 
by a Gaussian averaging over the wave number $k$ with a width 
$\gamma=0.6/R_0$ that emphasizes the gross-shell structure. We see that 
the agreement is nearly perfect. This figure also demonstrates that already 
a very simple cavity model with one degree of freedom ($c$, keeping
$h=\alpha=0$ fixed) is capable of yielding the main features of the 
double-humped fission barrier.

Let us now look at the influence of left-right asymmetric shapes with
$\alpha \neq 0$ on the shell-correction energy and, in particular, on 
the height of the second fission barrier. In Fig.\ \ref{sembar}, the 
semiclassical result of $\delta E$ is shown in a perspective view as 
a function of elongation $c$ and left-right asymmetry $\alpha$, taken
along $h=0$ in the region of the isomer minimum and the outer fission 
barrier. We see how the outer fission barrier is lowered for left-right 
asymmetric shapes. Instead of the heigher barrier obtained for left-right 
symmetric shapes with $\alpha=0$ (arrow labeled ``symm.''), the nucleus 
can go towards fission over a lower saddle when asymmetric shapes are 
allowed (arrow labeled ``asymm.''). To the left, we see the shapes 
corresponding to the three points A (fission isomer), B and C (along the 
asymmetric fission path) in the deformation energy surface. The vertical 
lines indicate the planes in which the POs are situated (solid lines for 
stable and dashed line for unstable POs). 

The instability of the outer 
fission barrier towards left-right asymmetric deformations, known from the 
quantum-mechanical SCM calculations \cite{asymn,asyplb}, can thus be 
described semiclassically using the POT, indeed. Hereby only the shortest
primitive POs are relevant due to the fast convergence of the PO sum for
the semiclassical shell-correction energy $\delta E$, as discussed above.

This convergence is also demonstrated in Fig.\ \ref{fftasy}. It shows the 
Fourier transform of the level density obtained for the asymmetric shape 
with $c$=1.5, $h$=0, $\alpha$=0.12, corresponding to the point B shown in 
Figs.\ \ref{sembar} and \ref{contours} below, coarse-grained as in Fig.\ 
\ref{dgfis}. The insert depicts the corresponding shape. The solid line is 
the quantum-mechanical and the dashed line the semiclassical result; the
two are seen to agree very well. We notice that the most significant peaks 
correspond to the lengths $L/R_0$ of the primitive orbits ($p,1$)AQ with $p=$ 
2, 3, 4, and 5 reflections. Some of the small peaks at $L>6R_0$ correspond to 
their second repetitions, the other come from meridional orbits. These do, 
however, not affect the gross-shell structure (cf.\ also Fig.\ 4 in \cite{tsuk}).

In Fig.\ \ref{contours}, we compare the old quantum-mechanical
results of SCM calculations \cite{fuhi} with realistic deformed 
Woods-Saxon potentials \cite{dws} with the semiclassical POT results 
using the present simple fission cavity model. Shown are contour plots 
of $\delta E$ versus $c$ and $\alpha$ for two values of the neck
parameter $h$. We see that the semiclassical results (r.h.s.)
reproduce the gross-shell structure of the quantum results (l.h.s.) 
very well. The correct topology is obtained, displaying the lowering 
of the outer barrier for left-right asymmetric shapes. Also, the
amplitudes of the shell effects on both sides are comparable, which 
justifies our calculating only the gross-shell structure using the 
shortest periods on the semiclassical level. Of course, a detailed 
quantitative agreement cannot be expected for the two calculations 
using such different potentials as the sophisticated smooth Woods-Saxon 
potential including pairing, spin-orbit, and Coulomb interactions 
on one side (left), and the simple fission cavity model without 
these extra interactions on the other side (right). The more gratifying 
is the good overall good qualitative agreement of the gross-shell 
structure. This agreement demonstrates, by the way, an experience made 
from quantum-mechanical SCM calculations using realistic nuclear shell
model potentials: the gross-shell features of the fission barriers are 
much less sensitive to the radial dependence of the potentials than to 
its deformation. Hence the success of a simple cavity model that is 
very schematic, but uses the realistic $c,h,\alpha$ deformations.

The white dashed lines in the r.h.s.\ panels of Fig.\ \ref{contours}
show the loci of constant classical actions $S_{\rm PO}$ of the leading POs. 
They follow exactly the valleys of minimal energy in the $(c,\alpha)$ planes 
which define the adiabatic fission paths. Thus, as it was already observed in 
\cite{strd} and seen in Fig.\ \ref{dele2}, the condition for minimizing 
the shell-correction energy is semiclassically given by a least-action 
principle: $\delta S_{\rm PO} = 0$. 

We should emphasize that in Figs.\ \ref{sembar} and \ref{contours}
only the shell-correction energy $\delta E$ is shown. The complete fission
barrier is obtained by adding its smooth LDM part, according to \eq{etot}, 
which for $^{240}$Pu in the $(c,h,\alpha)$ parametrization occurs \cite{fuhi} 
at $c\simeq 1.45 - 1.5,\, h=\alpha=0$. Since the LDM barrier is rather 
smooth around its maximum, the relative heights of the isomer minimum 
and the outer barrier are not affected much by it. However, for $c 
\simg 1.6$ the LDM barrier is already going steeply down. Therefore in 
the total energy, the minimum around $c\simeq 1.65 - 1.7$ seen in Figs.\ 
\ref{sembar} and \ref{contours} along $h=0=\alpha=0$ (like also in Fig.\ 
\ref{dgfis}) vanishes in the steep slope of the total fission barrier, 
as shown schematically in Fig.\ \ref{asymbar}.

It is also interesting to note, as demonstrated explicitly in \cite{chaofi}, 
that the quantum-mechanical probability maxima of those s.p.\ states which 
microscopically are responsible \cite{gumni} for the asymmetry effect in 
the SCM approach (see the schematic plot on the r.h.s.\ of Fig.\
\ref{asymbar}) lie exactly in the planes perpendicular to the symmetry
axis that contain the classical POs. This constitutes a nice 
quantum-to-classical relationship. It was also shown in \cite{chaofi} 
that the classical dynamics of the nucleons with small angular momenta
$L_z$ is more than 90\% chaotic in the region of the outer barrier. A very 
small phase-space region of regular motion is thus sufficient to create 
the shell effect that leads towards the asymmetric fission of the nucleus!

We emphasize once more that the fission cavity model, 
in its present form without spin-orbit and Coulomb interactions, 
is not suitable for predicting fission barriers for a larger range of 
nuclear isotopes and deformations. The present 
semiclassical calculation should be taken as a model study of a typical 
actinide fission barrier, demonstrating that the POT in principle is 
capable of explaining the existence of a double-humped barrier, and also 
the onset of mass asymmetry around the outer barrier, in terms of a few
short classical periodic orbits. It was in no way meant as
a substitute for the quantum-mechanical SCM for calculations of 
static fission barriers. Its aim was rather to provide, as 
suggested by the late Strutinsky, a qualitative physical understanding
of a sophisticated quantum shell effect by means of simple classical 
pictures.

\subsection{SPHEROIDAL CAVITY}

As one of the main subjects in this report, we apply the semiclassical
theory to a 3D spheroidal (prolate) cavity \cite{strd,spheroid_pr, spheroid},
which may be taken as a simple model for a large deformed nucleus or a 
(highly idealized) deformed metallic cluster \cite{book}. We shall investigate 
the role of orbit bifurcations in the shell structure in connection with the 
superdeformation (corresponding to the fission isomer of heavy nuclei).
Although the spheroidal cavity is integrable, it exhibits all the difficulties 
mentioned above (i.e., bifurcations and symmetry breaking) and therefore gives
rise to an exemplary case study of a nontrivial 3D system. We apply the 
ISPM (see sect. 3.1) for the bifurcating orbits and succeed in 
reproducing the superdeformed shell structure by the POT, hereby observing a 
considerable enhancement of the oscillation amplitudes of the level density 
and the energy shell corrections near the bifurcation points and for larger 
deformations.

\subsubsection{{\it CLASSICAL DYNAMICS}}

We consider a cavity system having a spheroidal shape boundary with
the lengths of minor and major axis $a$ and $b$, respectively,
with a conserved volume $a^2b=R^3$. The spheroidal deformation is 
therefore described by a single parameter, which we take to be the axis 
ratio $\eta=b/a$. The classical dynamics of particles in the spheroidal 
cavity can be solved in terms of action-angle variables related to the 
elliptic coordinates $\kappa=v,u,\varphi$ in this Hamiltonian system. 
The Hamiltonian, $H(I_v,I_u,I_\varphi)$, depends only on the partial 
actions \cite{strd,spheroid},
\bea\l{actions}
I_v &\propto& \int_{v_{min}}^{v_{max}} \d v \sqrt{\cosh^2v -I_1-
I_2/\sinh^2v}\;,\nonumber\\ 
I_u &\propto& \int_{-u_{max}}^{u_{max}} \d u \sqrt{I_1 - 
\sin^2u - I_2/\cos^2u}\;, \qquad I_\varphi \propto \sqrt{I_2}\;,
\eea
where $I_1$ and $I_2$ are new single-valued separation constants, 
introduced for convenience instead of the two remaining free actions 
for the fixed energy by using the energy conservation of the particle motion:
$~~\vareps=H(I_u,I_v,I_\varphi)$, and the boundaries of the 
spatial part of
the phase-space occupied by classical trajectories are 
shown explicitly in the integration
limits. (These two constants are identical to
$\sigma_1$ and $\sigma_2$ in the notation of Refs.\ \cite{strd,spheroid}). 
The periodic-orbit (SPM) equations (\ref{spmcond}) for the stationary 
points $I_1^\ast$ and $I_2^\ast$ can be written as a resonance condition of 
commensurability of the partial frequencies:
\be\l{percondfreq}
\om_v:\om_u:\om_\varphi=n_v:n_u:n_\varphi\;,\qquad \mbox{with}\qquad 
\om_\kappa=\partial H/\partial I_\kappa\;.
\ee
For definiteness, we shall consider only prolate spheroidal deformations,  
$\eta>1$, following the suggestions in \cite{strd}.

The POs in the spheroidal cavity can be classified as follows:
a) meridional (elliptic, 2DE) orbits, denoted by $M\left(n_v,n_u\right)$ 
($n_v \geq 2 n_u$), and b) equatorial (EQ) orbits, denoted by 
$M\left(n_v,n_\varphi\right)$ ($n_v \geq 2 n_\varphi$). Hereby $M$ is 
the repetition number of the primitive orbits which are characterized 
by the positive integers $n_v$ = number of corners and 
$n_u, n_\varphi$ = winding numbers), see Figs.\ \ref{fig4} and \ref{fig5}, 
respectively. Both these types of orbits exist at all deformations $\eta > 1$.
c) There are also specific families of orbits 
which exist only at specific deformations larger the critical (bifurcation) 
value, $\eta > \eta_{bif}$, where they are born from the EQ POs 
in pitchfork-like bifurcations. Examples of such bifurcated orbits 
$M(n_v,n_u,n_\varphi)$ are the meridional hyperbolic orbits (2DH), 
e.g., the butterfly (4,2,1) (see the fourth orbit in Fig.\ \ref{fig4}), 
which bifurcates from the doubly repeated EQ diameter 2(2,1)
(cf.\ the first orbit in Fig.\ \ref{fig5}) at the deformation 
$\eta_{bif}=\sqrt{2}$ and exists for all deformations $\eta > \sqrt{2}$ 
along with the diameter 2(2,1). Another important example is the  
bifurcation of the equatorial star orbit (fourth orbit in Fig.\ 
\ref{fig5}) at $\eta=1.618...$ where the new 3D orbits (5,2,1) emerge from
the parent orbit (see the first orbit in Fig.\ \ref{fig6}); again both 
exist at all larger deformations. One also finds a bifurcation of the 
doubly repeated EQ triangle 2(3,1) which bifurcates at $\eta=1.732$,
whereby the new 3D orbits (6,2,1) emerge.

For the following it is useful to note the classical degeneracies $\cK$
of the above-mentioned POs \cite{strm,ellipse,spheroid,maf}. 
The meridional (planar) orbits (denoted by ``2D'' 
henceforth for both elliptic 2DE and hyperbolic 2DH POs) and all 3D orbits 
have the degeneracy $\cK=2$, so that the largest phase-space volume is covered 
by these PO families. The POs in the equatorial plane (denoted by ``EQ'') have 
the degeneracy $\cK=1$ (except at the critical bifurcation points where they
are degenerate with the newborn orbit families having $\cK=2$).  
There is also an isolated ($\cK=0$) diametric orbit along the symmetry 
axis for $\eta>1$, but its contribution can be neglected for sufficiently 
large deformations as compared to all other orbits.

In the following subsections, we will show more explicitly that bifurcations 
of POs are not just a technical obstacle which appears in the semiclassical
POT, but that they leave signatures in the quantum-mechanical results
of the system and thus cause physical phenomena like in Sect.\ 3.2.

\subsubsection{{\it TRACE FORMULAE AND SHELL STRUCTURE ENHANCEMENT}}

The contributions of the above periodic orbits are in the following 
separately considered according to their degeneracies $\cK$ discussed
above. According to Eq.\ (\ref{dgsc}), the oscillating 
part of the level density is thus given semiclassically by
\begin{gather}\l{totdeltaden}
\delta g_{\rm scl}(E)=\delta g^{(2)}_{\rm 3D}(E)
 +\delta g^{(2)}_{\rm 2D}(E)+\delta g^{(1)}_{\rm EQ}(E), \\
\delta g^{(\cK)}(E)
 =\Re\sum_{po}\cA^{(\cK)}_{po}
 \exp\left(ikL_{po}-
i\frac{\pi}{2}\sigma^{(\cK)}_{po}-i\phi_d^{(\cK)}\right).
\l{delsum}
\end{gather}
For the explicit analytical expressions of the amplitudes
$\cA^{(\cK)}_{\rm po}$, constant phases $\sigma^{(\cK)}_{po}$ and 
$\phi_d^{(\cK)}$, 
see Ref.\ \cite{spheroid}.

Figs.\ \ref{fig11} and \ref{fig12} show the characteristic 
enhancement of the semiclassical amplitudes near the bifurcations 
of the EQ orbits (5,2) and 2(3,1), respectively.
As clearly seen from these Figures, the ISPM amplitude (solid lines) 
${\cA}_{EQ}$ has a pronounced but finite maximum instead of the divergence
of the SSPM amplitude (dashed lines) at the corresponding 
critical points 1.618... and 1.732... . 
The SSPM amplitudes ${\cA}_{3D}$ for the bifurcated 3D families (5,2,1) 
and (6,2,1) have discontinuities at these deformations. Their ISPM 
amplitudes are continuous curves with an enhancement in the form of 
a rather wide maximum slightly to the right of the critical
deformations. The relative enhancement near the bifurcations is of order 
$\hbar^{-1/2}$ [or $(k_F L_{po})^{1/2}$ at the Fermi energy 
$\vareps=\vareps_F$]. As noted in sect.\ 3.1, it is due to one more 
exact integration in the improved SPM, as compared to the standard SPM.
This enhancement occurs around the critical deformation (on both sides) 
for the parent orbits and to the right of the bifurcation point for the
newborn POs. For any integrable and non-integrable systems, one has an 
increase of the classical degeneracy of the parent orbits locally at the 
bifurcation point. The newborn POs form 
families having the larger degeneracy ($\cK=2$) at the
critical point and all larger deformations, as noted above. 
This discrete behavior of the classical degeneracies is ``smeared
out'' semiclassically in the ISPM over the regions around the bifurcation
point (or slightly on its right) where the enhancement is found in the 
amplitudes of Figs.\ \ref{fig11} and \ref{fig12} for the general reasons 
discussed in Sect. 3.1.

We observe also some smaller oscillations in these amplitudes further
away from the bifurcation points. They can be neglected with respect to 
the enhancement at the order of the semiclassical parameter discussed above. 
In any case, to improve the SPM with respect to its standard version (SSPM), 
we have to use finite limits in the integration over a phase-space volume 
occupied by CT trajectories as given in the {\it more exact} expression
(\ref{pstrace}) for the level density (before applying further approximations
to the trace of the semiclassical Green function).  

Within the simplest ISPM, using the expansion of 
the action phase near the stationary point up to second-order terms like in 
(\ref{errfuns}) (``2nd-order ISPM''), one can extend the integration 
limits to $\pm \infty$ for deformations asymptotically far from the 
bifurcations in order to obtain the SSPM asymptotics.
For more exact higher-order expansions of the action (\ref{ispmexp}),
for example up to third-order terms (``3rd-order ISPM''), the convergence 
for the extension of the integration limits to an infinite region can be 
sufficiently good, even near the bifurcations, see \cite{spheroid}. 
As shown from the comparison of the 2nd and 3rd order ISPM
approaches in Figs.\ 11 and 12 in \cite{spheroid}, the convergence of this 
expansion around the bifurcation points is rather fast and, therefore, it 
can be stopped at the 2nd order ISPM, but with finite limits.

\subsubsection{\it COMPARISON WITH QUANTUM RESULTS}

We now compare numerically our semiclassical results for the spheroidal cavity 
with exact quantum-mechanical results. In Figs.\ \ref{fig13} and \ref{fig14} 
we show the oscillating part of the level density $\delta g_\gamma(k)$ versus 
the wave number $k$ for six charecteristic values of the deformation parameter 
$\eta$. Both the quantum-mechanical (solid lines) and the semiclassical 
(dotted 
lines) level densities have been averaged over a Gaussian width of 
$\gamma=0.3/R$, in order to emphasize some finer shell structure besides the 
gross-shell oscillations. In Figs.\ \ref{fig15} and \ref{fig16} we show the
results for the energy shell-correction $\delta E$. The quantum results (solid 
lines) are obtained by the shell-correction method, see (\ref{dele}), from the 
quantum spectrum of the spheroidal cavity. The semiclassical results are 
obtained
using (\ref{desc}) which, in analogy to \eq{totdeltaden}, becomes
\be\l{deltaE}
\delta E_{scl} = \Re
\sum_{\cK,po} \left(\frac{\hbar}{t^{}_{po}}\right)^2
\cA_{po}^{(\cK)}(E_F)
\exp\left(ik_FL_{po}-
i\frac{\pi}{2}\sigma_{po}^{(\cK)}-i\phi_d^{(\cK)}\right),
\ee
where the sum is over the same three orbit families as in 
\eq{totdeltaden}
and $t_{po} = m L_{po}/p_F$ are their periods, taken at the Fermi momentum
$p_F=\hbar k_F=\sqrt{2mE_F}$ which is fixed for each particle number $N$ 
through (\ref{timepo}).

A close examination of contributions of individual POs at the deformation 
$\eta=1.2$, see Figs.\ \ref{fig13} and \ref{fig15}, shows that a good 
agreement between the semiclassical (ISPM or SSPM) and quantum calculations 
is obtained by including only short (period-one) elliptic 2D (2DE) and EQ 
orbits. More specifically, we find a rather good convergence of the PO sums 
by using the 2DE orbits with $n_v \leq 12$, $n_u=1$ and the EQ orbits with 
maximum vertex and winding numbers $M n_v=14$ and $M n_\varphi=1$, 
respectively. 
For small and medium deformations up to $\eta\simeq 1.4$, one has 
a clear dominance of the gross-shell structure by the shortest meridian and
equatorial POs, similar to the situation in the fission cavity model (see
Fig.\ \ref{dele2}) up to $c\simeq 1.4$ (see also the discussion below in
connection with Figs.\ \ref{figperiod1} and \ref{sce_spheroida}). 

At $\eta=\sqrt{2}$, there occurs a bifurcation of the doubly repeated EQ 
diameter 2(2,1) with the emergence of the butterfly orbits (4,2,1)2DH, and 
the improved SPM becomes important in comparison with the quantum results, 
see the second panels of these figures. However, even at these
deformations ($\eta \siml 1.4$), as shown in \cite{spheroid}, these 
longer orbits do not influence much the gross-shell structure due to the 
leading period-one orbits mentioned above. Noticeable ISPM contributions 
of the bifurcating (4,2,1)2DH orbits are found in the lowest panels of Figs.\ 
\ref{fig13} and \ref{fig15} for $\eta=1.5$, in agreement with the enhancement 
of their amplitudes to the right of their bifurcation point. At $\eta = 1.5$, 
these contributions become comparable in magnitude with respect to those of the 
period-one orbits, although they have similar periods, and they become even 
larger with increasing deformation and particle number. Thus, for $\eta 
\simg 1.5$ one has a transition from period-one to period-two dominating PO 
contributions; the latter are strongest at $\eta=2$ and beyond.

Figs.\ \ref{fig14} and \ref{fig16} show the cases of superdeformed states 
exactly at the bifurcation points $\eta=1.618...$, $1.732...$ and $2.0$. The 
quantum results are here nicely reproduced by the semiclassical ISPM results 
(like for $\eta=1.414...$ above) including the dominant contributions from 
the bifurcating orbits. Their important role is illustrated in Fig.\ 
\ref{fig18} for the same superdeformations $\eta \simg 1.6$, where only a 
few bifurcated period-two orbits are used in the ISPM results, which are
again compared to the quantum results. It is seen
from this Figure that the main features (periods and amplitudes) 
of the shell and supershell structures are in fairly good agreement which 
becomes the better the larger particle number. Some of the remaining 
discrepancies are still due to the (much smaller) contributions of
period-one orbits and some other missing period-two orbits. 

The importance of the bifurcating period-two orbits at these larger 
deformations is further stressed in the following two figures.
In Fig.\ \ref{figperiod1}, the thin 
solid and dotted lines are again the quantum and semiclassical ISPM results, 
like in Figs.\ \ref{fig15}, \ref{fig16} and \ref{fig18},
showing a good agreement. The heavy line shows the ISMP result which is
obtained if one uses only the shortest (period-one) orbits. We see that with
increasing values of $\eta$ and $N^{1/3}$, the resulting gross-shell 
structure does not reproduce the basic features mentioned above. At the 
superdeformation $\eta=2$, the shortest alone orbits fail, in fact, completely 
to reproduce the gross-shell structure. At $\eta=1.732...$, the
shell structure is dominated mainly by the (4,2,1)2DH and (5,2,1)3D orbits, 
which have emerged from bifurcations at $\eta_{bif}=1.414...$ and $1.618...$, 
respectively, and by the 2(3,1)EQ and (6,2,1)3D orbits bifurcating exactly at 
$\eta_{bif}=1.732...$. At $\eta_{bif}=2$, the (6,2,1)2DH orbits bifurcate from 
the 3(2,1)EQ orbits. Besides these and the above-mentioned POs, also other new 
bifurcating period-three orbits like (7,3,1) and (8,3,1) 3D -- besides the 
3(2,1)EQ orbits which have enhanced amplitudes at $\eta=2$ as usually -- 
are determining the semiclassical shell-correction energies.

These findings are also transparently supported in 
Fig.\ \ref{fourier_SD} by the quantum Fourier spectra at the same 
deformations as above (see a more detailed discussion of the Fourier spectra  
below). Many high peaks in the superdeformed region $\eta=1.6-2.0$
remarkably confirm that the period-two POs are mainly responsible for
the shell structure here and dominating above the period-one POs, 
especially clearly so at $\eta \sim 1.7-2.0$.

We see, therefore, that in the superdeformed region, the period-one 
orbits alone cannot explain the shell structure. The reason why the bifurcated 
period-two orbits become so
important lies in their enhanced semiclassical amplitudes (cf.\ Figs.
\ref{fig11} and \ref{fig12} above) which 
significantly overcome the suppression due to the $1/L_{po}^2$ factor 
in (\ref{deltaE}). One important ingredient of this enhancement
is the larger phase-space volume covered by the newborn orbits due to
their degeneracy $\cK=2$ (see Sect. 3.1 for a more general discussion
of the origins of the enhancement).  

In Figs.\ \ref{sce_spheroida} and \ref{sce_spheroidb} we plot the 
quantum-mechanical shell-correction energy $\delta E$ as function of 
deformation $\eta$ and particle number $N^{1/3}$ in order to study
its shell structure. The basic gross-shell structure of the energy 
valleys of minima is nicely described by the constant-action curves of the
leading shortest POs at deformations smaller than the superdeformed
shapes, as shown in Fig.\ \ref{sce_spheroida}. The main features 
are rather similar to those seen in Fig.\ \ref{figshells} for the Woods-Saxon 
potential with spheroidal shapes, taken from \cite{strd}, and also
similar to Fig.\ \ref{dele2} for relatively small deformations 
in the more realistic fission cavity model. 
In the region of small deformations, the dominant period-one orbits are the 
meridional 2DE orbits (3,1,1) and (4,1,1) whose constant-action lines
follow precisely the down-going ground-state valleys in all three models.
At larger intermediate deformations $1.3 \siml \eta\siml 1.5$, the slopes 
of the upgoing second-minimum valleys are roughly followed by the constant-action 
lines of the period-one orbits (2,1)EQ and (3,1)EQ, which is similar again
as in Fig.\ \ref{dele2} for the fission cavity model for $1.25 \siml c \siml 1.5$. 
However, for $\eta \simg 1.6$ the secondary minimum valleys have a trend
to bend down towards smaller values of $N^{1/3}$ with increasing 
$\eta~$,\footnote{Note that this trend can also be seen in Fig.\ \ref{figshells} 
for the diffuse Woods-Saxon potential with spheroidal deformations.} which is 
not explained by the primitive EQ orbits. This trend can only be explained by 
the influence of the longer orbits which bifurcate from the higher repetitions 
of the EQ orbits. As seen from Fig.\ \ref{sce_spheroidb}, the constant-action
lines of the newborn orbits (4,2,1), (5,2,1) and (6,2,1) do, in fact, follow 
the downbending of the second-minimum valleys very nicely, especially
for $\eta \simg 1.7$ in good agreement with Fourier peaks
in Fig.\ \ref{fourier_SD}. In the region where the downbending
begins, $\eta = 1.4-1.5$, the dominance of the 2(2,1)EQ and the newborn 2DH 
orbits born at $\eta_{bif}=\sqrt{2}$ reflects the enhancement of their
semiclassical amplitudes at and beyond the critical deformation. This confirms 
the conclusions drawn above in connection with 
Figs.\ \ref{figperiod1} and \ref{fourier_SD}.
In the region of superdeformations $\eta \simg 1.6$, the newborn 
period-two and period-three orbits dominate the gross-shell structure 
of the shortest (period-one) EQ orbits completely, as clearly seen also from 
Fig.\ \ref{fourier_SD}.

[Note that similar properties of the superdeformed shell structure
owing to the amplitude enhancement of the dominating bifurcated period-two
(and higher period) orbits were shown in more detail, including the comparison 
of the maps of the energy shell corrections $\delta E(\eta, N)$ for
quantum and semiclassical calculations (with and without bifurcating POs), 
in Fig.\ 20 of \cite{ellipse}. In particular, the ISPM results for $\delta E$ 
were in nice agreement with the quantum results at all critical 
characteristic deformations. The bifurcating period-two orbits were important
to reproduce the superdeformed shell structure at $\eta \simg 1.6$ like
for the present spheroid cavity.]

We point out that the type of the bifurcations causing the downbending of 
the isomer valleys in the fission cavity model, seen in Fig.\ \ref{dele2}, 
is a different one: there it is the period-one bifurcation of the primitive 
EQ orbits at $c=1.49$ which gives birth to the AQ orbits (also of period one) 
whose constant-action lines follow correctly the downbending for $c \simg 1.5$.
It is a subject of our ongoing research to investigate how bifurcations of 
further period-one and period-two orbits in the fission cavity model will affect 
finer details of the shell structure and, in particular, the ``quantization'' 
of the various local minima along the isomer valleys (see also the discussion 
in Sect.\ 4).

Note that in the axially-symmetric deformed 3D harmonic oscillator,
discussed in \cite{strd}, a smooth upgoing behaviour of the valley of minimal 
$\delta E$ is more pronounced due to the shortest EQ orbits along their 
constant-action lines (following the gross-shell structure). It is accompanied 
by deeper local minima along the valleys (see Fig.\ 2d in \cite{strd}). 
They are associated with the amplitude enhancement owing to the longer-3D orbit
bifurcations causing finer shell structure details.  These local minima are 
related to the bifurcations of the simplest 3D orbits, for instance (3,2) at 
the deformation $\eta=\omega_\perp/\omega_z=1.5$ and (2,1) at $2.0$, where 
$\omega_\perp$ and $\omega_z$ are the HO frequencies perpendicular and 
parallel to the symmetry axis. The bifurcation amplitude enhancement near the 
intermediate deformation $\eta \approx 1.5$ there does not destroy the 
leading gross-shell 
structure valleys, like in the two cavity models discussed here. However, the 
decreasing slopes at superdeformations $\eta \approx 2$ are much less 
pronounced in the deformed HO model. We emphasize the important difference 
of the cavity models with respect to the deformed HO that the enhancement 
of the amplitudes of the bifurcating orbits occurs for all deformations larger
than the bifurcation point, see \cite{strd}. We may also point out that the 
downgoing ground-state deformation valleys, due to the meridional orbits in  
the present cavity models, are absent in the harmonic oscillator and, in fact, 
degenerated into one spherical point $\eta=1$. Each of the bifurcating 3D 
orbits appears and exists only at one deformation in the HO, whereas they 
exist in both cavity models at all deformations larger than their bifurcation 
points. 

The above discussed properties of the gross and fine shell structure related to 
the shortest and the bifurcating period-two orbits in the spheroidal cavity 
are nicely confirmed also by Fig.\ \ref{fig7}, where we show the modulus 
of the Fourier 
transform $|F(L)|$, see Eq.\ (\ref{Fourier_cavity}), of the quantum spectrum 
of the spheroidal cavity, plotted versus the orbit-length variable $L$ and 
deformation $\eta$. It exhibits clear quantum signatures of the amplitude 
enhancement owing to the newborn orbits above their bifurcation points (marked 
by black dots). Besides the enhancement of $|F(L)|$ at the PO lengths
(3,1), (4,1) (and others) at the spherical shape ($\eta=1$), we
observe the enhancement also above the bifurcation of the butterfly orbit 
at $\eta=\sqrt{2}$ and the superdeformations 1.618..., $\sqrt{3}=1.732...$  
etc., with the specific PO lengths $L=L_{(5,2,1)}$, $L_{6,2,1)}$, etc.
The enhanced Fourier signals, which correspond to the highest peaks in Fig.\ 
\ref{fourier_SD}, follow precisely the various lines in Fig.\ \ref{fig7}
which show the lengths of the classical POs (as explained in the insert
on the upper right). This is again a nice quantum-to-classical correspondence,
showing that classical periodic orbits and their bifurcations do have a
physical relevance in the quantum spectrum of this system. 

It is worth noting that the semiclassical origins of the deformed shell
structures discussed above are related with bridge orbits \cite{bridge},
at least for potentials with finite diffuseness which can be approximately
described by the power-law potential discussed in the following subsection.  
In a power-law potential with spheroidal shape, we have strong enhancement of 
shell effects due to the bridge orbit bifurcation $(n:m)$ which emerges from a 
period-$n$-tupling bifurcation of equatorial orbits and submerges into a 
symmetry-axis orbit through a period-$m$-tupling bifurcation, forming the 
``bridge'' between those POs, (see \cite{bridge} for details). 
The meridional 2DE orbits $(n,1,1)$ responsible for the 
deformed shell structure at small deformations correspond to (1:1) bridges, 
and the meridional 2DH and 3D orbits $(n,2,1)$ responsible for the 
superdeformed shell structures correspond to (2:1) bridges, and so on. 
In the next subsection, we discuss the power-law potential with spherical 
shape, which also shows bifurcation enhancement effects at certain values 
of the diffuseness parameter.

\subsection{RADIAL POWER-LAW POTENTIALS}

The idea of \cite{aritapap} is that the spherical Woods-Saxon (WS) 
potential, known as a realistic mean-field potential model for spherical
nuclei and metallic clusters, is nicely approximated (up to a constant shift
and without the spin-orbit term) by much a simpler power-law 
potential\footnote{Note: The potential power parameter $\alpha$ here should 
not be confused with the asymmetry deformation parameter of Sect.\ 3.2.}
which is proportional to a power of the radial coordinate $r^\alpha$
\begin{equation}\label{ramodra}
V(r)=U(r/R)^\alpha.
\end{equation}
With a suitable choice of the parameters $U$ and $\alpha$, the approximate 
equality
\[
V_{WS}(r) \approx V_{WS}(0) + U (r/R)^\alpha
\]
holds up to $r\lesssim R$, where $R=r_0 A^{1/3}$ with $r_0=1.2$~fm represents
the nuclear radius for a given mass number $A$.
Thus one finds nice agreement of the quantum 
spectra up to Fermi energy.

The potential \eq{ramodra} includes the limits of the harmonic oscillator 
$(\alpha=2)$ and the cavity $(\alpha\to\infty)$; realistic nuclear potentials
with steep but smooth surfaces correspond to values in the range $2<\alpha<\infty$.
The advantage of this potential is that it is a homogeneous function of the 
coordinates, so that the classical equations of motion are invariant under the 
scale transformations:
\begin{equation}\l{scaling}
\bm{r}\to s^{1/\alpha}\bm{r}, \quad
\bm{p}\to s^{1/2}\bm{p}, \quad
t \to s^{1/2-1/\alpha}t \quad
\mbox{for} \quad E\to sE\;.
\end{equation}
Therefore, one only has to solve the classical dynamics once at a fixed
energy, e.g.\ $E=U$; the results for all other energies are then simply 
given by the scale transformations (\ref{scaling}) with $s=E/U$.
This highly simplifies the POT analysis \cite{aritapap,bridge}.

With this, we are able to apply the POT to more realistic potentials, 
accounting for a finite diffuseness of the potential along the radial
variable. Note that the definition \eq{ramodra} can also be generalized 
to include deformations (see, e.g., \cite{bridge}).

\subsubsection{{\it PERIODIC ORBITS IN THE RADIAL POWER-LAW POTENTIALS}}
\label{poisson}

The definition \eq{ramodra} can be used in arbitrary dimensions,
as long as $r$ is the corresponding radial variable. In practice, we
are interested only in the 2D and 3D cases.  The spherical
3D and the circular 2D potential models have common
sets of periodic orbits.  They are labeled by integers $M(n_r,
n_\varphi)$, where $n_r$ and $n_\varphi$ are mutually commensurable
integers representing the number of oscillations in radial
direction and the number of rotations around the origin, respectively,
and $M$ is again the repetition number.
For the harmonic oscillator ($\alpha=2$), all the classical orbits are
periodic with (degenerate) ellipse shapes. By slightly varying $\alpha$
away from 2, the diameter and circular orbits remain as the shortest
periodic POs.  With increasing $\alpha$, the circular orbit and its
repetitions cause successive bifurcations, generating various new periodic 
orbits $(n_r,n_\varphi)$ $(n_r > 2n_\varphi)$ (see Appendix A2 for details).
Fig.\ \ref{po_power} shows some of the shortest POs. The only difference 
between the periodic orbits in the 3D and 2D systems is their degeneracies.
The shortest PO is the diameter which has the degeneracy $\cK=1$ in the
2D system and $\cK=2$ in the 3D case at $\alpha>2$. Other polygon-like 
orbits have, respectively, $\cK=1$ and $\cK=3$ at $\alpha > \alpha^{}_{bif}$ 
where $\alpha^{}_{bif}$ is the bifurcation value (\ref{bifeqra1}).  
The circular orbit, having maximum angular momentum, is isolated ($\cK=0$) 
for the 2D system and has $\cK=2$ in the 3D case (except at the bifurcation 
points).

\subsubsection{{\it TRACE FORMULA FOR 2D CIRCULAR POTENTIAL}}

In the 2D circular potential, it is useful to take the angular momentum $\ell$ 
as one of the integration variables in the phase-space trace formula
(\ref{pstrace}). The integration limits $\ell=\pm \ell_c$, given in 
(\ref{rcLcd2Fcra}), correspond to the circular orbits moving anticlockwise (+)
and clockwise ($-$).

For the contributions of the one-parametric families of the diameters and
polygon-like POs, the integration over the cyclic angle variable 
$\varphi$ conjugate to $\ell$ is easily carried out, and the remaining 
integration over $\ell$ is done within the simplest ISPM with finite limits 
$-\ell_c \leq \ell \leq \ell_c$. Thus, one obtains the ISPM trace formula in 
terms of error functions, as explained in sect.\ 3.1. Extending the integration 
limits to $\pm\infty$, one will arrive at the asymptotic Berry-Tabor 
trace formula \cite{bt}, like in the cavity model.
It is also worth noting that the limits $\alpha\to\infty$ and $\alpha\to 2$
of the derived ISPM trace formula successfully reproduce their 
asymptotic limits obtained previously for the circular
billiard (disk) and the 2D isotropic harmonic oscillator, respectively.

The contribution of the isolated $({\cal K}=0)$ circular orbits are
derived by adopting the ISPM for two integrations over suitable 
(radial phase-space) variables. It yields a product of two error functions 
as shown in (\ref{errfuns}). Extension of the finite integration limits 
in both error functions to infinity (i.e., to $\pm \infty$ after their 
transformation to the corresponding Fresnel integrals of positive or 
negative real arguments) leads to the (SSPM) Gutzwiller amplitudes 
for the considered isolated POs \cite{gutz}.

The total semiclassical level density (\ref{glev}) is thus given by
\begin{equation}
g_{scl}(E)=g_{TF}(E)+\delta g_{scl}^{(1)}(E)
+\delta g_{scl}^{(0)}(E)
\label{totdensc}
\end{equation}
with ${\widetilde g}\,(\vareps) \approx g_{TF}(E)$ of the Thomas-Fermi 
model and
\begin{equation}
\delta g_{scl}^{(\cK)}(E)=\Re\sum_{po}\cA_{po}^{(\cK)}(E)
\exp\left[\frac{i}{\hbar}S_{po}(E)-\frac{i\pi}{2}\sigma^{(\cK)}_{po}
-i\phi_d^{(\cK)}\right].
\label{dgrake}
\end{equation}
The analytic expressions for the amplitudes
$\cA_{po}^{(\cK)}$ and phases $\sigma^{(\cK)}_{po}$ and $\phi^{(\cK)}_d$
are given in Appendix A3.
Using the scale invariance (\ref{scaling}), one may factorize 
the action integral
$S_{po}(E)$ as
\begin{equation}
S_{po}(E)=\oint_{po(E)}\bm{p}\cdot d\bm{r}
=\left(\frac{E}{U}\right)^{\frac12+\frac{1}{\alpha}}
\oint_{po(E=U)}\bm{p}\cdot d\bm{r}
\equiv\hbar\varepsilon\tau_{po}.
\end{equation}
In the last equation, we define dimensionless scaled energy
$\varepsilon$ and scaled period $\tau_{po}$ by
\begin{equation}
\varepsilon=\left(\frac{E}{U}\right)^{\frac12+\frac{1}{\alpha}},
\qquad
\tau_{po}=\frac{1}{\hbar}\oint_{po(E=U)}\bm{p}\cdot d\bm{r}.
\label{eq:scaled_E_and_T}
\end{equation}
To realize the advantage of the scaling
invariance, it is helpful to use the scaled
energy (period) in place of the corresponding 
original variables.
For the  harmonic oscillator one has $\alpha=2$, and the 
scaled energy and period are proportional to the unscaled quantities.
For the cavity potential ($\alpha\to\infty$), they are proportional 
to the momentum $p$ and length $L_{po}$, respectively.
The energy-scaled level density is written as
\begin{equation}
\delta g_{scl}^{(\cK)}(\varepsilon)
=\der{E}{\varepsilon}\delta g^{(\cK)}(E)
=\Re\sum_{po}\cA_{po}^{(\cK)}(\varepsilon)
\exp\left[i\varepsilon\tau_{po}-\frac{i\pi}{2}\sigma^{(\cK)}_{po}
-i\phi^{(\cK)}_d\right].
\label{dgrakeps}
\end{equation}
This simple form of the phase function enables us to make use of 
the Fourier transformation technique again. The Fourier transform of the
semiclassical energy-scaled level density with respect to the scaled
period  $\tau$ is given, like in Eq.\ (\ref{Fourier_cavity}), by
\begin{equation}
F(\tau)=\int \d \varepsilon\; g(\varepsilon)e^{i\varepsilon\tau}
\sim F_0(\tau)+\sum_{po}\widetilde{\cA}_{po}
\delta(\tau-\tau_{po})\;,
\label{fourier_power}
\end{equation}
which exhibits peaks at periodic orbits $\tau=\tau_{po}$. $F_0(\tau)$ 
represents the Fourier transform of the smooth TF level density and 
has a peak at $\tau=0$ related to the direct (zero-action) trajectory, 
as discussed in Sect.\ 3.1. Thus, from the Fourier transform of the 
energy-scaled quantum-mechanical level density,
\begin{equation}
F(\tau)=\int \left[\sum_n\delta(\varepsilon-\varepsilon_{n})\right]
e^{i\varepsilon \tau}d\varepsilon
=\sum_n e^{i\varepsilon_{n}\tau}, \qquad
\varepsilon_{n}=\left(\frac{E_n}{U}\right)^{\frac12+\frac{1}{\alpha}},
\end{equation}
one can directly extract information about the classical PO contributions.

\subsubsection{{\it AMPLITUDE ENHANCEMENT AND 
COMPARISON WITH QUANTUM RESULTS}}
 
Fig.\ \ref{fig23} shows the typical enhancement
phenomena owing to bifurcations of POs, e.g.\ for the
case of the birth of the triangle-like PO $(3,1)$ from the simplest
(primitive) circular orbit $(1,1)$. The scaled amplitudes $|\cA_{sc,\;po}|$
are presented as functions of the power parameter $\alpha$. 
In Fig.\ \ref{fig23}, the enhancement of the diameter amplitudes 
$|A_{sc,\;}(2,1)|$ and those $|A_{sc,\;(1,1)}|$ for the primitive 
circular orbit $(1,1)$ in the HO
limit $\alpha \rightarrow 2$ are clearly shown. The second sharp peak in
the circular PO amplitude $\cA_{sc,\;(1,1)}$  near the
bifurcation point $\alpha=7$ is related to the bifurcation of the PO $(3,1)$
from the parent orbit $(1,1)$.  Note that a rather wide maximum 
appears in the ISPM amplitude $|A_{sc\; (3,1)}|$ of the triangular-like 
orbit to the right of the bifurcation $\alpha=7$ in Fig.\ \ref{fig23}, even 
larger than for 3D orbits in the spheroidal cavity (see Sect.\ 3.3).
[The third maximum in $|\cA_{sc,\;(1,1)}|$ for a birth of the simplest 
quadrangle-like PO (4,1) 
at the bifurcation $\alpha=14$ of the same parent-circular orbit 
is also seen, because the convergence of the ISPM triangle (3,1)
amplitude $|\cA_{sc,\;(3,1)}|$ to its SSPM asymptotics is realized within such 
a very wide region up to about $\alpha=20$.] 
The ghost part of the $|A_{sc,\;(3,1)}|$ curve
for $\alpha$ to the left of the bifurcation point $\alpha=7$ is suppressed
by rapid oscillations of the phase of a complex amplitude 
 $A_{sc,\;(3,1)}$ with a frequency going to zero
in the HO limit. It is similar like 
for all 3D PO ghosts in the spheroidal cavity in Sect.\ 3.3 (see also
\cite{spheroid}), and for such one-parametric family POs in 
the integrable version of the 2D H\'enon Heiles
potential (see Fig.\ 4 in \cite{maf} for the arguments of the 
complex amplitude). 

Divergences of $|A_{sc,\;(1,1)}|$ and discontinuities of 
$|A_{sc,\;(3,1)}|$ of the SSPM amplitudes are seen in Fig.\ \ref{fig23}, 
too. The continuous results of the improved SPM through the bifurcations 
(or symmetry-breaking points) are also seen with the characteristic 
enhancement. 

Fig.\ \ref{ffig_power} shows the Fourier transform of the energy-scaled
quantum-mechanical level density (\ref{fourier_power}). For a smaller 
$\alpha=2.1$, the diameter (2,1) orbit makes the dominant contribution 
to the gross-shell structure
as the shortest POs (see the peak at $\tau\sim5.0$).
With increasing $\alpha$, the amplitude of the circular
orbit becomes again larger and shows a prominent enhancement around the
bifurcation point ($\tau\sim6.2$ at $\alpha_{bif}=7.0$).
Near the bifurcation, the contribution of the newborn 
triangular-orbit
family (3,1), having a higher degeneracy $\cK=1$, becomes also important
 and dominating for larger $\alpha >\alpha_{bif}$. 
[The newborn (3,1) peak again cannot be distinguished from the 
parent circular (1,1) orbit near the bifurcation point $\alpha_{bif}$,
like the diameter and circular orbits at $\alpha$ close to the HO limit.]

Fig.\ \ref{fig25} shows a nice agreement of the fine-resolved semiclassical
and the quantum level densities $\delta g_\gamma(\varepsilon)$ as functions 
of the scaled energy $\varepsilon$ at the critical bifurcation points 
$\alpha=7$ and $4.25$ for the births of the triangle-like $(3,1)$ and 
the star-like $(5,2)$ orbits. For the semiclassical trace formula, see 
(\ref{deltadenav}), we used
\begin{equation}
\delta g^{}_{\gamma,\;scl}(\varepsilon)=\sum_{{\cal K}=0}^1  
\delta g_{\gamma,\;scl}^{({\cal K})}(\varepsilon) =
\sum_{{\cal K}=0}^1 \sum_{po} 
\delta g_{po}^{({\cal K})}(\varepsilon)\,
\exp{\left(-\tau_{po}^2 \gamma^2/4\right)}\;,
\label{avdenra}
\end{equation}
where $\delta g_{po}^{({\cal K})}(\varepsilon)$ is given by (\ref{dgrakeps}).
For the quantum calculations we used the standard Strutinsky averaging
(over the scaled energy $\varepsilon$), finding a good plateau around the 
Gaussian averaging width ${\widetilde \gamma}=2-3$ with curvature-correction 
degree ${\cal M}=6$. According to Eq.\ (\ref{desc}), for the corresponding
shell correction energy $\delta E_{scl}$ one obtains
\be\l{enra1ra}
\delta E_{scl}/U
=\sum_i \varepsilon_{i}\delta n_i
\simeq
\frac{4\alpha}{\alpha+2}\;\varepsilon^{(3 \alpha-2)/[2(\alpha+2)]}
\sum_{{\cal K}=0}^1 \sum_{po} 
\frac{1}{\tau^2_{po}}\;
\delta g_{po}^{({\cal K})}(\varepsilon^{}_{F})\;.
\ee
Here, $\delta n_i$ are the variations of the occupation numbers 
defined by the Strutinsky smoothing procedure (see, e.g., \cite{fuhi}). 
Fig.\ \ref{fig26} shows the scaled energy shell corrections $\delta E_{scl}$ 
(\ref{enra1ra}) as functions of the particle number $N^{1/2}$ by using the 
standard relationship $N=N(\varepsilon^{}_{F})$, see (\ref{timepo}),
\be\l{partnum2}
N=2\int_0^{\varepsilon^{}_{F}} g_{scl}(\varepsilon) 
\d \varepsilon\;.
\ee
A good plateau for the quantum calculations of the scaled energy shell 
corrections, see the first equation in (\ref{enra1ra}), is realized 
near the same averaging parameters ${\widetilde \gamma}$ and ${\cal M}$ mentioned above.
In the semiclassical calculations for the relation $N(\varepsilon^{}_{F})$, 
we integrated in (\ref{partnum2}) the semiclassical level density
\be\l{scldenstyra}
g_{\gamma,\;scl}(\varepsilon)= g^{}_{TF}(\varepsilon) +
\sum_{{\cal K}=0}^1
\delta g^{({\cal K})}_{\gamma,\;scl}(\varepsilon)\;,
\ee
where $g^{}_{TF}(\varepsilon)$ is the scaled 
average part obtained within the TF approximation \cite{book},
\be\l{scldenstyraTF}
g^{}_{TF}(\varepsilon)=\frac{\alpha}{\alpha+2}\;\varepsilon\;,
\ee
and $\delta g^{({\cal K})}_{\gamma,\;scl}(\varepsilon)$ 
of the ISPM was determined as discussed above.
For the relation $N=N(\varepsilon^{}_{F})$ we used specifically 
an averaging in small $\gamma=0.1$
because there is almost no sensitivity of this integral characteristics within the
interval of smaller $\gamma$ ($\gamma=0.02-0.1$). 
The PO sums converge for the averaging width $\gamma=0.2$ of a fine resolution 
of the shell structure and energy shell corrections
with taking into account the major simplest
POs [about 4 repetition numbers for the circular and diameter ($M=4$) and
a few first simplest other $\cK=1$ POs like
(3,1), (5,2), (7,3) and (8,3)] which are important at $\alpha=7 $. 
As seen from Figs.\ \ref{fig25} and \ref{fig26}, we obtain a nice 
agreement between the semiclassical (ISPM, solid) and quantum 
(QM, dotted) results. Notice that
the dominating contributions in these semiclassical results 
at the bifurcation point $\alpha=7$ are coming 
from the bifurcating circular 
$(1,1)$ and newborn (3,1) orbits, while the bifurcating circular 
$2(1,1)$ and star-like $(5,2)$
orbits are important for $\alpha=4.25$, as expected from enhancement of 
their amplitudes seen in Fig.\ \ref{fig23}. 
The POs (3,1) and (5,2) yield more 
contributions near the bifurcation values of $\alpha$ and even more on their 
right in a wide region of $\alpha$ as mentioned above. 
The orbits (1,1), (3,1) [or (2(1,1),(2,5)] give essential ISPM 
contributions of about the same order
of magnitude and phase in the upper (or lower) panels of 
Figs.\ \ref{fig25} and \ref{fig26}, respectively. 
Notice that the newborn POs (5,2), (7,3) and (8,3) 
give comparable contributions in this Figure, 
in nice agreement with the quantum Fourier
spectra Fig.\ \ref{ffig_power}.

The nice beating seen in these figures is explained
by the interference of these orbits with the 
simple diameters of the same order in magnitude but with different phases.
The latter ISPM contributions are close to the SSPM asymptotic ones
near the bifurcation points $\alpha=7$ and 4.25, because
they are far enough away from their only symmetry-breaking point at 
the harmonic oscillator value $\alpha=2$.

\section{SUMMARY AND CONCLUSIONS}

In the first part (Sect.\ 2) of this article we have given a short review of 
Strutinsky's shell-correction method (SCM). It was conceived as a practicable tool 
for numerical realizations of the theory of finite interacting fermion systems, 
developed by Migdal and collaborators, in particular
at times when computers were not sufficiently fast to perform fully selfconsistent 
microscopic calculations. The SCM, primarily applied to atomic nuclei, combines the
average binding (or deformation) energies from the classical liquid drop model with
a microscopic shell-correction energy $\delta E$ that is extracted from the 
quantum-mechanical energy spectra of realistic (deformed) nuclear shell models. It was
and still is used world wide for very successful computations, in close agreement with 
experimental data, of nuclear masses and deformation energies, in particular fission 
barriers.
 
In a second and larger part (Sect.\ 3), we have presented a semiclassical theory of quantum
oscillations, based upon Gutzwiller's trace formula which connects the level density of a 
quantum system to a sum over periodic orbits (POs) of the corresponding classical system.
This periodic orbit theory (POT) was extended by Strutinsky and collaborators and applied 
to express the shell-correction
energy $\delta E$ of a finite fermion system in terms of POs. The semiclassical
trace formulae for $\delta E$ exhibit a rapid convergence of the PO sum, due to an
inverse dependence of the individual orbit contributions on the squares of their
periods (lengths) (see Sect.\ 3.1). This allows one often to express significant 
features of the shell structure in terms of a few short periodic orbits. In many cases, 
the shortest POs are sufficient to describe the gross-shell features in $\delta E$. 

Bifurcations of periodic orbits can have a significant influence on a fermionic 
quantum system and leave signatures in its energy spectrum (visualized, e.g., by
its Fourier transform) and hence its shell structure. We have (in Sect.\ 3.1) presented
a general method to incorporate bifurcations in the POT, employing an improved
stationary-phase method (ISPM) based on the theory of Fedoryuk and Maslov, and hereby
overcoming the divergence of the semiclassical amplitudes of the Gutzwiller theory
at bifurcations. The improved semiclassical amplitudes typically exhibit a clear
enhancement near a bifurcation and on that side of it where new orbits 
emerge (see, e.g., Figs.\ \ref{fig11}, \ref{fig12} and \ref{fig23}),
which is of the order $\hbar^{-1/2}$ in the semiclassical parameter $\hbar$.
This, in turn, leads to enhanced shell structure effects.

In Sect.\ 3.2, we have presented a semiclassical calculation of a typical actinide 
fission barrier using the POT, employing a fission cavity model that uses a realistic 
description of the three principal axially-symmetric deformations (elongation, neck 
formation and left/right asymmetry) occurring in the (adiabatic) fission process (see
Fig.\ \ref{chalpha}). 
The characteristic (static) double-humped barrier (Fig.\ \ref{asymbar}) and, 
in particular, the sensitivity of the outer barrier to left/right-asymmetric deformations 
can be qualitatively well described by the POT (Fig.\ \ref{sembar}). The loci of minimal 
quantum shell-correction energies 
$\delta E$, both in particle number vs.\ deformation space (Fig.\ \ref{dele2}) and in
two-dimensional deformation space (Fig.\ \ref{contours}), are correctly followed
by the constant-action loci of the shortest POs. Hereby we observe a clear
signature of period-one bifurcations of the shortest equatorial orbits
which here were
treated semiclassically using normal forms and uniform approximations.

In Sect.\ 3.3, we have summarized the results of the spheroidal cavity model
which had early been used in POT calculations for semiclassical investigations
of nuclear deformations (see Fig.\ \ref{figshells}). This integrable
system allows for analytical calculation of the PO properties. We
show that the bifurcations of longer orbits, semiclassically treated
with the ISPM, have an important influence on the shell structure of this 
system. More particularly, the bifurcated period-two and period-three orbits start
to dominate the shell structure at medium (axis ratio $\eta\sim 1.5$) and larger
deformations. In the superdeformed region $\eta\sim 1.6-2$, they dominate the 
minima of the shell-correction energy $\delta E$, the more the larger
the particle number (especially at $\eta \sim 2$), due to their
enhanced semiclassical amplitudes, and the shortest (period-one)
orbits have only a little influence (Figs.\ \ref{fig18} and \ref{figperiod1}). 
The quantum-mechanical Fourier spectra of the spheroidal cavity
exhibit a nice quantum-to-classical correspondence, in that enhanced
Fourier signals follow exactly the lengths of the semiclassically enhanced 
classical POs (Fig.\ \ref{fig7}). This correspondence appears also in the correct 
description of the loci of large deformations in particle number vs.\ 
deformation space by the constant-action lines of the bifurcated 
period-two and -three orbits (see Fig.\ \ref{sce_spheroidb}). 
An important reason for their strong enhancement at large deformations, 
besides the general argument given above (and explained in Sect.\ 3.1), 
is also the fact that the new bifurcated orbits have a larger classical
degeneracy ($\cK=2$) than their parent orbits and the period-one  
orbits ($\cK=1$) (except near the bifurcations). This is not so
in the fission cavity model (Sect.\ 3.2), where all POs (and all POs
bifurcating from them) have the degeneracy $\cK=1$. (Another reason
for differences between the two cavity models is, of course, the
different deformation space used by them, which leads, amongst others, 
to a shift of critical deformations to larger values in the spheroidal cavity.)

In Sect.\ 3.4 we have, finally, presented a class of radial power-law potentials $V(r) \propto 
r^\alpha$, which turn out as good approximations to the popular Woods-Saxon potential
in the spatial region where the particles are bound. The advantage of the power-law
potential is that, in spite of its diffuse surface, the classical dynamics scale 
with simple powers of the energy, which makes it particularly easy for POT calculations. 
The quantum Fourier spectra yield directly the lengths of the leading classical POs  
(Fig.\ \ref{ffig_power}). We have developed semiclassical trace formulae for this
potential and study various limits of the power $\alpha$ (harmonic oscillator
potential for $\alpha=2$, cavity potential for $\alpha\to\infty$). Bifurcations
are, again, treated in the ISPM leading to semiclassical enhancement of the orbit
amplitudes. The trace formulae are shown numerically to give good agreement with the
quantum-mechanical level density oscillations (Fig.\ \ref{fig25}) and shell-correction 
energies (Fig.\ \ref{fig26}).

In conclusion, we may state that the semiclassical theory (POT) is
well capable of explaining the main features of quantum shell structure 
in terms of a few classical periodic orbits. Bifurcations of POs are
not simply an obstacle of the semiclassical theory, but they leave
clear signatures both in the quantum Fourier spectra and the locations
of minima of the shell-correction energy $\delta E$ plotted versus
particle number and deformations.

In our future work we intend to further study finer shell structures
due to longer orbits, applying both the ISPM and uniform approximations 
to treat their bifurcations. In particular, we want to understand the
``quantization'' into isolated minima that occurs along the valleys of
both the ground-state and the secondary (isomer) minima and, in 
particular, we want to better understand the superdeformed shell 
structure. In the
region of superdeformation in the spheroidal cavity, we see clearly
from Fig.\ \ref{sce_spheroidb} that this is due to the bifurcated
period-two and -three orbits. It will be interesting to study the
analogue situation in the fission cavity model by investigating
the higher-period bifurcations occurring there. One complication
in that model is that we have to vary three deformation parameters
($c,h,\alpha$), instead of only one ($\eta$) for the spheroid, in 
order to fully cover the fission barrier landscape. This will 
naturally lead to a much larger variety of possible bifurcations.
Another object of our studies will be to understand semiclassically
the transition regions between the downgoing ground-state valleys
and the initially upgoing secondary-minimum valleys (see Figs.\
\ref{dele2} and \ref{sce_spheroida}). They are well described
separately by the shortest meridional and equatorial orbits,
respectively, but in the transition regions ($1.2\siml c\siml 1.35$
and $1.2\siml \eta \siml 1.5$, respectively) we expect further
bifurcations and corresponding new orbits to play a role.
As a more technical aspect, we also want to perform detailed
comparisons between the results obtained with the ISPM and the
uniform approximations for specific bifurcations.

On a longer time scale, we intend also to include pairing and spin-orbit 
interactions into the POT calculations, in order to come closer
to a realistic description of nuclei, and to extend the POT
towards dynamics (e.g., studying inertial moments, friction
coefficients, etc.). We keep in mind and emphasize, however, that
the POT is not intended as a substitute for a fully quantum-mechanical
theory, but rather a tool for a better understanding of its results 
in terms of classical pictures.

\section*{\centering{\large Acknowledgement}}

We are very grateful to our late teachers, Profs. A. B. Migdal and 
V. M. Strutinsky, for so many creative ideas and collaborations. 
We also thank Profs. K. Matsuyanagi and E. E. 
Saperstein for fruitful collaborations and many useful discussions.

\appendix
\setcounter{equation}{0}
\renewcommand{\theequation}{A\arabic{equation}}
\renewcommand{\thesection}{{\large A\arabic{section}.}}

\newpage

\section*{\centering{\large Appendix A: \\[-8pt]
2D RADIAL POWER-LAW POTENTIAL}}

\section{{\large Classical Dynamics in Generic 2D Circular Hamiltonians}}
\label{appA}

For an isotropic 2D potential $V(r)$, the Hamiltonian $H$
in the canonical phase-space variables $\{r,\varphi,p_r,p_\varphi\}$  writes
\begin{equation}
H= \frac{1}{2m}\,\left(p_r^2 + \frac{p_\varphi^2}{r^2}\right)+ V(r)
= \vareps\;. \label{Hsphra}
\end{equation}
Here, $p_\varphi=\ell$ is the angular momentum, 
$\varphi$ is the cyclic variable,
\begin{equation}
\varphi=\omega_\varphi t + \varphi'\;,
\label{thetaphi}
\end{equation}
$\varphi'=\varphi(t=t'=0)$ is the initial condition,
\begin{equation}
p_r(r)\equiv \sqrt{p^2(r) -
  \frac{\ell^2}{r^2}}=\sqrt{2m\left[\vareps-V(r)\right] - 
\frac{\ell^2}{r^2}} = m \dot{r},\qquad 
p(r)=\sqrt{2m\left[\vareps-V(r)\right]}\;.
\label{prra}
\end{equation}
In particular, we shall consider the power-law potential 
(\ref{ramodra}) from \cite{aritapap}.

Integrating the first equation in (\ref{prra}) 
one has also the radial trajectory $r(t)$,
\begin{equation}
m \int_{r'}^{r} \frac{{\rm d}\rho }
{\sqrt{2m\left[\vareps-V(\rho)\right] - \ell^2/\rho^2}}=t\;,
\label{rtra}
\end{equation}
where $r'=r(t=t'=0)$ as the initial condition for the radial coordinate
$r$. 
For the radial action $I_r$ in polar action-angle variables over one
period ($t=T$) with $\varphi'=\varphi''$ ($r'=r''$) one has
\bea\l{actionvarra}
I_r&=&\frac{1}{2\pi}\,\oint p_r \,\d r=
\frac{1}{\pi}\,\int_{r_{min}}^{r_{max}} \d r\,
\sqrt{2m\left[\vareps-V(r)\right] - 
\frac{\ell^2}{r^2}}=I_r(\vareps,\ell),\nonumber\\
I_\varphi&=&\frac{1}{2\pi}\,\oint p_\varphi \,{\rm d}\varphi=\ell\;,
\eea
where $r_{min}$ and $r_{max}$ 
are the turning points for a given energy $E$, which are solutions 
of the equation:
\begin{equation}
{\cal F}(r,\ell) \equiv 2m\left[\vareps-V(r)\right] - \frac{\ell^2}{r^2}=0\;.
\label{turneqra}
\end{equation}
For frequencies one has
\begin{equation}
\omega_r=\frac{\partial H}{\partial I_r}=
\frac{1}{\partial I_r/\partial \vareps}
=\left(\frac{m}{\pi} \,\int_{r_{min}}^{r_{max}}\frac{\d r}
{\sqrt{2m\left[\vareps-V(r)\right] - \ell^2/r^2}}\right)^{-1}\;,
\label{freqra}
\end{equation}
where $I_r=I_r(\vareps,\ell)$ is the surface energy (\ref{actionvarra}).
Thus, the periodic-orbit equation writes
\begin{equation}
f(\ell) \equiv \frac{\omega_\varphi}{\omega_r} \equiv 
-\frac{\partial I_r(\vareps,\ell)}{\partial \ell} \equiv
\frac{\ell}{\pi} \,\int_{r_{min}}^{r_{max}}\frac{\d r}
{r^2\sqrt{2m\left[\vareps-V(r)\right] - \ell^2/r^2}} =
\frac{n_\varphi}{n_r}\;,
\label{percondra}
\end{equation}
$n_\varphi$ and $n_r$ are co-primitive integers. The energy surface
$I_r=I_r(\vareps,\ell)$ is simplified to function of one variable $\ell$. 
The solutions
to the PO equation (\ref{percondra}) for $\ell=\ell^\ast(n_r,n_\varphi)$ 
define the  ${\cal K}=1$ families  
of orbits $M(n_r,n_\varphi)$, see \cite{strm}. 

\section{{\large Circular Orbit in Power-Law Potential}}

In the radial power-law potential (\ref{ramodra}), the 
isolated circular
orbits cause successive bifurcations with increasing $\alpha$.
At the energy $E$, the radius of a circular orbit
is given by
\begin{equation}
r_c=R\left(\frac{2\vareps}{(2+\alpha)U}\right)^{1/\alpha},\qquad
\ell_c=p(r_c)r_c,\qquad  
{\cal F}_c''=-\frac{4m\,\vareps\alpha}{r_c^2} < 0, 
\label{rcLcd2Fcra}
\end{equation}
 ${\cal F}_c$ is defined by (\ref{turneqra}). 
The radial frequency $\Omega_c=\omega_r(r_c)$ is given by
\begin{equation}
\Omega_c=\sqrt{\frac{\alpha(2+\alpha)U}{mR^2}}\,
\left[\frac{2\vareps}{(2 + \alpha)U}\right]>0.
\label{Omrcra}
\end{equation} 
From (\ref{freqra}) at $\ell=\ell_c$ by using (\ref{rcLcd2Fcra}) for
$r_c$ and $\ell_c$ one finds \cite{aritapap}
\begin{equation}
\omega_c=\omega_{\varphi}(\ell=\ell_c)=\sqrt{\frac{\alpha U}{mR^2}}\,
\left(\frac{2\vareps}{(2+\alpha)U}\right)^{1/2-1/\alpha}\;.
\label{omtcra}
\end{equation} 
From (\ref{Omrcra}) and (\ref{omtcra}), 
one has the equation, 
\begin{equation}
\frac{\Omega_c}{\omega_c}=\frac{n_r}{n_\varphi}\;,
\label{bifeqra0}
\end{equation}
 given explicitly by \cite{aritapap}
\begin{equation}
\sqrt{2+\alpha}=\frac{n_r}{n_\varphi}\;.
\label{bifeqra}
\end{equation}
Thus, for the bifurcation diffuseness parameter $\alpha^{}_{bif}$ 
where the one-parametric orbits $(n_r,n_\varphi)$ are born 
one arrives at the explicit expression:
\be\l{bifeqra1}
\alpha^{}_{bif}=n_r^2/n_\varphi^2-2\;.
\ee

\section{{\large Trace formula for 2D Circular 
 Power-Law Potential Model}}

From the ISPM procedure presented in sect. 3.1, we obtain  
contributions of the
$\cK=1$ families and isolated circular orbits into the 
semiclassical level
density.  For a $\cK=1$ family, we choose the 
angular momentum $\ell$ and angle
$\varphi$ conjugate to $\ell$ for integration variables in
the phase-space trace formula (\ref{pstrace}).  The integration over
a cyclic angle $\varphi$ generates only the factor $2\pi$.  
The remaining
integration over $\ell$ is carried out within the simplest ISPM by
expanding a phase function $\Phi$ in powers of $\ell-\ell^{*}$ 
up to quadratic terms. [The expansion
of pre-exponent factor of integrand in powers of $\ell-\ell^*$ 
is cut at zero order in this approximation.]

Thus, we obtain the amplitude $\cA^{(1)}_{po}$ of Eq.\ (\ref{dgrake}) as
\begin{equation}
\cA_{po}^{(1)}=\frac{T_{po}}{\pi\hbar^{3/2}\sqrt{Mn_r
\left|K_{po}\right|}}\; \erf(\cZ_{po}^-,\cZ_{po}^+)
\label{ampra1}
\end{equation}
with the period
\begin{equation}
T_{po}=\frac{2\pi n_r}{\omega_r}=\frac{2\pi n_\varphi}{\omega_\varphi}\;,
\label{tpora1}
\end{equation}
$K_{po}$ is the curvature  
of the energy surface,
\begin{equation}
K=\frac{\partial^2 I_r}{\partial I_\varphi^2}= 
\frac{\partial^2 I_r(E,\ell)}{\partial \ell^2}= 
-\frac{\partial f(\ell)}{\partial \ell},
\label{curvra}
\end{equation}
$f(\ell)$ is the ratio of frequencies defined in the first
three equations of (\ref{percondra}).
The arguments of the generalized complex error function 
in (\ref{ampra1}), see also (\ref{ispmlimits}), is given by
\begin{equation}
\cZ_{po}^{\pm}=\sqrt{-i\pi Mn_r\left|K_{po}\right|/\hbar}\,
b_{po}\;(\ell_{\pm}-\ell_{po})\;, 
\label{limra1}
\end{equation}
where $b_{po}=1$ for all $\cK=1$ polygon-like PO families, 
except for the diameters for which one has (see \cite{maf})
\begin{equation}
b^{}_{M(2,1)}= 1-\frac12\;\exp\left[-\left(\frac{\alpha-2}{2 \Delta_D}
\right)^2\right], \qquad\qquad \Delta_D=\frac{1}{\sqrt{\pi M n_r K_D/\hbar}},
\label{bcoef}
\end{equation}
$K_D$ is the diameter curvature. For simplicity, the finite integration 
interval of the angular momenta was split
into two parts, $-\ell_{c} \leq \ell \leq 0$ and 
$0 \leq \ell \leq \ell_c$, where $\ell_c$ is the
angular momentum of a circular orbit as mentioned above. 
There are the symmetric stationary points
$\pm |\ell^*|$ related to the clockwise and anticlockwise
motion of the particle along the PO in these two parts of the phase space. 
They give equivalent contributions to the amplitude owing to the independence 
of the Hamiltonian of time. Thus, we have reduced the integration region
to $0 \leq \ell \leq \ell_c$ accounting for this time-reversibility symmetry
simply by a factor 2 in (\ref{ampra1}). Therefore, for the polygon-like 
$(n_r>2)$ POs, 
one has $\ell_=0$ and $\ell_{+}=\ell_c$.
In the HO limit $\alpha\rightarrow 2$, only the diameter (2,1) 
and the circle (1,1) (with repetitions) survive 
and form $\cK=2$ families in the HO potential.
It is natural to assume that the whole phase-space in the HO limit 
can be split into the two equivalent parts, $0<\ell<\ell_c/2$ for the
diameter and $\ell_c/2<\ell<\ell_c$ for the circular orbits,
for the angular momenta and correspondingly for the conjugate
angles. Thus, we arrive at the integration limits (\ref{limra1}) with
(\ref{bcoef}). 
The same prescription has been successfully applied to the
integrable version of the H\'enon-Heiles potential for 
contributions of orbits B
and A in the same HO limit \cite{maf}.
For the Maslov index of the 
considered $\cK=1$ family of POs, one obtains
\begin{equation}
\sigma_{po}^{(1)}=2Mn_r,
\label{maslra1}
\end{equation}
and for the constant phase independent of the orbits one has 
$\quad \phi_d=-\pi/4$.

Extension of the limit $\ell_{\pm}\to \pm \infty$ exactly 
reproduces the
asymptotic Berry-Tabor formula \cite{bt,book}
\begin{equation}
\cA_{po}^{(1)}=\frac{d_{po}T_{po}}{\pi\hbar^{3/2}\sqrt{
Mn_r\left|K_{po}\right|}}\;,
\label{btra1}
\end{equation}
where $d_{po}$ takes into account the discrete degeneracy, $d_{po}=1$
for diameters $M(2,1)$ $(n_r=2n_\varphi)$ and 2 for all other POs
$(n_r>2n_\varphi)$ \cite{book}.
In the circular billiard limit ($\alpha\to\infty$), the action
$S_{po}(E)$
is given by 
\begin{equation}
S_{po}(E)=pL_{po}, \qquad p=\sqrt{2mE}\;,
\label{actra1}
\end{equation}
and the curvature $K_{po}$ from (\ref{curvra}) is obtained explicitly by
\begin{equation}
K_{po}=\frac{1}{\pi pR\sin\phi}\;.
\label{kpobt1}
\end{equation}
  Substituting these quantities, Eqs.\ (\ref{actra1})
and (\ref{kpobt1}), into (\ref{btra1}), from (\ref{dgrake}) and 
(\ref{maslra1}) one
obtains the Balian-Bloch trace formula for the circular billiard \cite{book}.

For the contribution of circular orbits, one should perform 
the two
exact phase-space integrations over radial variables within ISPM.  
Finally, we arrive at the result:
\begin{equation}
\cA_{c,M}^{(0)}=\frac{T_c}{2\pi\hbar\sqrt{F_{c,M}}}
\erf(\cZ_{pc}^{(-)},\cZ_{pc}^{(+)})
\erf(\cZ_{rc}^{(-)},\cZ_{rc}^{(+)}),
\label{ampra0}
\end{equation}
where $T_c$ is the period of a primitive circular orbit.
$F_{c,M}$ is its stability factor \cite{gutz} given by
\begin{equation}
F_{c,M}=4\sin^2\left[\pi M\sqrt{\alpha+2}\right].
\end{equation}
The arguments of the complex error functions in (\ref{ampra0}) are 
given by
\begin{eqnarray}
{\cal Z}_{p\;c}^{(\pm)}&=&\sqrt{-\frac{i}{\hbar}\,
\pi\,M\,\sqrt{\alpha+2}\,\Big|K_c\Big|}
\,\left(\ell_{\pm}-\ell_c\right),\qquad \ell_{+}=\ell_c,
\qquad \ell_{-}=\ell_c/2\;, \nonumber\\
{\cal Z}_{r\;c}^{(\pm)} &=& \sqrt{\frac{i\,F_{c,M}}
{4 \pi\,M\, \hbar\,\sqrt{\alpha+2}\;\Big| K_c\Big|}}\;
\Theta_r^{(\pm)}\;,\qquad \Theta_r^{(+)}=\pi,\qquad \Theta_r^{(-)}=0\;,
\label{argerrorC0ra}
\end{eqnarray}
where $K_c$ is the circular-orbit curvature,
\begin{equation}
K_c = -\frac{(\alpha+1)(\alpha-2)}{12 \,(\sqrt{\alpha+2})^3\,
  \ell_c}\;. 
\label{curvraCra}
\end{equation}
The time-reversibility of the Hamiltonian was similarly 
taken into account as explained above. For the Maslov index 
one has $\sigma^{(0)}_{po}=4M$ and $\phi_d=0$.
(The detailed derivation will appear elsewhere.)
In the asymptotic limit of the finite non-zero integration 
boundaries $\ell_{-}\rightarrow  -\infty$ and 
$\Theta_r^{+} \rightarrow \infty$, i.e. far from any 
bifurcations
$\alpha^{}_{bif}$ (also from the harmonic oscillator symmetry breaking
at $\alpha=2$), 
the expression (\ref{ampra0}) tends to the amplitude of the Gutzwiller
trace formula for isolated orbits \cite{gutz,book}.


\clearpage


\clearpage

\begin{figure}
\hspace{0.5cm}
\begin{center}
\vspace{-0.9cm}
\includegraphics[width=\columnwidth,clip=true]{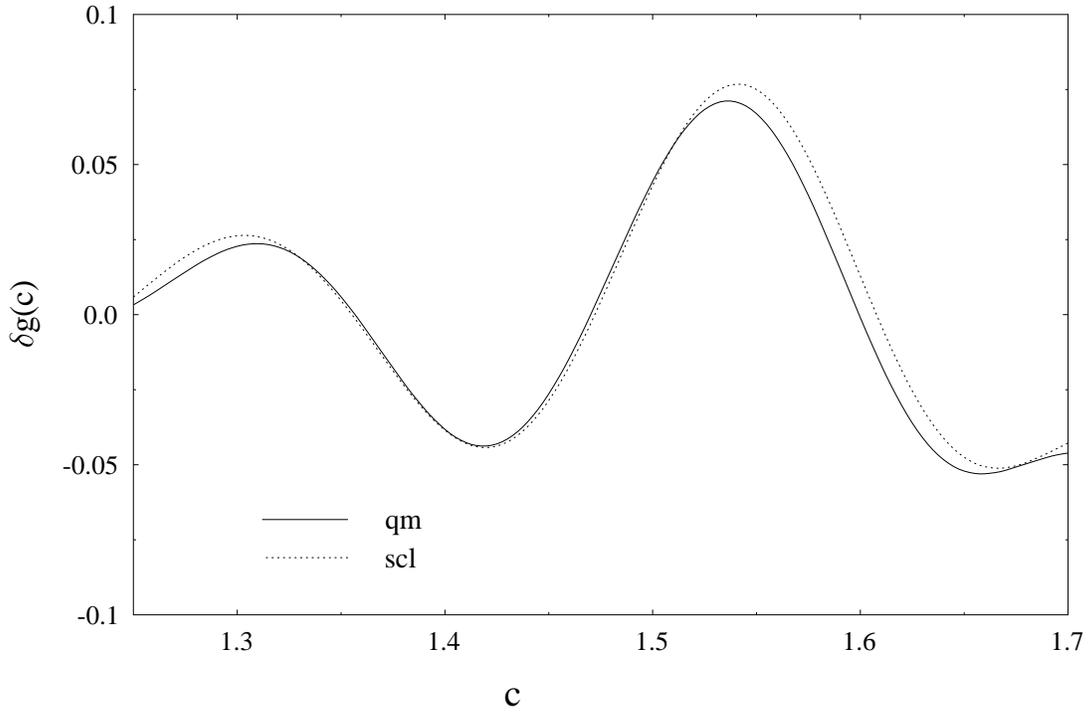}
\caption{\label{dgfis} 
Quantum-mechanical (solid line) and semiclassical (dashed line)
level density $\delta g$ versus elongation $c$ (along $h=\alpha=0$) 
in the fission cavity model taken at the Fermi wave number $k_F=12.1/R_0$.
(Gaussian averaging over wave number $k$ with width $\gamma=0.6/R_0$.)
}
\end{center} 
\end{figure}

\begin{figure}
\begin{center}
\includegraphics[width=0.65\columnwidth,clip=true]{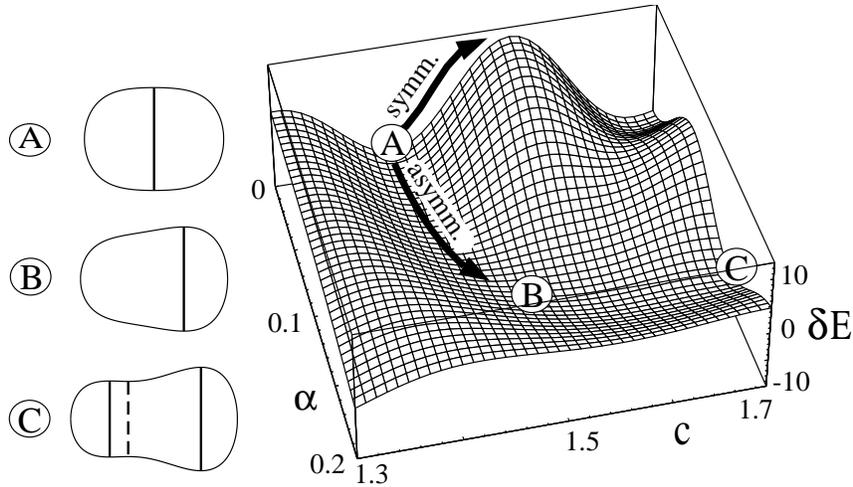}
\caption{
Perspective view of the semiclassical outer fission barrier versus
elongation $c$ and left-right asymmetry $\alpha$ for $h=0$. To the
left, the shapes corresponding to the points A, B and C in the
deformation energy surface are displayed; the vertical solid 
(dashed) lines indicate the planes containing the stable (unstable)
POs.
}
\label{sembar}
\end{center} 
\end{figure}

\begin{figure}
\begin{center}
\vspace{-0.9cm}
\includegraphics[width=\columnwidth,clip=true]{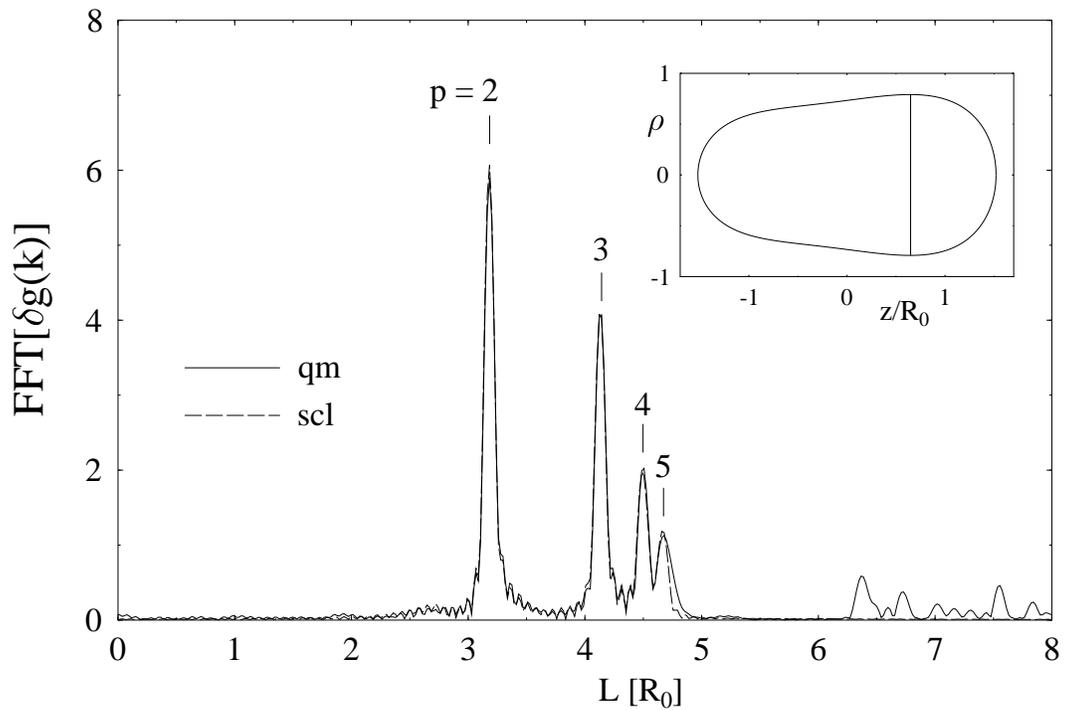}
\vspace{-0.7cm}\caption{
Squared amplitudes of the Fourier transforms of $\delta g(k)$ 
obtained (as in Fig.\ \ref{dgfis}) quantum mechanically (solid line) 
and semiclassically (dashed line) for $c$=1.5, $h$=0, $\alpha$=0.12. 
The insert shows the cavity shape and the location (vertical line) 
of the plane containing the shortest POs with reflection number $p$.
The abscissa shows the length $L$ of the POs in units of the 
spherical-cavity radius $R_0$; see the text for further details. 
}
\label{fftasy}
\end{center} 
\end{figure}

\begin{figure}
\begin{center}
\vspace{-0.9cm}
\includegraphics[width=0.7\columnwidth,clip=true]{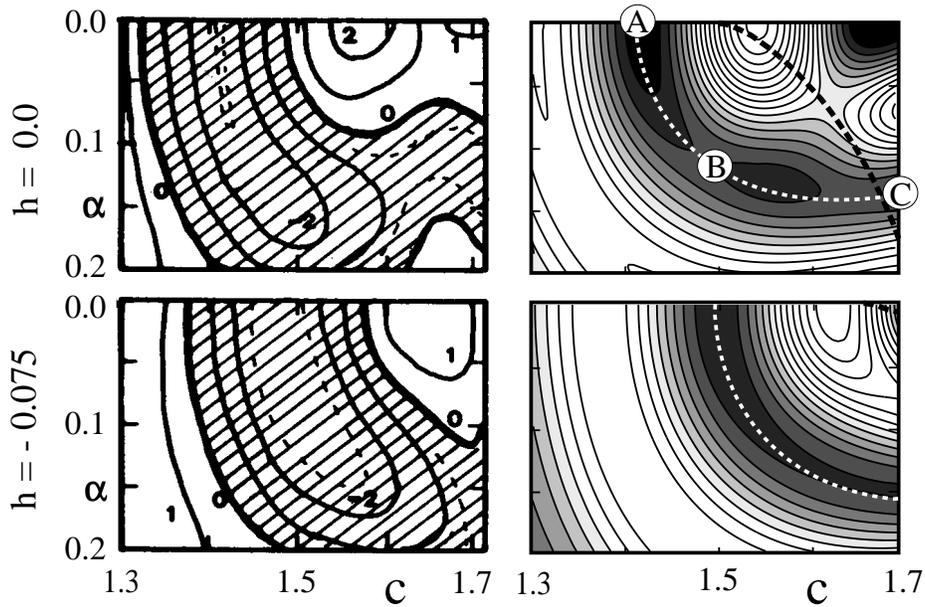}
\vspace{-0.5cm}\caption{
Contour plots of the shell-correction energy $\delta E$ versus
$c$ and $\alpha$. {\it Upper panels:} for $h=0$, {\it lower panels:}
for $h=-0.075$. {\it Left panels:} results of quantum-mechanical
SCM calculations with realistic nuclear shell model potentials 
\cite{fuhi} (shown is the shell correction of the neutrons); 
{\it right panels:} semiclassical POT results with the fission 
cavity model described above.
}
\label{contours} 
\end{center} 
\end{figure}

\clearpage

\begin{figure}
\begin{center}
\includegraphics[width=.7\textwidth,clip]{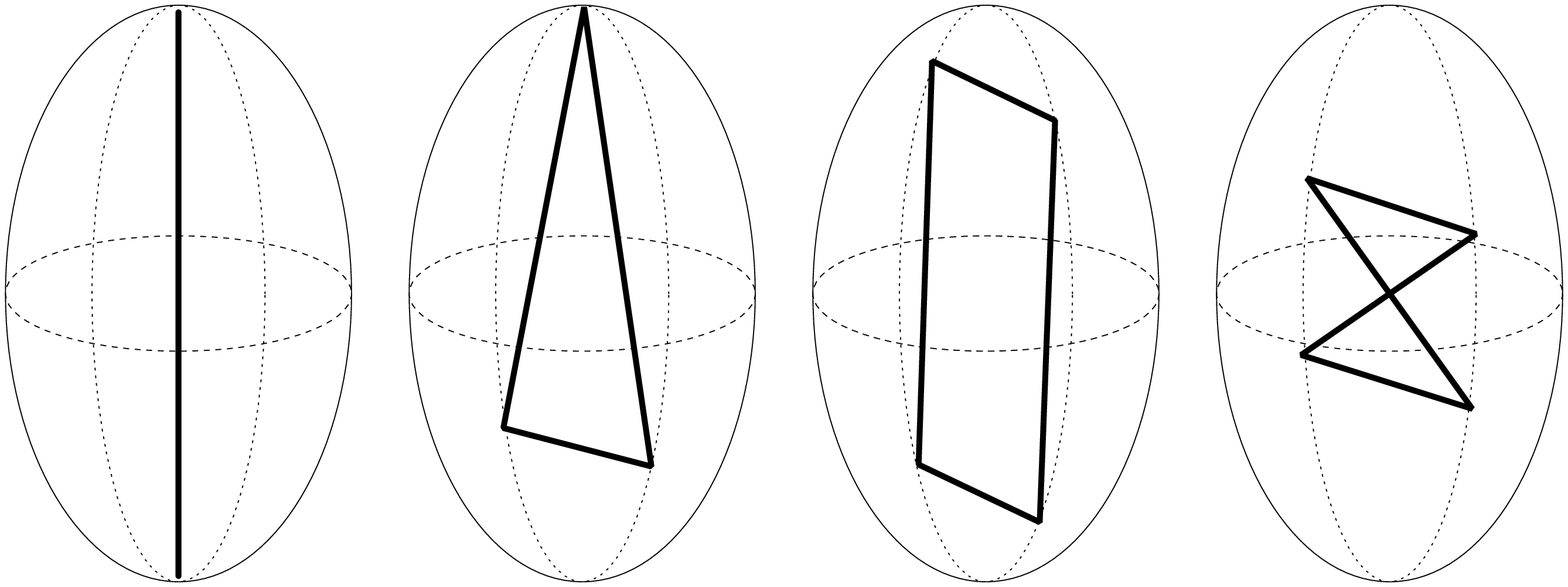}
\end{center}
\caption{POs $M(n_u,n_v,n_\varphi)$ in the spheroidal
cavity model: 2D orbits.
}
\label{fig4}
\end{figure}
\begin{figure}
\begin{center}
\includegraphics[width=.7\textwidth,clip]{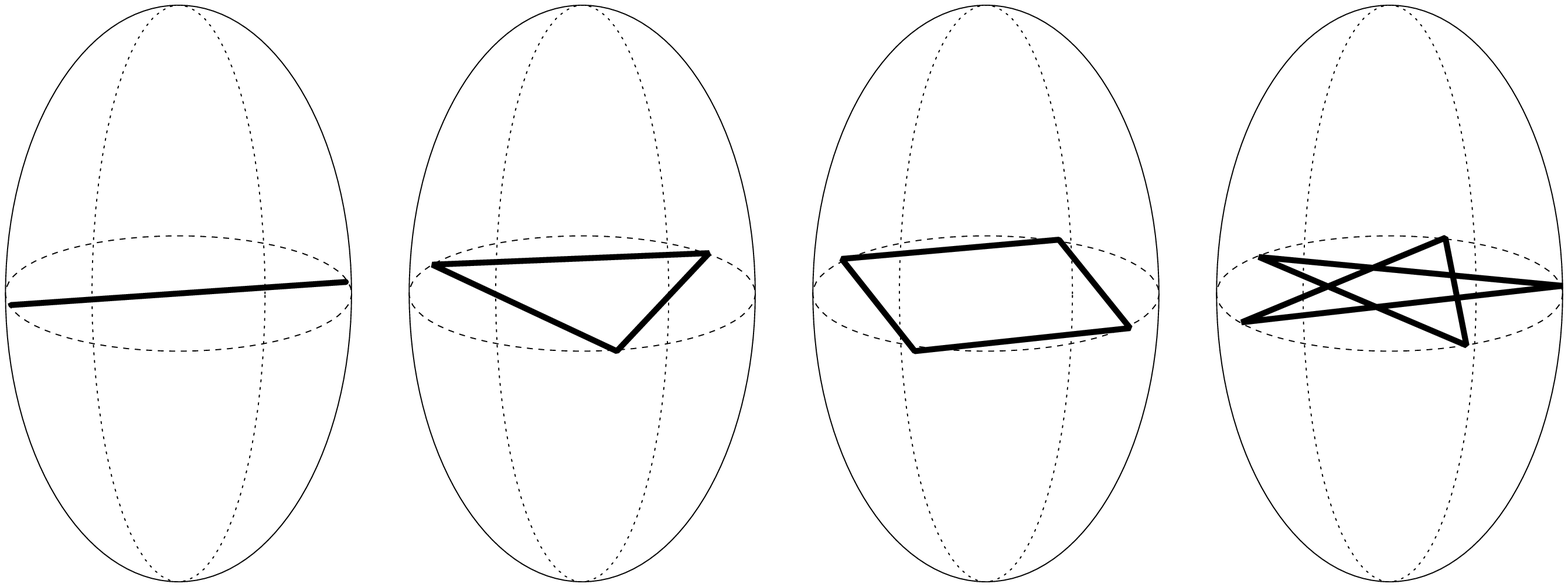}
\end{center}
\caption{ POs $M(n_u,n_v,n_\varphi)$ in the spheroidal
cavity model: EQ orbits.
} 
\label{fig5}
\end{figure}
\begin{figure}
\begin{center}
\includegraphics[width=.35\textwidth,clip]{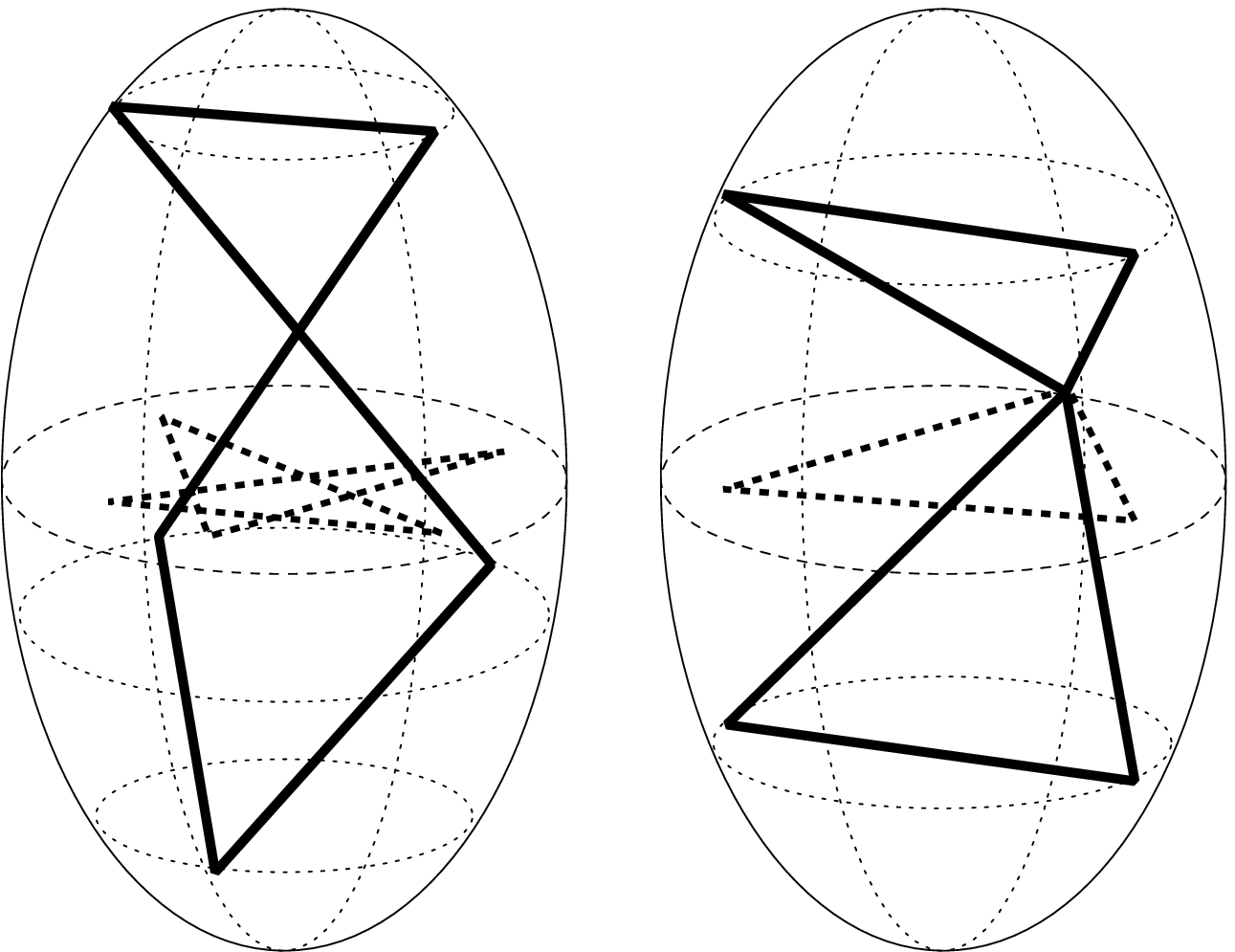}
\end{center}
\caption{ POs $M(n_u,n_v,n_\varphi)$ in the spheroidal 
cavity model: 3D orbits.
}
\label{fig6}
\end{figure}
\begin{figure}
\begin{center}
\includegraphics[width=.7\textwidth,clip]{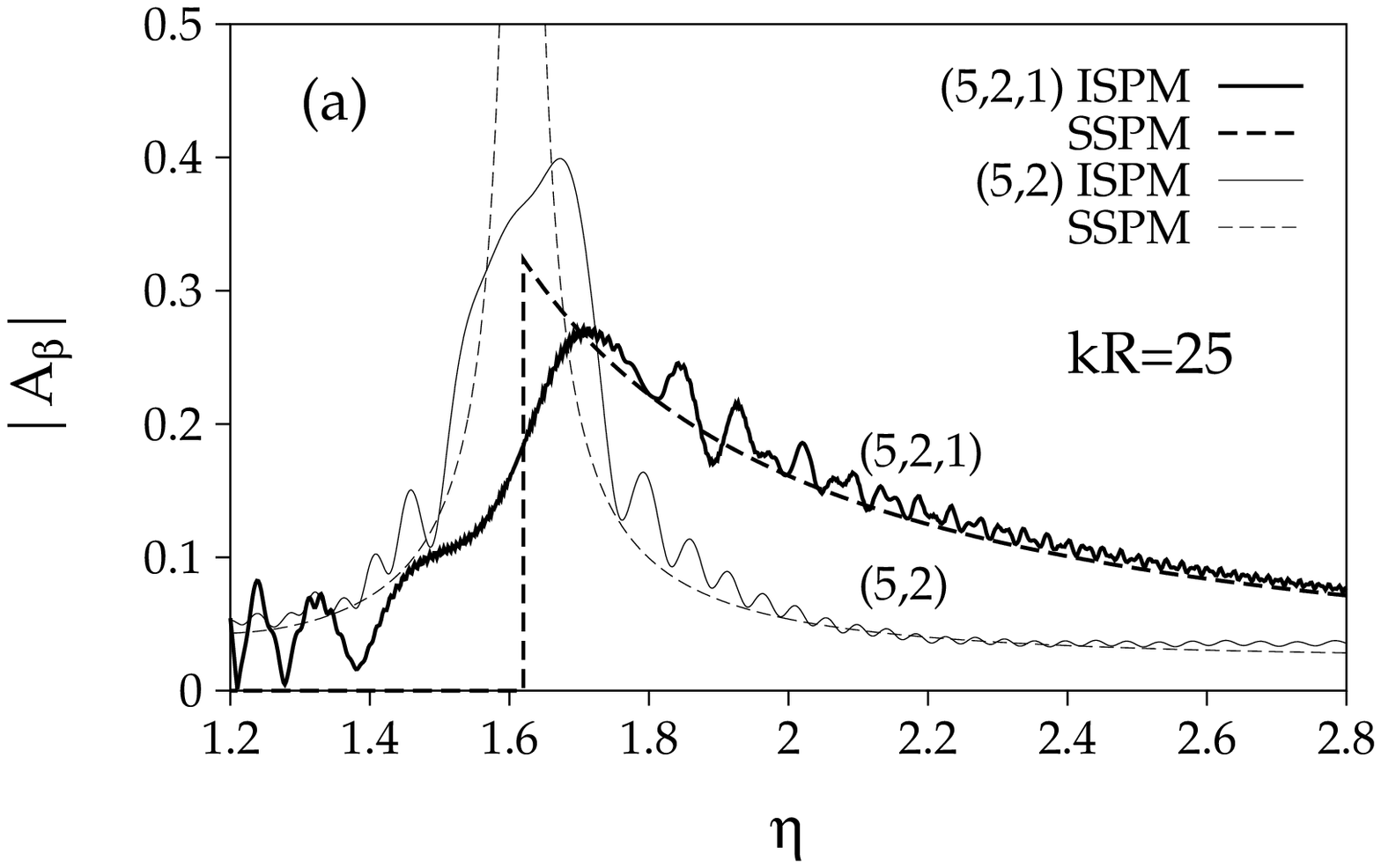}
\end{center}
\caption{ Amplitudes ${\cal A}_{5,2,1}$. Solid and dotted lines 
show the quantum and semiclassical ISPM results, respectively.
}
\label{fig11}
\end{figure}
\begin{figure}
\begin{center}
\includegraphics[width=.7\textwidth,clip]{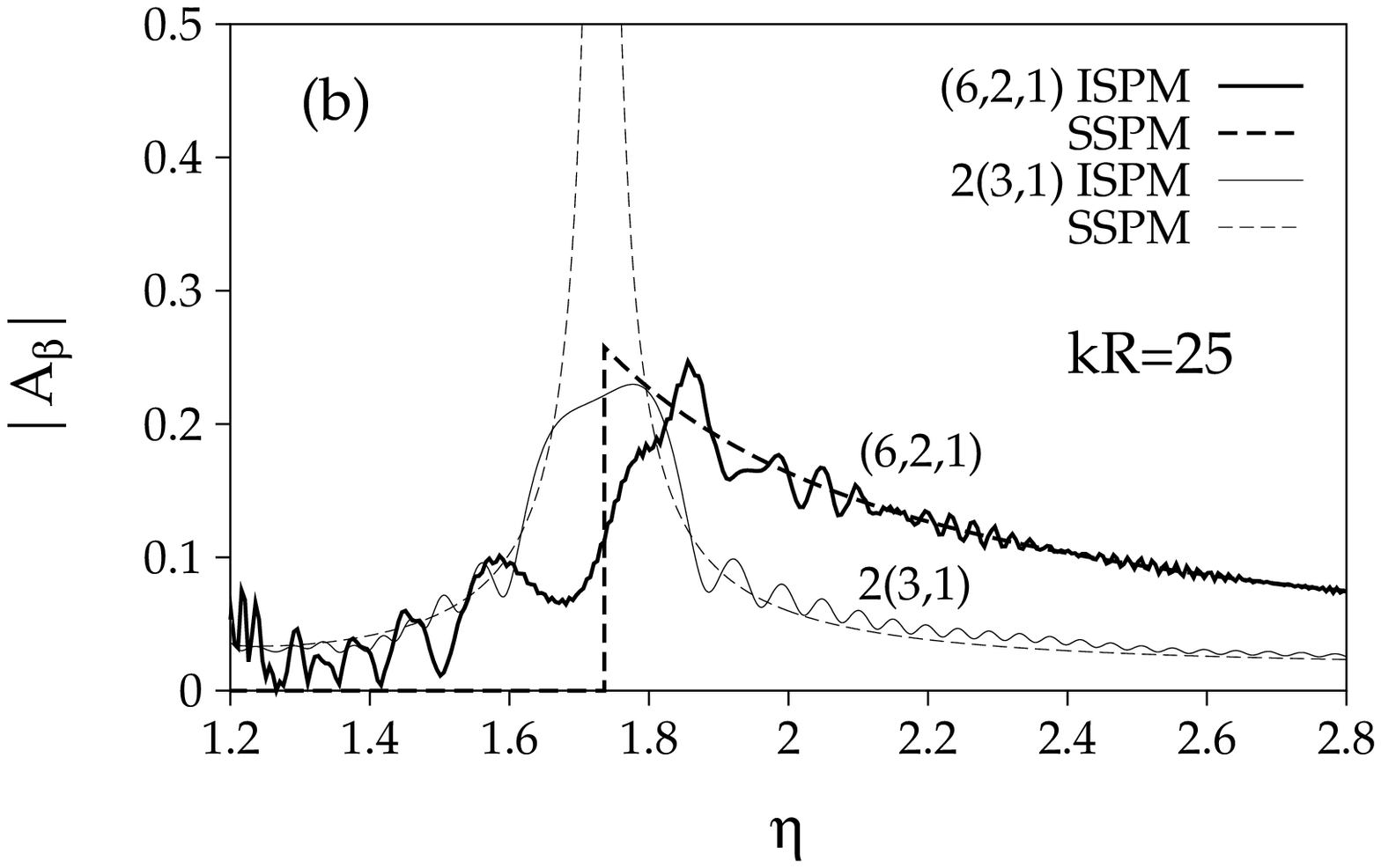}
\end{center}
\caption{ Amplitudes ${\cal A}_{6,2,1}$. Solid and dotted lines 
show the quantum and semiclassical ISPM results, respectively.
}
\label{fig12}
\end{figure}
\begin{figure}
\begin{center}
\includegraphics[width=0.9\textwidth,clip]{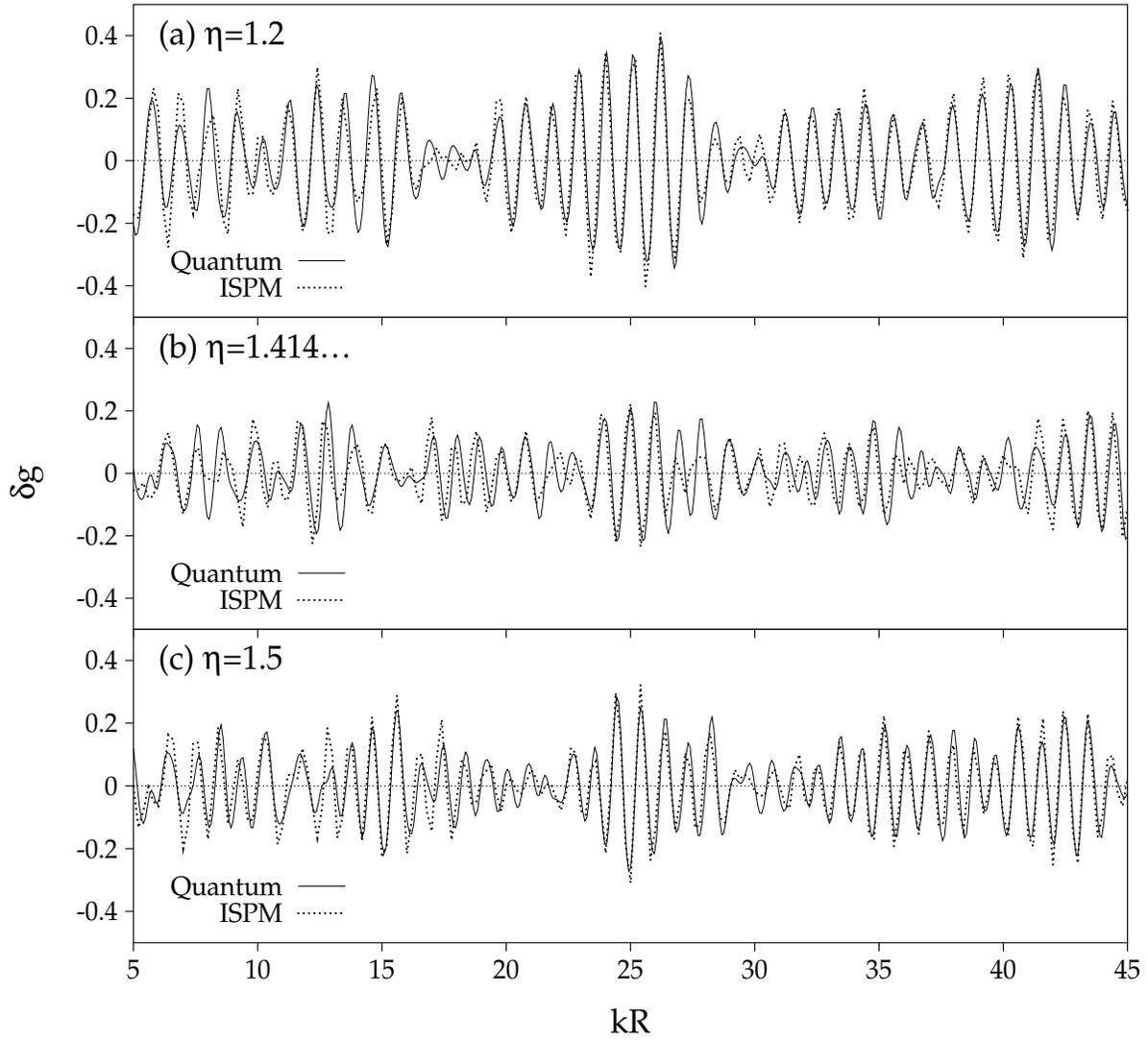}
\end{center}
\caption{ Oscillating part of level density $\delta g(k)$ in the
spheroidal cavity (in units of $2mR^2/\hbar^2$, where $R$ is the radius
of the cavity at $\eta=1$), plotted versus $kR$, for three specific
deformations $\eta$. The Gaussian averaging parameter is $\gamma=0.3/R$.
}
\label{fig13}
\end{figure}
\begin{figure}
\begin{center}
\includegraphics[width=0.9\textwidth,clip]{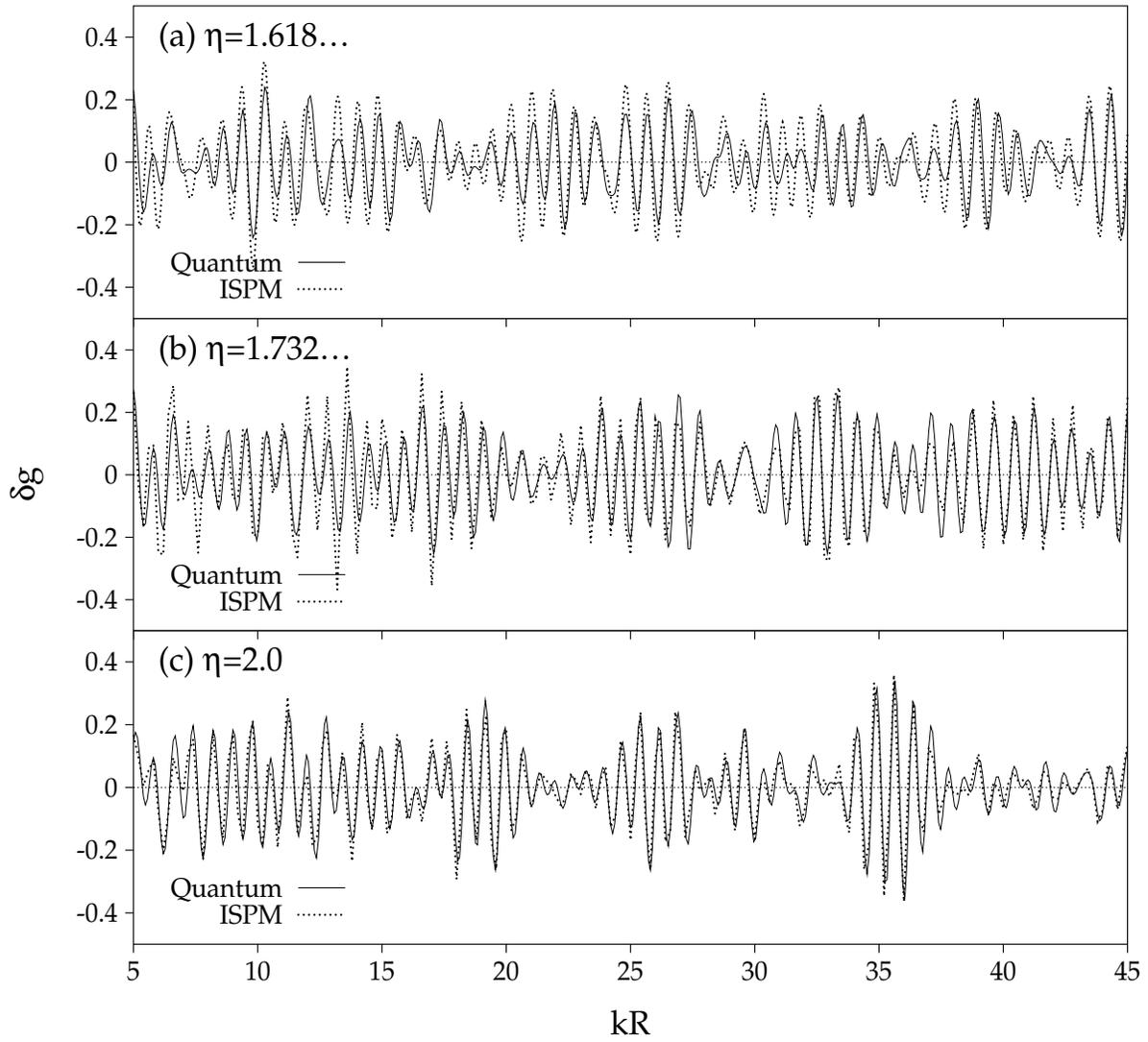}
\end{center}
\caption{ The same as in Fig.\ \ref{fig13} but for three 
deformations in the superdeformed region.
}
\label{fig14}
\end{figure}
\begin{figure}
\begin{center}
\includegraphics[width=0.9\textwidth,clip]{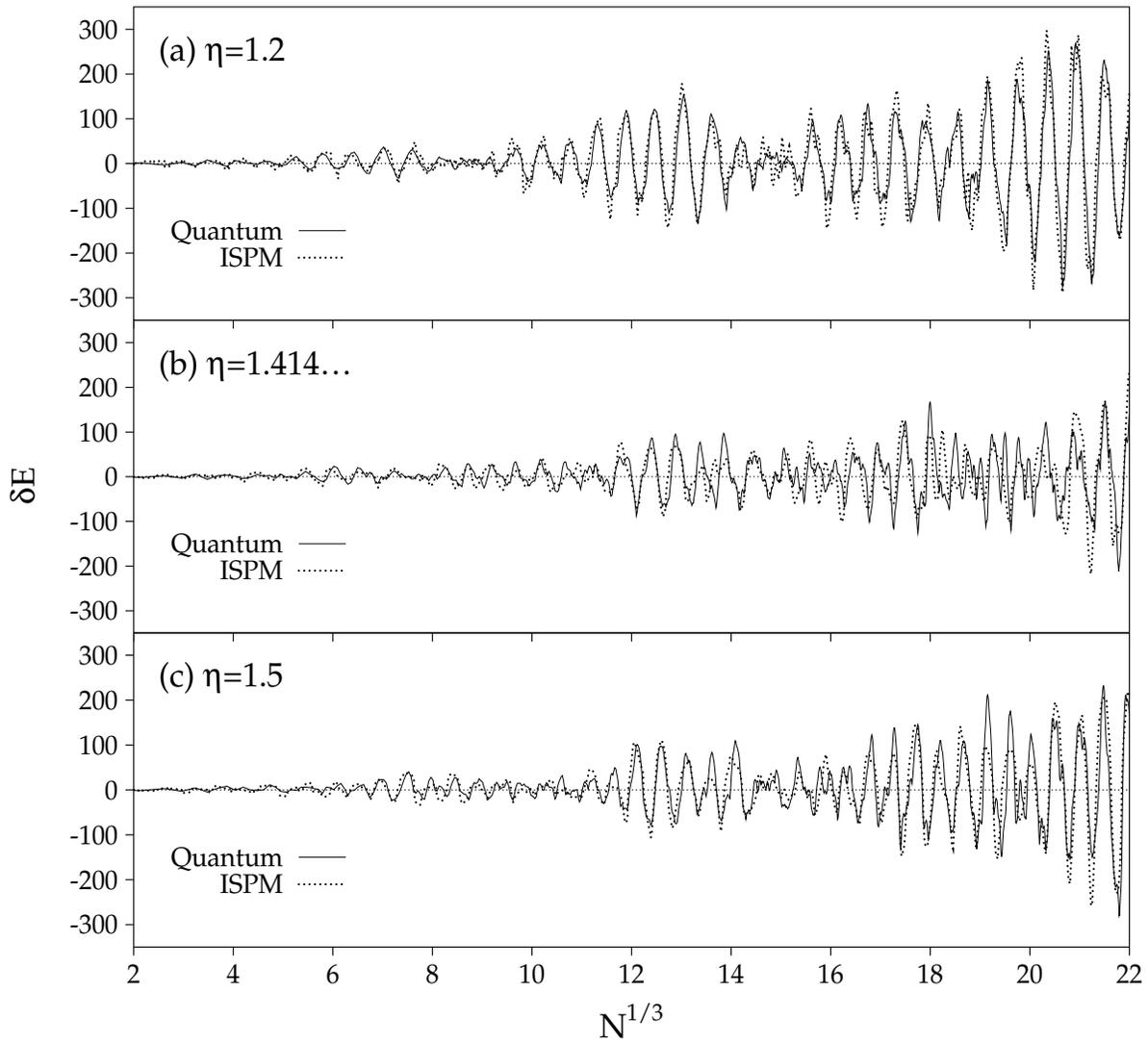}
\end{center}
\caption{Shell-correction energy $\delta E$ (in units of $\hbar^2/2mR^2$)
vs. cube root of particle number $N^{1/3}$ for the
same three deformations as in Fig.\ \ref{fig13}.
}
\label{fig15}
\end{figure}
\begin{figure}
\begin{center}
\includegraphics[width=0.9\textwidth,clip]{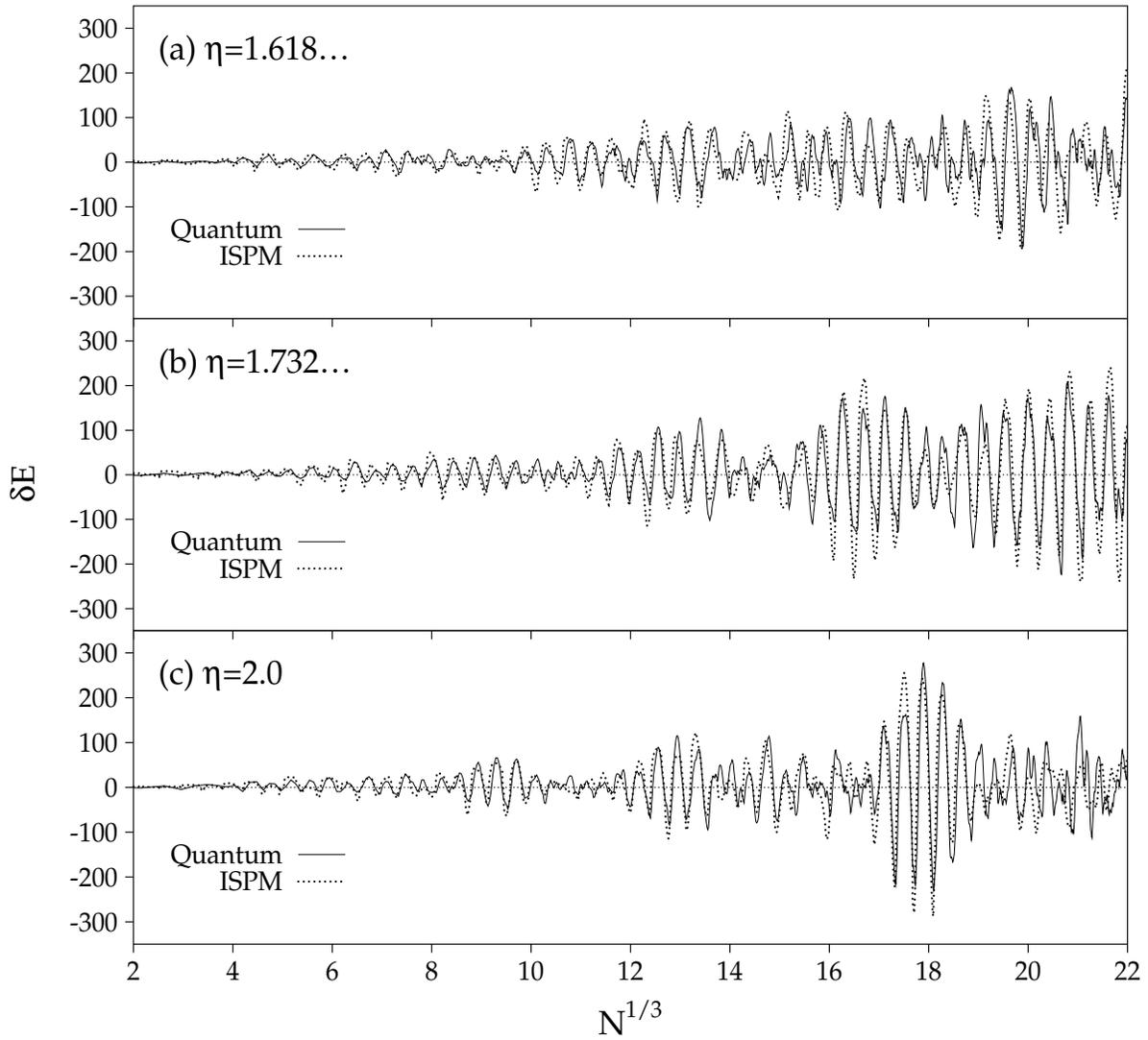}
\end{center}
\caption{The same as in Fig.\ \ref{fig15} but for the three 
deformations of Fig.\ \ref{fig14}.
}
\label{fig16}
\end{figure}
\begin{figure}
\begin{center}
\epsfig{
figure = 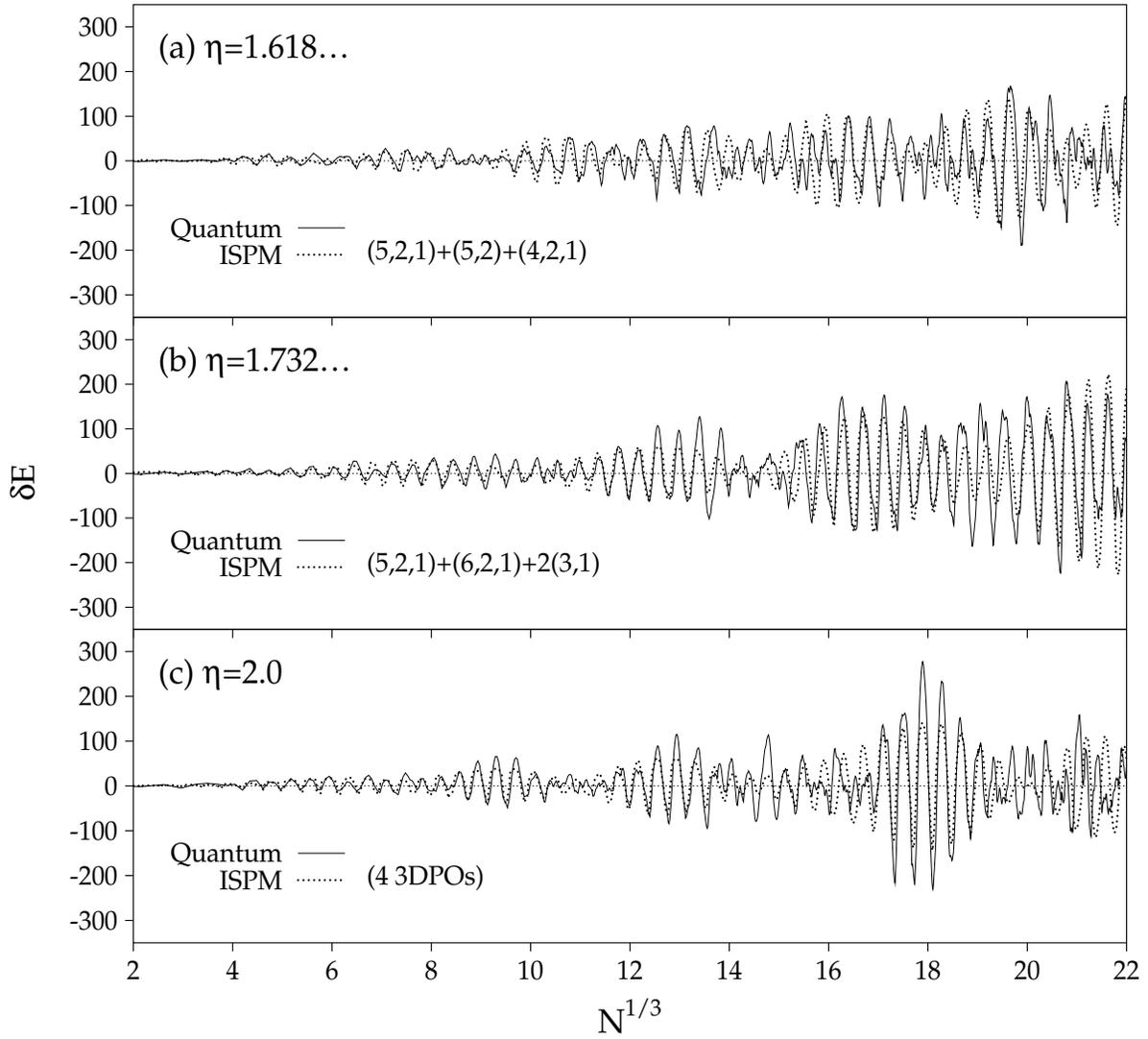,width = 0.9\textwidth,clip}
\end{center}
\caption{The same as in Fig.\ \ref{fig16} in the superdeformation 
region, but the semiclassical results (ISPM) here include only 
the bifurcated period-two and -three orbits specified near
the dotted lines ``ISPM''.
} 
\label{fig18}
\end{figure}
\begin{figure}
\begin{center}
\includegraphics[width=0.8\textwidth,clip]{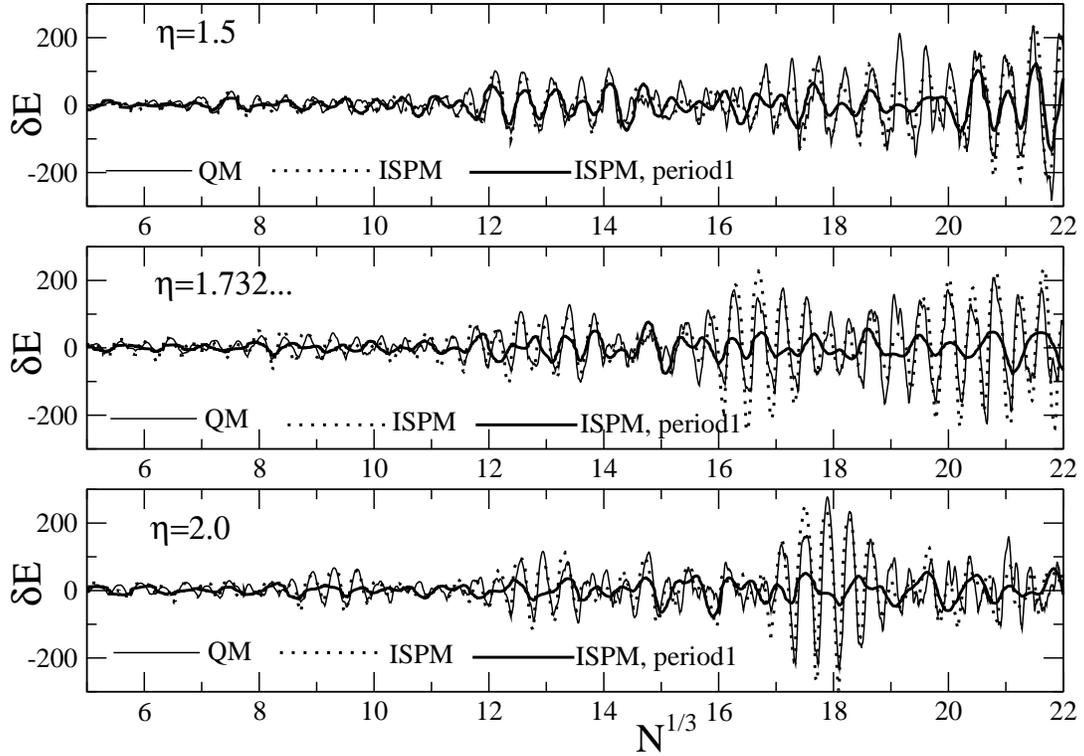}
\end{center}
\caption{Shell-correction energy $\delta E$ (in units of $\hbar^2/2mR^2$)
vs.\ cube root of particle number $N^{1/3}$ for the
characteristic deformations $\eta=1.5$, $1.732...$ and $2.0$. 
Solid lines are the quantum results, dots represent the ISPM results including
the bifurcating period-two orbits. The thick solid lines show the 
ISPM results obtained using only the shortest EQ and meridional orbits
(see text for discussion).
}
\label{figperiod1}
\end{figure}
\begin{figure}
\begin{center}
\includegraphics[width=.85\textwidth,clip]{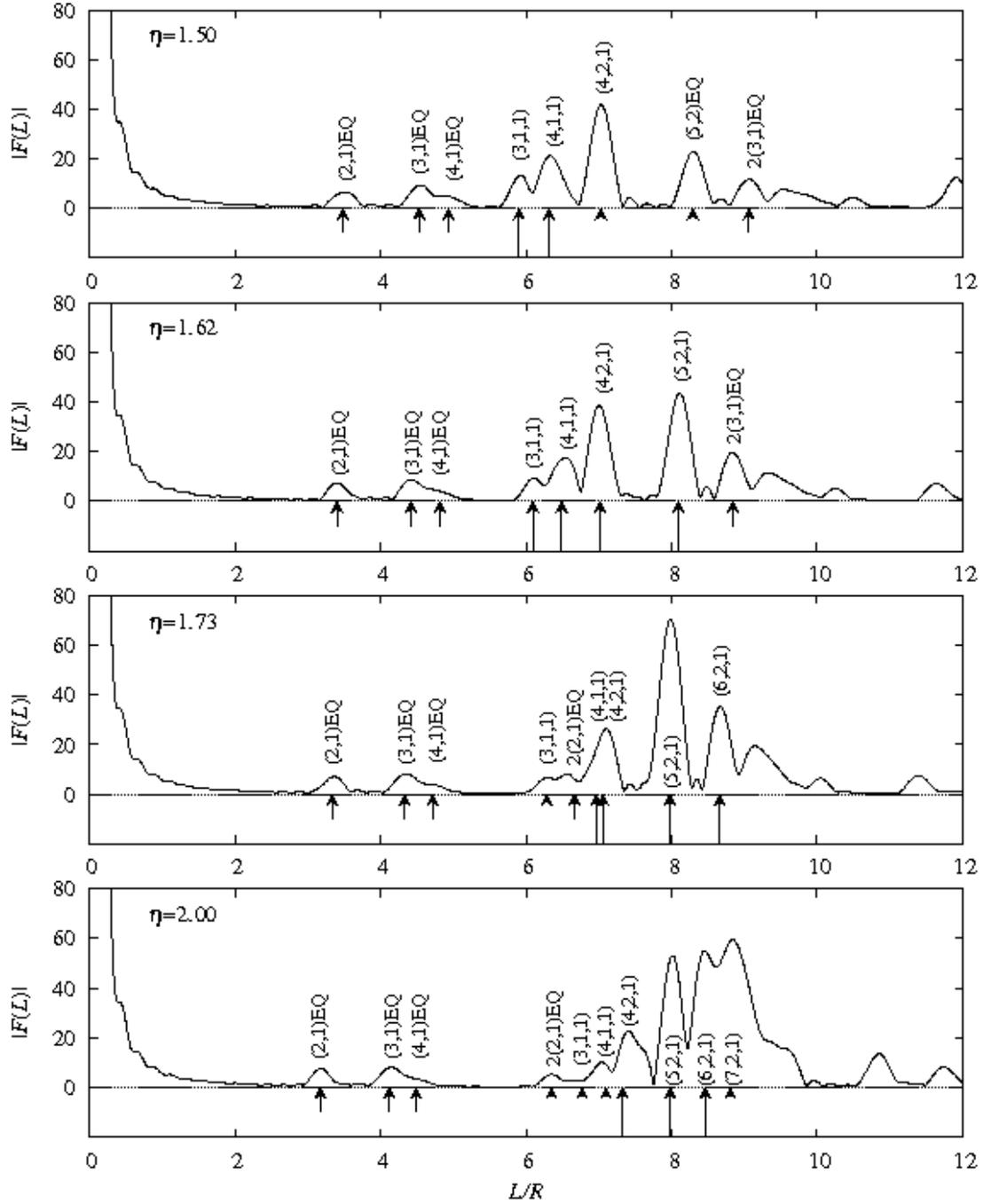}
\end{center}
\caption{Fourier transform $|F(L)|$ (\ref{Fourier_cavity}) 
of the quantum level density $g(k)$ of the spheroidal cavity, 
plotted  vs.\ length variable $L$ at the critical deformations 
$\eta=1.5$, $1.62$, $1.73$ and $2.0$.
It shows a correspondence to the lengths of classical periodic
orbits and clear dominance of longer bifurcated period-two POs 
above the shortest period-one POs: arrows near the peaks 
mark the lengths of the most important POs; some peaks at the 
bifurcation deformations show the sum of both newborn and parent 
bifurcating orbits like in Fig.\ \ref{fourspec}. 
}
\label{fourier_SD}
\end{figure}
\begin{figure}
\begin{center}
\includegraphics[width=0.9\textwidth]{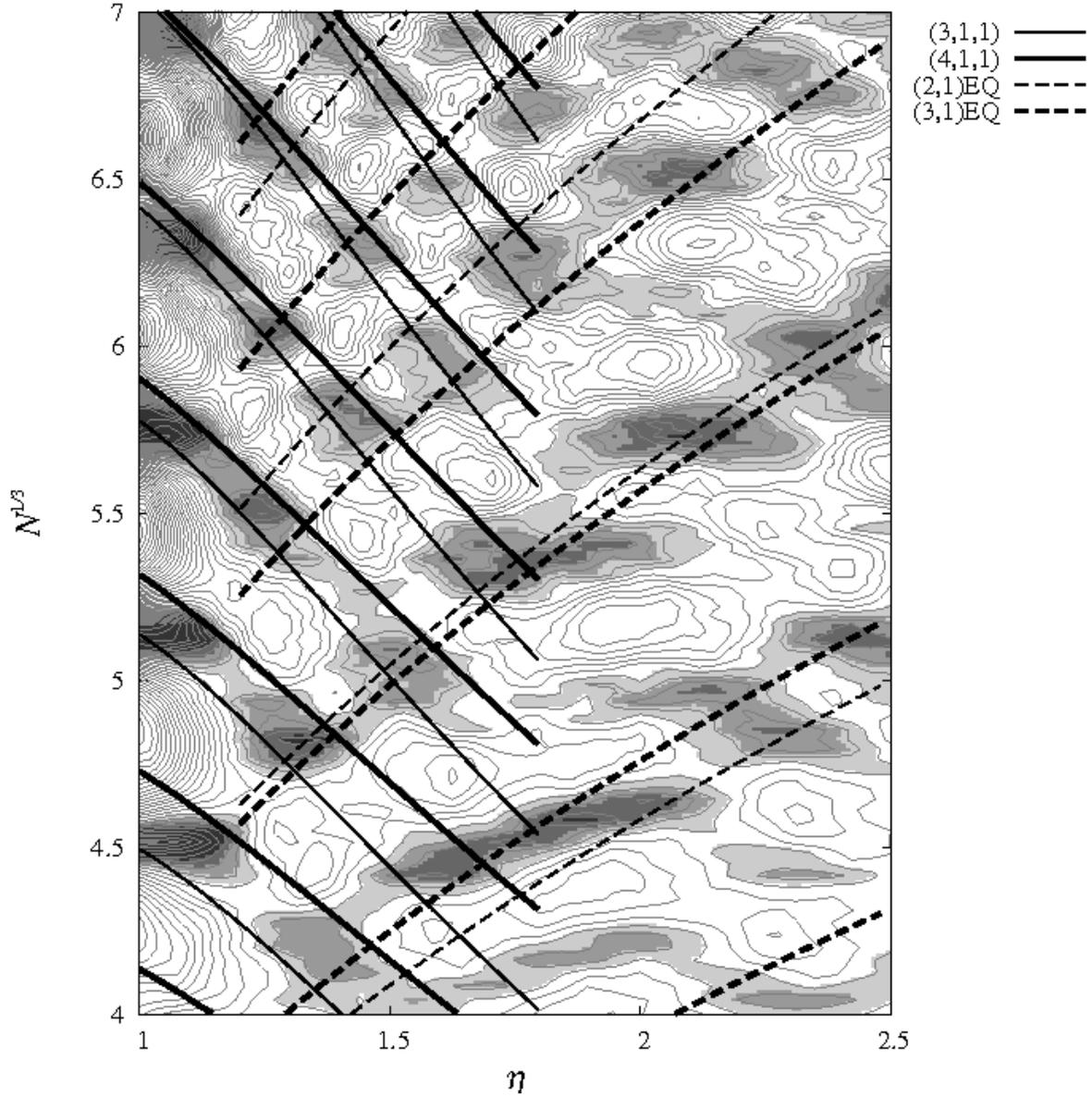}
\end{center}
\caption{ Contour map of shell correction energy $\delta E$
of the spheroidal cavity versus
deformation $\eta$ and cube-root of particle number $N^{1/3}$.
Thick and thin solid lines represent constant-action curves of the classical
shortest meridian POs; thick and thin dashed lines show the shortest EQ
POs;  all are specified on right and are dominating the periodic-orbit
sum for the gross-shell structure.
}
\label{sce_spheroida}
\end{figure}
\begin{figure}
\begin{center}
\includegraphics[width=0.9\textwidth]{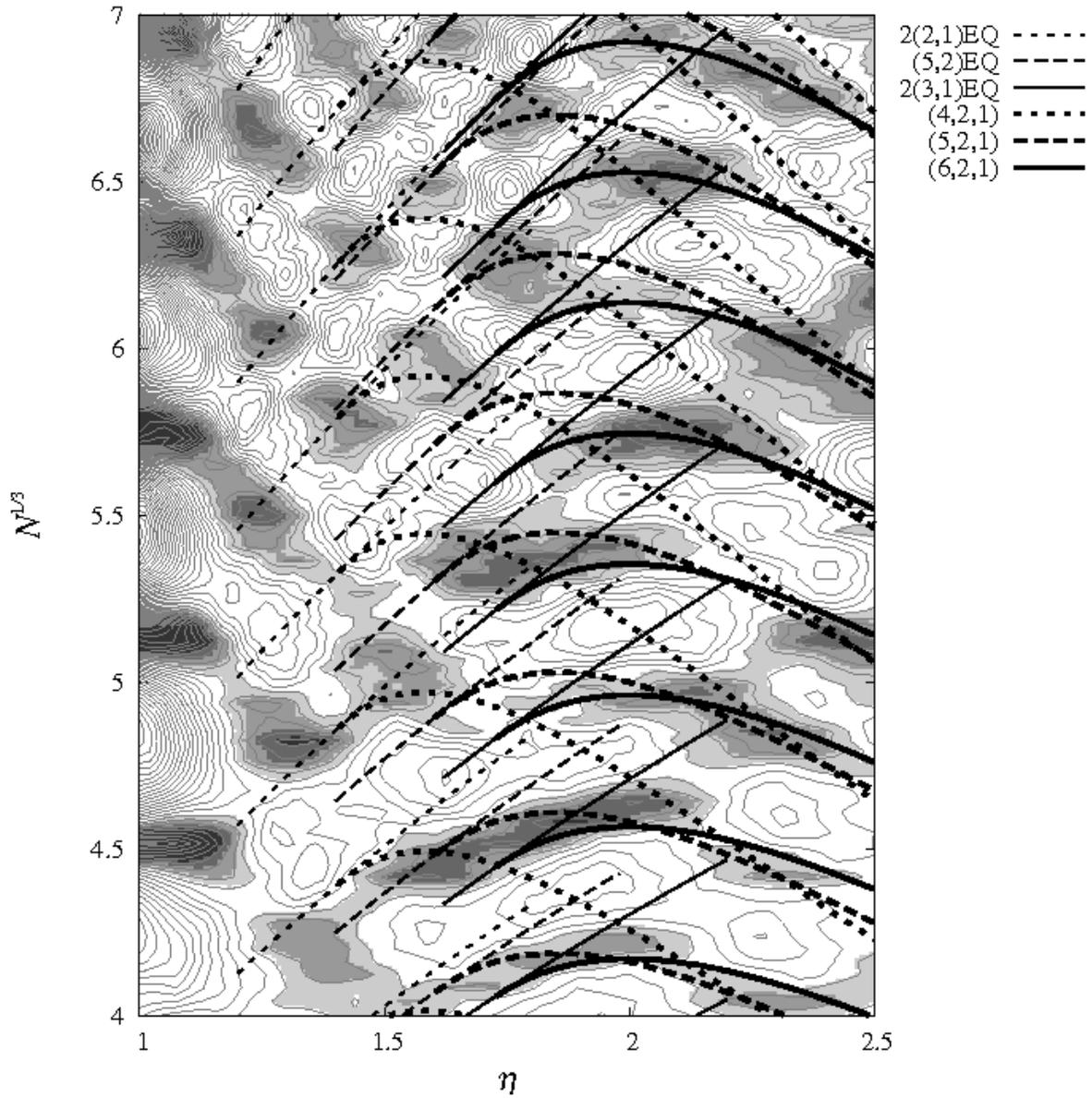}
\end{center}
\caption{ The same as in Fig.\ \ref{sce_spheroida} but with 
constant-action lines of the bifurcating POs. Three kinds of curves
(dotted, dashed and solid) are shown:
Normal and heavy dotted lines present the constant-action curves
of the bifurcating 2(2,1)EQ and newborn (4,2,1)2DH orbits; similar dashed 
lines the (5,2)EQ and (5,2,1)3D orbits, and solid lines show the 2(3,1) 
and (6,2,1)3D orbits, respectively. All newborn 2DH and 3D orbits 
dominate the fine shell structure at large enough deformations.
}
\label{sce_spheroidb}
\end{figure}
\begin{figure}
\begin{center}
\includegraphics[width=.7\textwidth,clip]{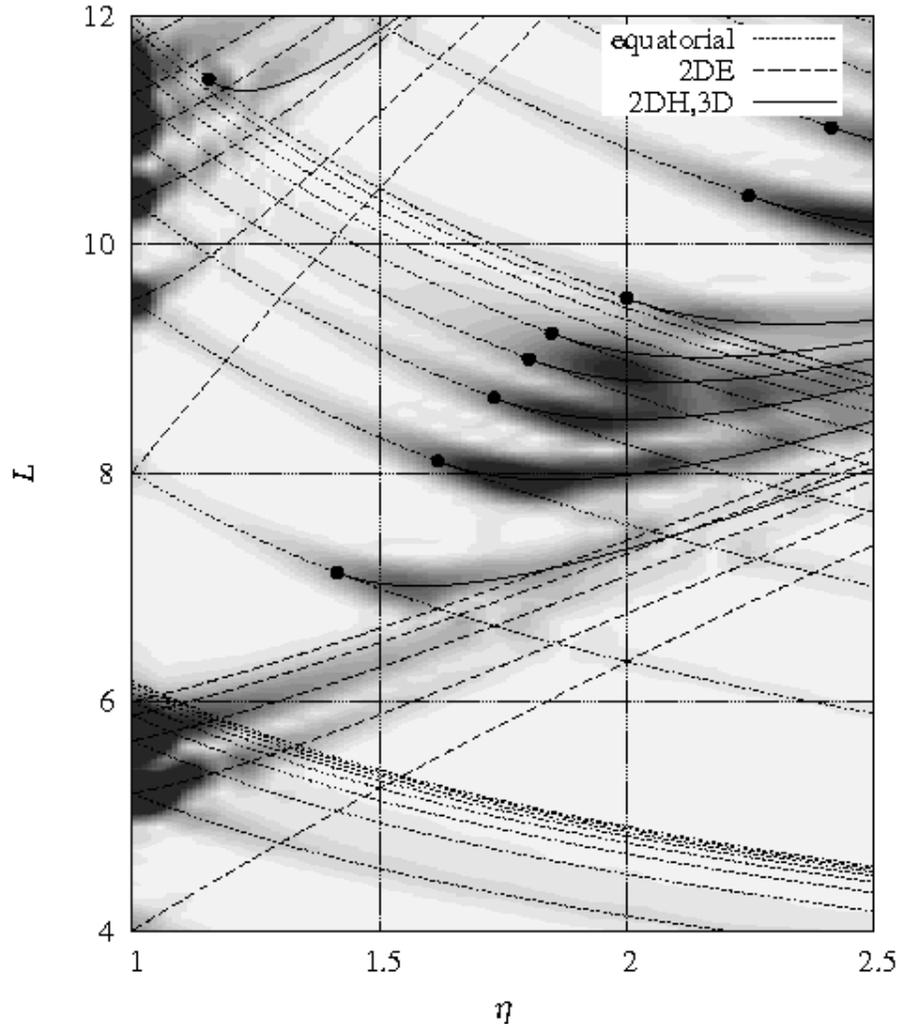}
\end{center}
\caption{The Fourier transform $|F(L)|$ (\ref{Fourier_cavity}) 
of the quantum level density $g(k)$ is plotted in grayscale.
It shows a clear correspondence to the lengths of classical periodic
orbits: dotted curves represent EQ orbits, and dashed curves stand for
meridional orbits with elliptic caustics 2DE.  Solid curves are either
meridional orbits with hyperbolic caustics 2DH or 3D orbits,
which are generated via period-multiplying bifurcations of EQ
orbits. The black dots show the bifurcation points.
}
\label{fig7}
\end{figure}

\clearpage

\begin{figure}
\begin{center}
\includegraphics[width=.6\textwidth]{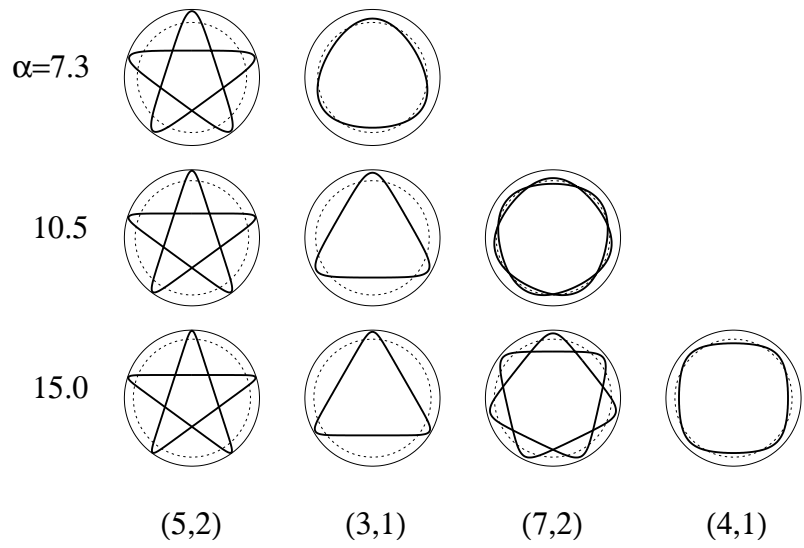}
\end{center}
\caption{\label{po_power}
Some short periodic orbits in the power-law potential 
\eq{ramodra} for several values of $\alpha$.
}
\label{orbitra}
\end{figure}

\begin{figure}
\begin{center}
\includegraphics[width=.8\textwidth,clip]{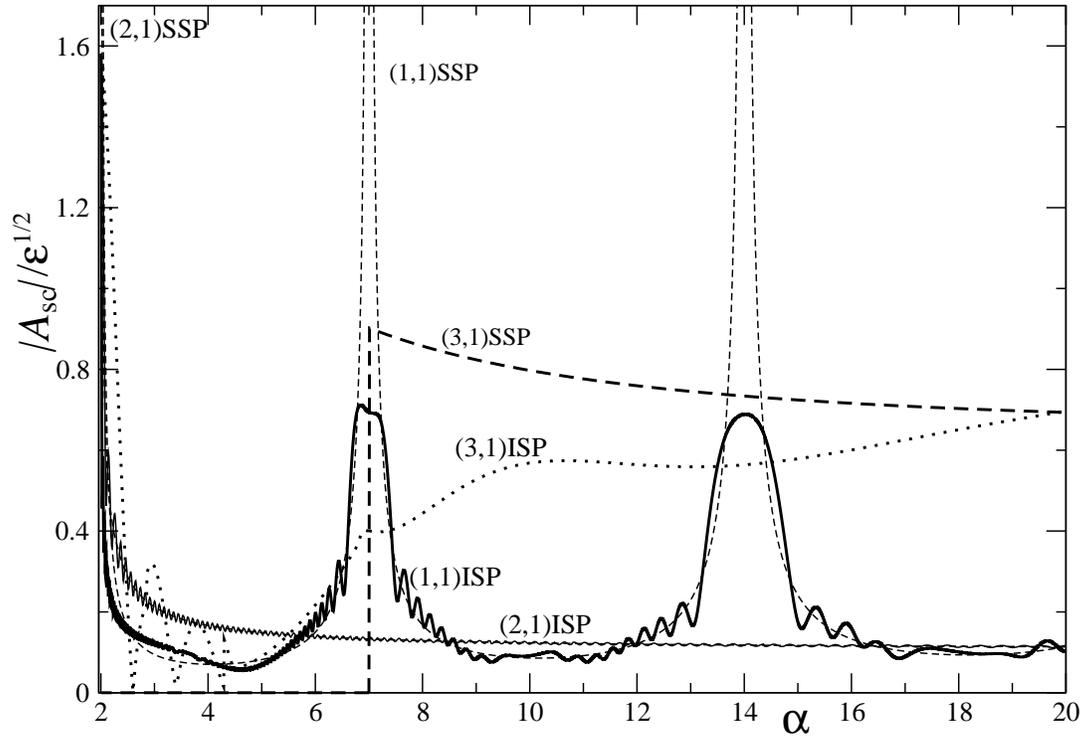}
\end{center}
\caption{ 
The scaled amplitudes $|{\it A}_{sc,po}|$ as functions of $\alpha$
for the once repeated  ($M=1$) circular [$(1,1)$, see (30)],  
 one-repeated ($M=1$) diameter $(2,1)$ and triangle PO $(3,1)$,
in units of $\varepsilon^{1/2}$ at the scaled energy $\varepsilon=10$. 
Dashed thin (circular (1,1)SSP, diameter (2,1)SSP) and thick 
(3,1)SSP lines are SSPM results, and solid (thick circular (1,1)ISP 
and thin diameter (2,1)ISP) as well as dotted (3,1)ISP lines 
are ISPM results.
} 
\label{fig23}
\end{figure}
\begin{figure}
\begin{center}
\includegraphics[width=.9\textwidth]{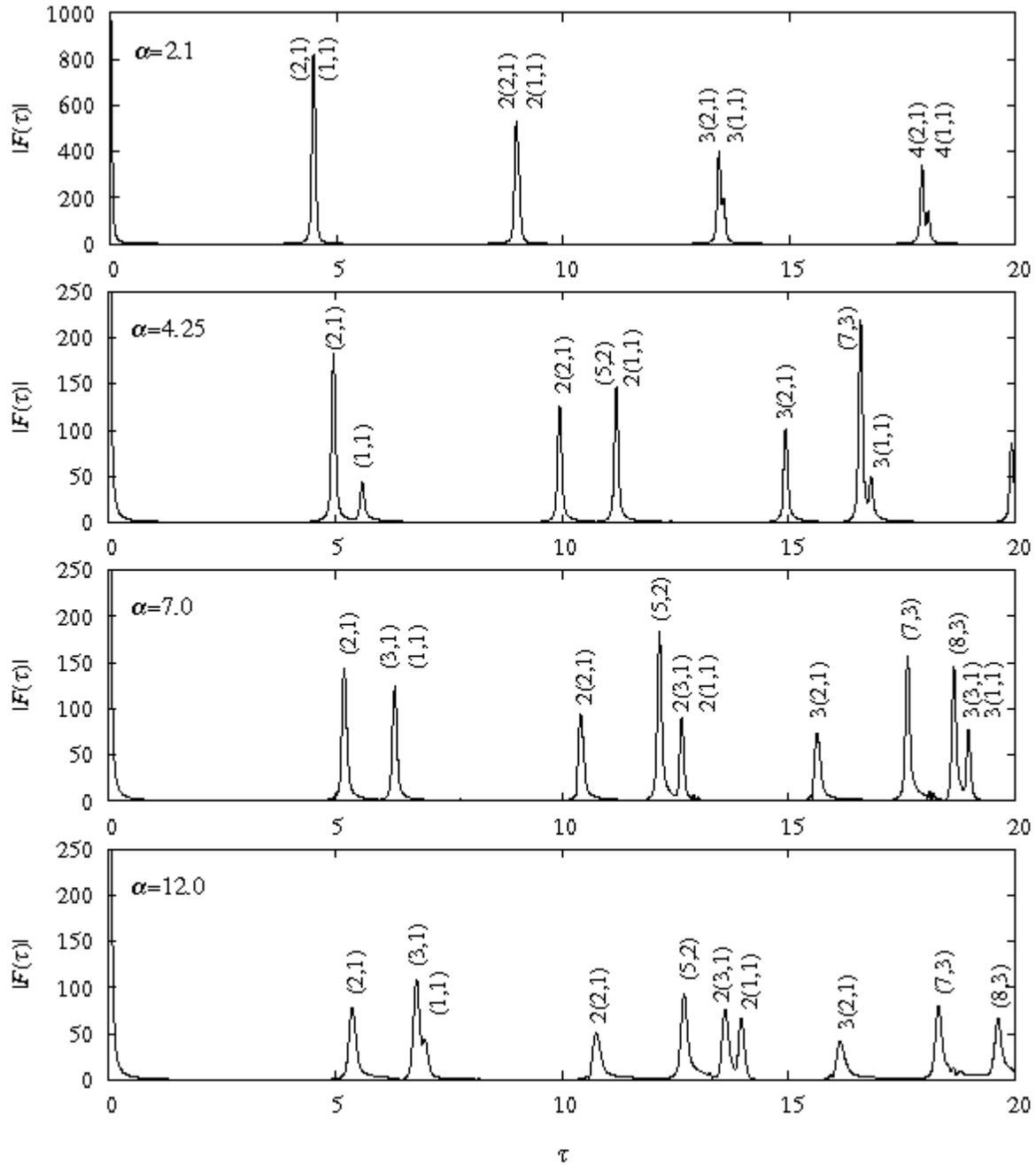}
\end{center}
\caption{Moduli of the Fourier transform $|F(\tau)|$ of 
the energy-scaled quantum level density (\ref{fourier_power}),
plotted for several values of $\alpha$.
}
\label{ffig_power}
\end{figure}
\begin{figure}
\begin{center}
\includegraphics[width=.8\textwidth,clip]{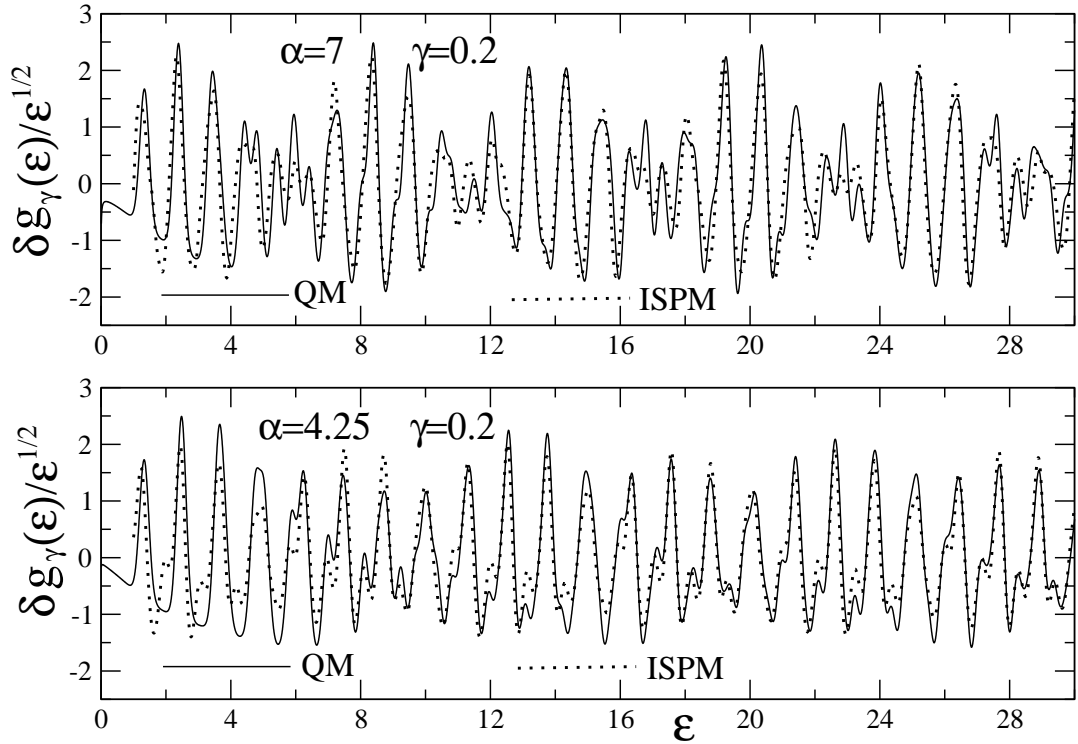}
\end{center}
\caption{ 
Oscillating part of scaled level density versus scaled energy 
$\varepsilon$ at the bifurcation points $\alpha=7$ and $4.25$ 
in units of $\varepsilon^{1/2}$. QM (solid line) is the 
quantum-mechanical result (using Strutinsky averaging with parameters 
${\widetilde \gamma}=3,\,{\cal M}=6$); ISPM (dotted line) is the semiclassical
result. In both cases, a (dimensionless) width $\gamma=0.2$ was used
in the Gaussian coarse-graining over the scaled energies.
} 
\label{fig25}
\end{figure}
\begin{figure}
\begin{center}
\includegraphics[width=.8\textwidth,clip]{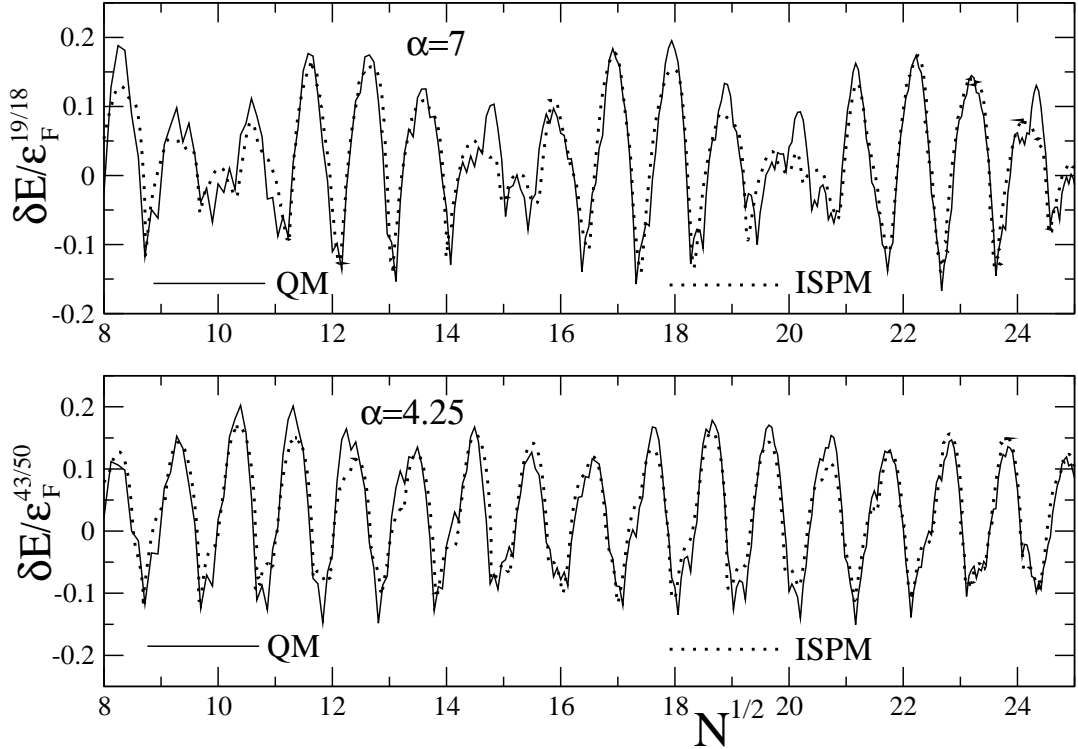}
\end{center}
\caption{
Shell-correction energy $\delta E$ 
vs.\ particle number parameter $N^{1/2}$ at $\alpha=7$ and $4.25$
in units $\varepsilon^{3(\alpha-2)/[2(\alpha+2)]}$.
QM (solid line) is the quantum-mechanical result using
the shell-correction method (Strutinsky averaging as in
Fig.\ \ref{fig25} above); ISPM (dotted line) is the semiclassical
result using a small coarse-graining width $\gamma=0.1$
in the calculation of $N(\varepsilon_F)$ by (\ref{partnum2}) 
and (\ref{totdensc}). 
} 
\label{fig26}
\end{figure}

\newpage 
~~~~~~~~

\end{document}